\documentclass[final,1p,times,authoryear]{elsarticle}

\usepackage{amssymb}
\usepackage{amsmath}
\usepackage{amsthm}
\usepackage{amsfonts,amssymb}

\usepackage{hyperref}
\hypersetup{colorlinks=true,linkcolor=blue,urlcolor=blue,citecolor=blue,bookmarks=false}
\usepackage{bm,array,setspace,booktabs}
\usepackage[labelfont=bf]{caption,subcaption}
\usepackage{ragged2e}
\usepackage{placeins}

\usepackage{longtable}
\usepackage{colortbl}
\usepackage[table,dvipsnames]{xcolor}

\usepackage[sectionbib]{bibunits}
\usepackage{natbib}
\defaultbibliographystyle{elsarticle-harv}
\defaultbibliography{bibliography_2026_03_19}

\newcommand{\iid}{\stackrel{\mathrm{iid}}{\sim}}

\newcommand{\expec}{\mathbb E}
\newcommand{\var}{\mathrm{var}}
\newcommand{\cov}{\mathrm{cov}}

\DeclareMathOperator*{\normal}{\mathcal{N}}

\newcommand\bmf{\bm f}

\newcolumntype{H}{>{\setbox0=\hbox\bgroup}c<{\egroup}@{}}
\newcolumntype{C}{>{$}c<{$}}
\newcolumntype{R}{>{$}r<{$}}
\newcolumntype{L}{>{$}l<{$}}

\usepackage{geometry}
\usepackage{alphalph}

\begin{document}

\begin{frontmatter}

\title{The Co-Pricing Factor Zoo\tnoteref{t1,t2}}
 \tnotetext[t1]{Nikolai Roussanov was the editor for this article. We are thankful for comments and suggestions from  the editor, an anonymous referee, Svetlana Bryzgalova, Mikhail Chernov, Magnus Dahlquist, Jiantao Huang, Byounghyun Jeon, Ian Martin, Mamdouh Medhat, Yoshio Nozawa, Andrew Patton, Paola Pederzoli, Lukas Schmid, Patrick Weiss, Paolo Zaffaroni, Irina Zviadadze, and seminar and conference participants at BTM KAIST College of Business, Cancun Derivatives and Asset Pricing Conference 2024, Chinese University of Hong Kong, EFA Bratislava, ESSFM Asset Pricing Gerzensee 2024, FIRS Berlin, 4th Frontiers of Factor Investing Conference Lancaster, Goethe University, HEC Lausanne, INQUIRE Europe Spring Seminar Brussels,  London School of Economics, NTU Taipei, Machine Learning and Finance Conference Oxford,  MFS Oslo,  SFS Cavalcade Atlanta, SFS Cavalcade Asia-Pacific Seoul, Tinbergen Institute, University of Essex, University of Surrey, WFA Honolulu, and the XXIV Brazilian Finance Meeting Curitiba. We gratefully acknowledge financial support from  INQUIRE Europe.
}
 \tnotetext[t2]{The companion website to this paper, \href{https://openbondassetpricing.com/}{openbondassetpricing.com}, contains updated corporate bond factor data. Full replication code is available on GitHub (\href{https://github.com/Alexander-M-Dickerson/co-pricing-factor-zoo}{\texttt{co-pricing-factor-zoo}}). An earlier version of this paper circulated under the title ``The Corporate Bond Factor Zoo.''}

\author[1]{Alexander Dickerson}
\ead{alexander.dickerson1@unsw.edu.au}

\author[2,4]{Christian Julliard}
\ead{c.julliard@lse.ac.uk}

\author[3]{Philippe Mueller\corref{cor1}}
\ead{philippe.mueller@wbs.ac.uk}

\cortext[cor1]{Corresponding author}

\affiliation[1]{organization={School of Banking \& Finance, The University of New South Wales},
                city={Sydney},
                citysep={},
                postcode={2052},
                state={NSW},
                country={Australia}}

\affiliation[2]{organization={London School of Economics and Political Science},
                addressline={Houghton St},
                city={London},
                citysep={},
                postcode={WC2A 2AE},
                country={United Kingdom}}

\affiliation[3]{organization={Warwick Business School, The University of Warwick},
                addressline={Scarman Rd.},
                city={Coventry},
                citysep={},
                postcode={CV4 7AL},
                country={United Kingdom}}

\affiliation[4]{organization={Centre for Economic Policy Research},
                addressline={2 Coldbath Square},
                city={London},
                citysep={},
                postcode={EC1R 5HL},
                country={United Kingdom}}

\begin{abstract}
We analyze 18 quadrillion models for the joint pricing of corporate bond and stock returns. Strikingly, we find that equity and nontradable factors alone suffice to explain corporate bond risk premia once their Treasury term structure risk is accounted for, rendering the extensive bond factor literature largely redundant for this purpose. While only a handful of factors, behavioral and nontradable, are likely robust sources of priced risk, the true latent stochastic discount factor is \emph{dense} in the space of observable factors. Consequently, a Bayesian Model Averaging Stochastic Discount Factor explains risk premia better than all low-dimensional models, in- and out-of-sample, by optimally aggregating dozens of factors that serve as noisy proxies for common underlying risks, yielding an out-of-sample Sharpe ratio of 1.5 to 1.8. This SDF, as well as its conditional mean and volatility, are persistent, track the business cycle and times of heightened economic uncertainty, and predict future asset returns.
\end{abstract}

\begin{keyword}
Bond-stock co-pricing \sep Corporate bonds \sep Factor zoo \sep Factor models \sep Bayesian methods \sep Macro-finance \sep Asset pricing.
\JEL G10 \sep G12 \sep G40 \sep C12 \sep C13 \sep C52.
\end{keyword}

\end{frontmatter}

\begin{bibunit}
\begin{flushright}
{\footnotesize 
\textit{
Wherever there is risk, it must be compensated to the lender by a higher premium or interest.\\
}
--- J. R. McCullough (1830, pp. 508--9) \nocite{McCullough_1830}
}
\end{flushright}

\vspace{-.2cm}

\noindent In their seminal paper, \cite{FamaFrench_1993} set themselves to ``examine whether variables that are important in bond returns help to explain stock returns, and vice versa.''  Thirty years later, the equity literature has produced its own, independent, `factor zoo' (\cite{Cochrane_2011_Pres}), while the corporate bond literature has effectively returned to square one with \citet*{DickersonMuellerRobotti_2023} showing that there is no satisfactory (observable) factor model for that asset class.\footnote{More precisely, they document that all low dimensional linear factor models in the previous literature add little spanning to a simple \emph{bond} version of the Capital Asset Pricing Model, the CAPMB. At the same time, they show that the CAPMB is in itself an unsatisfactory pricing model.} 
Hence, to date, a model for the \emph{joint} pricing of corporate bonds and stocks has escaped discovery---we fill this gap.

Generalizing recent methodological advances in Bayesian econometrics (\citet*{BryzgalovaHuangJulliard_2023}) to handle heterogeneous asset classes, we comprehensively analyze all observable factors and models proposed to date in the bond and equity literature. Our method allows us to not only study models or factors in isolation, but also consider all of their possible combinations, resulting in over 18 quadrillion models stemming from the joint zoo of corporate bond and stock factors. And we do so while relaxing the cornerstone assumptions of previous studies:  the existence of a unique, low-dimensional, correctly specified and well-identified factor model. 

Ultimately, this allows us to pinpoint the robust sources of priced risk in both markets, and a novel benchmark Stochastic Discount Factor (SDF) that prices both asset classes, significantly better than all existing models, both in- and out-of-sample. 
Remarkably, our analysis reveals that once corporate bonds' Treasury term structure risk is accounted for, stock and nontradable factors alone suffice to explain corporate bond risk premia---rendering the extensive bond factor literature largely redundant for this purpose.

First, we find that the `true' latent SDF of bonds and stocks is \emph{dense} in the space of observable bond and stock factors---literally dozens of factors, both tradable and nontradable, are necessary to span the risks driving asset prices. 
Yet, the SDF-implied maximum Sharpe ratio is not excessive, indicating that, as we confirm in our analysis, multiple bond and stock factors proxy for common sources of fundamental risk. 
Importantly, density of the SDF implies that the sparse models considered in the previous literature are affected by severe misspecification and, as we show, rejected by the data and outperformed by the most likely SDF components that we identify.

Second, a Bayesian Model Averaging Stochastic Discount Factor (BMA-SDF) over the space of all possible models (including bond, stock, and nontradable factors) explains (jointly and separately) corporate bond and equity risk premia better than all existing models and most likely factors, both in- and out-of-sample. 
Moreover, the BMA-SDF's conditional mean and volatility---hence, the implied conditional Sharpe ratio achievable in the economy---have clear business cycle patterns. In particular, the volatility of the SDF increases sharply at the onset of recessions and at times of heightened economic uncertainty. That is, the estimated SDF behaves as one would expect from the intertemporal marginal rate of substitution of an agent exposed to the risks arising from general economic conditions and market turmoil. 

Third, the predictability of the first and second moments of the SDF suggests time-varying risk premia in the economy and predictability of asset returns with lagged SDF information. We verify this by running predictive regressions of future asset returns on the conditional variance of the BMA-SDF, alone and interacted with the conditional mean of the SDF, as implied by the \cite{HansenJagannathan_1991} representation of the conditional SDF. We not only find that lagged SDF information is highly significant in predicting future asset returns, but also that the amount of explained time series variation in monthly and annual returns is much larger than what is achievable with canonical predictors. This result is remarkable for two reasons. First, the BMA-SDF is \emph{not} by construction geared toward predicting future returns: it is instead identified only under the restriction that a valid SDF should explain the cross-section of risk premia---not the time series of returns. Second, it offers an important validation of our estimation of the SDF: if risk premia are time-varying, future returns should be predictable with lagged SDF information, and that is exactly what our BMA-SDF delivers.

Fourth, we show theoretically that, to construct a tradable portfolio that captures the SDF-implied maximum Sharpe ratio achievable in the economy, one should focus on the posterior expectation of the market prices of risk of \emph{all} factors, rather than on the factors' posterior probabilities (or some ancillary selection statistic), which have been the focus of the previous literature. Such an approach can correctly recover the pricing of risk even if the observed factors are only noisy proxies of the true, yet latent, sources of risk priced in the market. \emph{In the data}, this yields a trading strategy with a time-series out-of-sample annualized Sharpe ratio of 1.5 to 1.8 (despite only yearly rebalancing) in an evaluation period (July 2004 to December 2022) that spans both the Global Financial Crisis and the COVID pandemic.

Fifth, we shed light on which factors, and which types of risk, are reflected in the cross-section of bond and equity risk premia. We find that only a handful of factors should be in the SDF with high probability. In particular, two factors meant to capture the bond and stock post-earnings announcement drift anomalies, PEADB and PEAD, respectively, are very likely sources of priced risk in the joint cross-section of bond and stock returns.\footnote{The post-earnings announcement drift phenomenon is the observation, first documented in equity markets, whereby firms experiencing positive earnings surprises subsequently earn higher returns than those with negative earnings surprises. See, e.g., \citet{HirshleiferTeoh_2003}, \citet{DellaVignaPollet_2009}, \citet{HirshleiferLimTeoh_2011} and \citet{NozawaQiuXiong_2025} for the microfoundations of this phenomenon.} In addition to these two behavioral sources of risk, the other most likely components of the SDF are all nontradable in nature, and are a proxy for the slope of the Treasury yield curve (YSP), the AAA/BAA yield spread (CREDIT), and the idiosyncratic equity volatility (the IVOL of \cite{CampbellTaksler_2003}). As we show, these factors alone are enough to price the cross-section of bonds and stocks better than canonical observable factor models. Nevertheless, the importance of \emph{individual} factors should not be overstated. Even excluding the most likely factors when constructing it, the BMA-SDF strongly outperforms these individual factors and \emph{all} low dimensional factor models---from the celebrated \cite{FamaFrench_1993} model to the latest arrival in the zoo (\cite{DickNielsenFeldhuetterPedersenStolborg_2024}). This superior performance occurs because the true latent SDF is dense and demands large compensations for risks that are not fully spanned by just a handful of individual observable factors. Furthermore, we find that both discount rate and cash-flow news are sources of priced risk, and yield sizeable contributions (albeit larger for the former) to the Sharpe ratio of the latent SDF.

Sixth, we demonstrate that a portion of corporate bond risk premia serves as compensation for their implicit Treasury term structure risk. Once this component is removed, the factors proposed in the tradable bond factor zoo have very little residual information content for characterizing the SDF: in this case, a BMA-SDF constructed only with stock and nontradable factors can explain the joint cross-section of bonds and stocks as well as our full BMA-SDF. This finding extends and explains the result in \citet{vanBinsbergenNozawaSchwert_2025}, who show that  once corporate bond returns are adjusted for duration risk, the equity CAPM has higher explanatory power for bond risk premia than benchmark bond models. Furthermore, we show that the empirical success of the bond factor zoo in the previous literature is largely driven by its ability to price the Treasury term structure risk---a component of bond risk premia that tradable stock factors do not capture.

Finally, we conduct extensive robustness checks. Most notably, we show that: (i) altering the priors regarding the relative importance of bond versus stock factors, or equivalently a potential `alpha mismeasurement' phenomenon in bond market data, has only a limited effect on the posterior probabilities of the factors and the pricing performance of the BMA-SDF;
(ii) a BMA-SDF estimated with a prior that imposes sparsity---overwhelmingly the focus of the previous literature---performs worse than our baseline BMA-SDF, yet still improves upon competing models;
(iii) as our theoretical results imply, \emph{removing} the most likely factors from the estimation---a challenging test for the method---leads to only minor deterioration in the performance of the BMA-SDF in- and out-of-sample;
(iv) all findings remain materially unchanged across \emph{hundreds} of sets of corporate bond and stock in-sample test assets---we identify a similar set of most likely factors, consistent market prices of risk, and stable in-sample asset pricing performance;
(v) out-of-sample, the pricing performance of the BMA-SDF is superior across \emph{millions} of alternative cross-sections of stocks and bonds;
(vi) lastly, the results are robust to extending, by dozens of factors, both the stock and bond factor zoos that we consider in our baseline estimation (to maximize the time-series sample size), to varying sample and subsample estimations, and to using a multiplicity of different corporate bond datasets.
The remainder of the paper is organized as follows. Below, we review the most closely related literature and our contribution to it. Section \ref{sec:data} describes the data used in our analysis, while Section \ref{sec:econ-method} outlines our Bayesian SDF method and its properties for inference, selection, and aggregation. Section \ref{sec:empirics} presents our empirical findings, and Section \ref{sec:robustness} contains extensive robustness checks. Section \ref{sec:conclusion} concludes. Additional details and results are reported in the Appendix and the Internet Appendix.

\paragraph{Closely related literature}
Our research contributes to the active and growing body of work that critically reevaluates existing findings in the empirical asset pricing literature using robust inference methods. Following \citet{HarveyLiuZhu_2016}, a large  literature has tried to understand which existing factors (or their combinations) drive the cross-section of returns. In particular, \citet{GospodinovKanRobotti_2014} develop a general method for misspecification-robust inference, while \citet{GiglioXiu_2018} exploit the invariance principle of PCA and recover the risk premium of a given factor from the projection on the span of latent factors driving a cross-section of returns. Similarly, \citet{DelloPreiteUppalZaffaroniZviadadze_2024} recover latent factors from the residuals of an asset pricing model, effectively completing the span of the SDF. \citet{FengGiglioXiu_2020} combine cross-sectional asset pricing regressions with the double-selection LASSO of \citet{BelloniChernozhukovHansen_2014} to provide valid  inference on the selected sources of risk when the true SDF is sparse. \citet{KozakNagelSantosh_2020}  use a ridge-based approach to approximate the SDF and compare sparse models based on principal components of returns. Our approach instead identifies a dominant pricing model---if such a model exists---or a BMA across the space of all models, even if the true model is not sparse in nature, hence cannot be proxied by a small number of factors. Furthermore, and importantly, our work focuses on the \emph{co-pricing} of corporate bond and stock returns, hence shedding light on both the common, as well as the market specific, sources of risk.

As \citet{Harvey_2017} stresses in his American Finance Association presidential address, the factor zoo naturally calls for a Bayesian solution---and we adopt one.  In particular, we generalize the Bayesian method of model estimation, selection, and averaging developed in \citet*{BryzgalovaHuangJulliard_2023} to handle heterogeneous asset classes.

Numerous strands of the literature rely on Bayesian tools for asset allocation, model selection, and performance evaluation. Our approach is most closely linked to \citet{PastorStambaugh_2000} and \citet{Pastor_2000} in that we assign a prior distribution to the vector of pricing errors, and this maps into a natural and transparent prior for the maximal Sharpe ratio achievable in the economy. \citet{BarillasShanken_2018} also extend the prior formulation of \citet{PastorStambaugh_2000} and provide a closed-form solution for the Bayes factors when all factors are tradable in nature. \citet{ChibZengZhao_2020} show that the improper prior formulation of \citet{BarillasShanken_2018} is problematic, and provide a new class of priors that leads to valid comparisons for tradable factor models. As in these papers, our model and factor selection is based on posterior probabilities, but our method is designed to work with both tradable and \emph{nontradable} factors---as we show, the latter are a first-order source of priced risk in the joint space of corporate bonds and stock returns. 

Our work is closely related to the literature that stresses the optimality of  Bayesian model averaging for a very wide set of optimality criteria (see, e.g., \citet{Schervish_1995} and \citet{Raftery_2003}).\footnote{In particular, BMA is ``optimal on average,'' i.e., no alternative method can outperform the BMA for all values of the true unknown parameters. Furthermore, a BMA-SDF can be microfounded thanks to the equivalence between an economy populated by agents with heterogeneous beliefs and a Bayesian representative agent setting (\citet{HeyerdahlLarsenIlleditschWalden_2023}).} We highlight that Bayesian model averaging \emph{over the space of models} can be expressed as model averaging \emph{over the space of factors}. This allows us to show that posterior factor probabilities (which the previous Bayesian asset pricing literature has overwhelmingly focused on) and posterior market prices of risk (across the space of models) have very different information content. In particular, as we demonstrate, it is the latter, not the former, that tells us how to construct tradable portfolios that achieve the BMA-SDF-implied maximum Sharpe ratio. In the data, this yields a trading strategy with an (annualized) out-of-sample Sharpe ratio of 1.5 to 1.8. Most importantly, our approach can deal with a very large factor space, is not affected by the common identification failures that invalidate inference in asset pricing (see, e.g., \citet*{KanZhang_1999gmm, KanZhang_1999two}, \citet{Kleibergen_2009}, and \citet{GospodinovKanRobotti_2019}), and provides an optimal method for aggregating the pricing information stemming from the joint zoo of corporate bond and equity factors even if only noisy proxies of the true fundamental risks are available.

In the complete market benchmark, the pricing measure should be consistent across asset classes, and equilibrium models normally yield nontradable state variables. Therefore, we focus on the co-pricing of corporate bonds and stocks, and consider jointly a very broad collection of potential sources of risk that extends well beyond the set of bond and stock tradable factors that have been studied in isolation in the previous literature. Hence, our paper speaks to the large literature on co-pricing, originated with the seminal work of \cite{FamaFrench_1993}, and market segmentation of bonds and stocks (see, e.g., \citet{ChordiaGoyalNozawaSubrahmanyamTong_2017}, \cite{ChoiKim_2018}, or \cite{Sandulescu_2022}). In particular, our paper is  related to the body of work that explores whether equity market risk proxies (see, e.g., \citet{BlumeKeim_1987} and \citet{EltonGruberAgrawalMann_2001}), equity volatilities (see, e.g., \citet{CampbellTaksler_2003} and \citet{ChungWangWu_2019}), and equity-based characteristics (see, e.g., \citet{Fisher_1959}, \citet{GieseckeLongstaffSchaeferStrebulaev_2011}, and \citet{GebhardtLeeSwaminathan_2001})  are likely drivers of corporate bond returns, and on the commonality of risks across markets (see, e.g., 
\citet{HeKellyManela_2017}, \citet{LettauMaggioriWeber_2014}, and \citet{ChenRoussanovWangZou_2024}).

Overall, we find that factors in both the corporate bond and equity zoos are needed for the joint pricing of both asset classes, and stock factors do carry relevant information to explain bond returns. Yet, there is substantial overlap between the risks spanned by these two markets. That is, multiple bond and stock factors are noisy proxies for common underlying sources of risk. Nevertheless, as we show, corporate bond risk premia include an implicit compensation for Treasury term structure risk---a risk that the bond factor zoo, and nontradable factors proposed therein, price very well, while equity factors do not. And once this term structure risk component is removed, tradable bond factors become largely unnecessary for the joint pricing of bonds and stocks.
 
Several theoretical contributions stress that real economic activity and the business cycle should be among the drivers of bond risk premia (see, e.g., \citet{BhamraKuehnStrebualev_2010}, \citet{KhanThomas_2013}, \citet{ChenCuiHeMilbradt_2018}, and \citet{FavilukisLinZhao_2020}). Echoing both the general equilibrium model predictions of \citet{GomesSchmid_2021} and the empirical findings of \citet{EltonGruberBlake_1995} and \citet{ElkamhiJoNozawa_2023}, we show that the BMA-SDF conditional first and second moments have a clear business cycle pattern and peak during recessions and at times of heightened economic uncertainty, and that nontradable factors (especially proxies of the economic cycle such as the slope of the yield curve), are salient components of the pricing measure.\footnote{\citet{EltonGruberBlake_1995} show that adding fundamental macro-risk variables (such as GNP, inflation and term spread measures) significantly improves  pricing performance relative to equity and bond market index models. \citet{ElkamhiJoNozawa_2023} show that the long-run consumption risk measure of \citet{ParkerJulliard_2003} yields a one-factor model with significant explanatory power for corporate bonds, and such an SDF, as documented in \cite{ParkerJulliard_2005}, has a very strong business cycle pattern.} Furthermore, we show that the business cycle properties of the BMA-SDF and its volatility are predictable, and predict---as  theory implies in this case---future asset returns, generating a substantial degree of time variation in conditional risk premia.

Our work also relates to behavioral biases and market frictions in asset pricing. In particular, complementing the evidence of \citet{DanielHirshleiferSun_2020} and \citet{BryzgalovaHuangJulliard_2023} for the equity market, we show that the post-earnings announcement drifts of both bonds (see \citet{NozawaQiuXiong_2025}) and stocks are extremely likely drivers of corporate bond and stock risk premia. Furthermore, we show that cash-flow and discount rate news (see, e.g., \cite{Vuolteenaho_2002},  \citet{CohenGompersVuolteenaho_2002}, \cite{Zviadadze_2021}, and \citet{DelaoHanMyers_2025}) are both important drivers of risk premia in the joint cross-section of bonds and stocks, but the latter are responsible for a larger share of the volatility of the co-pricing SDF.

\section{Data}\label{sec:data}
Our analysis relies on a combination of corporate bond and stock data, which we present below and in more detail in Internet Appendix~IA.1. As academic research relies on various sources for corporate bond data, we are careful to estimate our model across \emph{all} datasets available to us to ensure that our results are neither driven by the data source nor the choice of bond or stock test assets (see the discussion in Section~\ref{sec:robustness_data}).

\subsection{Corporate bond data and corporate bond returns}\label{sec:bond_returns} 

Our baseline results in the main text are based on the constituents of the corporate bond data set from the Bank of America Merrill Lynch (BAML) High Yield (H0A0) and Investment Grade (C0A0) indices  made available via the Intercontinental Exchange (\href{https://www.ice.com/fixed-income-data-services}{ICE}) from January 1997 to December 2022. For the period from January 1986 to December 1996, we augment the data using the Lehman Brothers Fixed Income (LBFI) database.\footnote{We follow \citet{vanBinsbergenNozawaSchwert_2025} and begin the LBFI sample in 1986. Prior to 1986, bonds in the LBFI database are predominantly investment grade (91\% of bonds) with 67\% of all bonds priced with matrix pricing (i.e., the prices are not actual dealer quotes).} We then merge these data with the Mergent Fixed Income Securities Database (FISD) to obtain additional bond characteristics. After merging the two datasets and applying the standard filters, our bond-level data spans 37 years, resulting in a total of 444 monthly observations. Our corporate bond sample is representative of the U.S. market and, once merged with CRSP equity data, covers 75\% of the total stock market capitalization of all listed firms on average (see Figure~IA.3 of the Internet Appendix).\footnote{See Internet Appendix~IA.1 for a detailed description of the databases and associated cleaning procedures. Therein, we also discuss the additional datasets used for robustness tests.}

In the baseline analysis, we use \textit{excess} bond returns defined as the total bond return minus the one-month risk-free rate of return.\footnote{We use the one-month risk-free rate from \href{https://mba.tuck.dartmouth.edu/pages/faculty/ken.french/data_library.html}{Kenneth French's website}.} In addition, we follow \citet{vanBinsbergenNozawaSchwert_2025} and repeat our analysis with \textit{duration-adjusted} returns, whereby we subtract the return on a portfolio of duration-matched U.S. Treasury bonds from the total bond return.  We do not further winsorize, trim, or augment the underlying bond return data in any way, avoiding the biases that such procedures normally induce (\citet{DuarteJonesMoKhorram_2024} and \citet{DickersonRobottiRossetti_2024}).

\subsection{The joint factor zoo}

We use all factors in published papers for which a monthly time series matching our sample is publicly available. Our bond-specific factor zoo includes 16 tradable bond factors. From the equity literature, we include an additional 24 tradable factors. This set is smaller than the tradable equity factor zoo in \citet{BryzgalovaHuangJulliard_2023} as for several of their 34 tradable factors, an updated series is not publicly available. Moreover, we exclude factors for which authors did not provide sufficient information for exact replication.\footnote{The excluded factors are all among the \textit{least} likely components of the equity SDF in \citet{BryzgalovaHuangJulliard_2023}. Nevertheless, we consider \emph{all} of their factors in our robustness analysis.} Our nontradable zoo comprises 14 factors, many of which have previously been used to study stock returns. 

Overall, in our baseline analysis, we consider 54 factors---40 tradable and 14 nontradable---yielding $2^{54} \approx$ 18 quadrillion models. In Section~\ref{sec:vary_zoo}, we extend this to include dozens of additional factors available over varying subsamples, for a grand total of 91 candidate pricing factors. Table~\ref{tab:table-001-appendix} of \ref{sec:factor_zoo} describes all factors.\footnote{All factors are publicly available from the authors' personal websites and public repositories, listed therein. We make our 16 tradable bond factors available on the companion website: \href{https://openbondassetpricing.com/}{openbondassetpricing.com}.} Internet Appendix~IA.1.3 analyzes the robustness of bond factors with respect to data source and calculation method.

\subsection{In-sample bond and stock test assets}\label{sec:test_portfolios_IS} 

For our in-sample (IS) estimation of the BMA-SDF, we construct a set of 50 bond portfolios that are sorted on various bond characteristics to ensure a sufficiently broad cross-section. The first 25 portfolios are double-sorted on credit spreads and bond size, while the remaining 25 portfolios are double-sorted on bond rating and time-to-maturity. All portfolios are value-weighted based on the market capitalization of the bond issue, defined as the bond dollar value multiplied by the number of outstanding units of the bond. For the stock test assets, we rely on a set of 33 portfolios and anomalies very similar to those used in \citet{KozakNagelSantosh_2020} and \citet{BryzgalovaHuangJulliard_2023}.\footnote{These are publicly available from \cite{ChenZimmermann_2021} and \citet{JensenKellyPedersen_2023}, and replicable using CRSP and Compustat. See  \href{https://jkpfactors.com/}{jkpfactors.com}.} 

In addition, we include the 40 tradable factors as \citet{BarillasShanken_2017} emphasize that factors included in a model should price any factor excluded from the model.  This restriction, along with the use of a nonspherical pricing error formulation (i.e., GLS) also imposes (asymptotically) the restriction of factors pricing themselves. For the estimation of the co-pricing BMA-SDF, we naturally include both bond and stock tradable factors, while we only include the respective bond and stock tradable factors to estimate the bond- and stock-specific BMA-SDFs. 

In summary, our baseline cross-section comprises a wide array of 50 bond and 33 stock portfolios, as well as the underlying 40 tradable factors, for a total of 123 IS test assets.

\subsection{Out-of-sample bond and stock test assets}\label{sec:test_portfolios_OOS}

To test the out-of-sample (OS) asset pricing efficacy of the BMA-SDF estimated on the IS test assets, we employ a broad cross-section of additional corporate bond, stock, and U.S. Treasury bond portfolios. For bonds, we use decile-sorted portfolios on: (i) bond historical 95\% value-at-risk, (ii) duration, (iii) bond value (\citet{HouwelingVanZundert_2017}), (iv) bond book-to-market (\citet{BartramGrinblattNozawa_2025}), (v) long-term reversals (\citet{BaliSubrahmanyamWen_2021}), (vi) momentum (\citet{GebhardtHvidkjaerSwaminathan_2005}), as well as the bond version of the 17 Fama-French industry portfolios---totaling 77 bond-based portfolios.

For stocks, we include decile-sorted portfolios on: (i) earnings-to-price, (ii) momentum, (iii) long-term reversal, (iv) accruals, (v) size (measured by market capitalization), (vi) equity variance, in addition to the equity version of the 17 Fama-French industry portfolios (following 
\citet{LewellenNagelShanken_2010}), also resulting in 77 stock-based portfolios.

For U.S. Treasury bonds, we use monthly annualized continuously compounded zero-coupon yields from \citet{LiuWu_2021}. We price the U.S. Treasury bonds each month using the yield curve data and then compute monthly discrete excess returns across the term structure as the total return in excess of the one-month Treasury Bill rate. Our set of OS U.S. Treasury portfolios consists of 29 portfolios, ranging from 2-year Treasury notes up to 30-year Treasury bonds in increments of one year. 

In summary, our baseline OS test assets comprise 154 bond and stock portfolios (77 each) from the 14 distinct cross-sections discussed above.\footnote{All are available from \href{https://mba.tuck.dartmouth.edu/pages/faculty/ken.french/data_library.html}{Kenneth French's webpage} and \href{https://sites.google.com/view/jingcynthiawu/yield-data}{Cynthia Wu's webpage}.}  We not only use the joint cross-section, but we also construct $2^{14}-1 = 16,383$ possible unique combinations of OS cross-sections.\footnote{Further details about factors and in- and out-of-sample test assets, as well as links to the data sources, can be found in Table~IA.II of the Internet Appendix.} For robustness, we conduct OS pricing tests with the \citet{JensenKellyPedersen_2023} and the \citet{DickNielsenFeldhuetterPedersenStolborg_2024} bond and stock anomaly data.

\section{Econometric method}\label{sec:econ-method}

This section introduces the notation and summarizes the methods employed in our empirical analysis. We consider linear factor models for the SDF and focus on the SDF representation since we aim to identify the factors that have pricing ability for the joint cross-section of corporate bond and stock returns.\footnote{Recall that a factor might have a significant risk premium even if it is not part of the SDF, just because it has non-zero correlation with the true latent SDF. Hence, in order to identify the pricing measure, focusing on the SDF representation is the natural choice.}

We first review the frequentist estimation and the inference problems that arise therein in the presence of weak identification caused by weak and useless factors. We then summarize the Bayesian method proposed by \citet*{BryzgalovaHuangJulliard_2023}  to address the weak identification problem, present our extension of the approach to handle different asset classes, and introduce a more flexible prior structure. Finally, we establish a set of important new properties for the Bayesian model averaging of the SDF, and illustrate its mechanics in finite samples with a simulation study.

\subsection{Frequentist estimation of linear factor models}\label{sec:frequentist}

We begin by introducing the notation used throughout the paper. The returns of $N$ test assets, which are long-short portfolios, are denoted by $\bm{R}_t= (R_{1t} \dots R_{Nt})^\top$, $t=1, \dots T$. We consider $K$ factors, $\bm{f}_t=(f_{1t} \dots f_{Kt})^\top$, $t=1, \dots T$, that can be either tradable or nontradable. A linear SDF takes the form $M_t = 1 - (\bm{f}_t-\mathbb{E}[\bm{f}_t])^\top \bm{\lambda_f}$, where $\bm{\lambda_f}\in \mathbb{R}^K$ is the vector containing the market prices of risk (MPRs) associated with the individual factors. Throughout the paper, $\mathbb{E}[X]$ or $\mu_{X}$ denote the unconditional expectation of an arbitrary random variable $X$.

In the absence of arbitrage opportunities, we have that $\mathbb{E}[M_t \bm{R_t}]=\bm{0_N}$; hence, expected returns are given by $\bm{\mu_R}\equiv\mathbb{E}[\bm{R}_t]=\bm{C_f} \bm{\lambda_f}$, where $\bm{C_f}$ is the covariance matrix between $\bm{R}_t$ and $\bm{f}_t$, and prices of risk, $\bm{\lambda_f}$, are commonly estimated via the cross-sectional regression
\begin{equation}\label{cs_reg}
\bm{\mu_R}=\lambda_c \bm{1_N}+\bm{C_f} \bm{\lambda_f}+\bm{\alpha} = \bm{C} \bm{\lambda}+\bm{\alpha},
\end{equation}
where $\bm{C} = (\bm{1_N}, \bm{C_f})$, $\bm{\lambda}^\top = (\lambda_c, \bm{\lambda_f}^\top)$, $\lambda_c$ is a scalar average mispricing (equal to zero under the null of the model being correctly specified),  $\bm{1_N}$ is an $N$-dimensional vector of ones, and $\bm{\alpha}\in \mathbb{R}^N$ is the vector of pricing errors in excess of $\lambda_c$ (equal to zero under the null of the model). 

Such models are usually estimated via GMM, MLE or two-pass regression methods (see, e.g., \citet{Hansen_1982}, \citet{Cochrane_2005}). Nevertheless, as pointed out in a substantial body of literature, the underlying assumptions for the validity of these methods (see, e.g., \citet{NeweyMcFadden_1994}), are often violated (see, e.g., \citet{KleibergenZhan_2020} and \citet{GospodinovRobotti_2021}), and identification problems arise in the presence of a \emph{weak} factor (i.e., a factor that does not exhibit sufficient comovement with any of the assets, or has very little cross-sectional dispersion in this comovement, but is nonetheless considered a part of the SDF). These issues, in turn, lead to incorrect inferences for both weak and strong factors, erroneous model selection, and inflate the canonical measures of model fit.\footnote{These problems are common to GMM (\citet*{KanZhang_1999gmm}), MLE (\citet{GospodinovKanRobotti_2019}), Fama-MacBeth regressions (\citet{KanZhang_1999two}, \citet{Kleibergen_2009}), and even Bayesian approaches with flat priors for risk prices (\citet{BryzgalovaHuangJulliard_2023}).} 

\subsection{The Bayesian solution}\label{sec:Bayesian}

Albeit robust frequentist inference methods have been suggested in the literature for specific settings, our task is complicated by the fact that we want to parse the entire zoo of bond and stock factors, rather than estimate and test an individual model. Furthermore, we aim to identify the best specification---\emph{if} a dominant model exists---or aggregate the information in the factor zoo into a single SDF if no clear best model arises. Therefore, we extend the Bayesian method proposed in \citet*{BryzgalovaHuangJulliard_2023} (BHJ), since it is applicable to both tradable and nontradable factors, can handle the entire factor zoo, is valid under misspecification, and is robust to weak inference problems. This Bayesian approach is conceptually simple, since it leverages the naturally hierarchical structure of cross-sectional asset pricing, and restores the validity of inference using transparent and economically motivated priors. 

Consider first the time-series layer of the estimation problem. Without loss of generality, we order the $K_1$ tradable factors first, $\bm{f}^{(1)}_t$, followed by $K_2$ nontradable factors, $\bm{f}^{(2)}_t$; hence $ \bm{f}_t \equiv(\bm{f}^{(1),\top}_t, \bm{f}^{(2),\top}_t)^\top$ and $K_1 + K_2 = K$.  Denote by $\bm{Y_t} \equiv \bm{f_t} \cup \bm{R_t}$ the union of factors and returns, where $\bm{Y_t}$ is a $p$-dimensional vector.\footnote{If one requires the tradable factors to price themselves, then $\bm{Y_t} \equiv (\bm{R_t}^\top, \bm{f}^{(2),\top}_t)^\top$ and $p = N + K_2$.} Modelling $\{\bm{Y_t} \}_{t=1}^T$ as multivariate Gaussian with mean $\bm{\mu_Y}$ and variance matrix $\bm{\Sigma_Y}$, and adopting the conventional diffuse prior $\pi(\bm{\mu_Y}, \bm{\Sigma_Y}) \propto |\bm{\Sigma_Y}|^{-\frac{p+1}{2}}$, yields the canonical Normal-inverse-Wishart posterior for the time series parameters $(\bm{\mu_Y}, \bm{\Sigma_Y})$ in equations (\ref{eq:muY}) and (\ref{eq:SigmaY}) of \ref{post_samp}.

The cross-sectional layer of the inference problem allows for misspecification of the factor model via the average pricing errors $\bm{\alpha}$ in equation (\ref{cs_reg}). We model these pricing errors, as in the previous literature (e.g., \citet{PastorStambaugh_2000} and \citet{Pastor_2000}), as $\bm{\alpha} \sim \normal(\bm{0_N}, \sigma^2 \bm{\Sigma_R})$, yielding the cross-sectional likelihood (conditional on  the time series parameters)
\begin{equation}\label{eq:xs-lf}
	p(\text{data}|\bm{\lambda},\sigma^2)=(2 \pi \sigma^2)^{-\frac{N}{2}} |\bm{\Sigma_R}|^{-\frac{1}{2}}   \exp \left\{- \frac{1}{2\sigma^2} (\bm{\mu_R}-\bm{C} \bm{\lambda})^\top \bm{\Sigma_R}^{-1}(\bm{\mu_R}-\bm{C} \bm{\lambda}) \right\},%
\end{equation}
where, in the cross-sectional regression, the `data' are the expected risk premia, $\bm{\mu_R}$, and the factor loadings, $\bm{C} \equiv (\bm{1_N}, \bm{C_f})$. The above likelihood can then be combined with a prior for risk prices (presented below) to obtain a posterior distribution that informs inference and model selection.

Note that the assumption of a Gaussian conditional cross-sectional likelihood in equation (\ref{eq:xs-lf}) is not strictly necessary, and we could, in principle, use an alternative formulation (albeit, in most cases, this would cause us to lose many of the closed-form results that make our method able to handle such high-dimensional models and parameter spaces). 
Nevertheless, there are two key reasons why Gaussianity is the most preferable assumption. First, the canonical quasi-maximum likelihood estimation property applies (\citet*{BollerslevWooldridge_1992}): that is, the likelihood in equation (\ref{eq:xs-lf}) yields consistent estimates even if the true distribution is not Gaussian. Instead, different distributional assumptions would yield consistency only if we ``guess'' the right distribution. Hence, Gaussianity is the \textit{robust} choice. Second, consider estimating the model $\bm{R}_t = \bm{C\lambda}+\bm{\varepsilon}_t$. Denoting with $\mathbb{E}_T$ the sample analogue of the unconditional expectation operator, we have $\mathbb{E}_T [\bm{R}_t] = \bm{C\lambda}+\mathbb{E}_T [\bm{\varepsilon}_t]$. This implies that the pricing error $\bm{\alpha}$ should be equal to $\mathbb{E}_T [\bm{\varepsilon}_t]$. But the latter, under very general central limit theorem conditions (see, e.g., \citet{Hayashi_2000}), follows (under the null of the model) the limiting distribution $\bm{\alpha} | \bm{\Sigma_R} \sim \mathcal{N}(\bm{0_N}, \frac{1}{T} \bm{\Sigma_R})$. Hence, the Gaussian likelihood encoding in equation (\ref{eq:xs-lf}) not only ensures consistent estimates but is also a natural choice that guarantees compatibility of our hierarchical Bayesian modeling with frequentist asymptotic theory. 

To handle model and factor selection, we introduce a vector of binary latent variables $\bm{\gamma}^\top=(\gamma_0,\gamma_1, \dots,\gamma_K)$, where $\gamma_j \in \{ 0,1 \}$. When $\gamma_j=1$, the $j$-th factor (with associated loadings $\bm{C_j}$) should be included in the SDF, and should be excluded otherwise.\footnote{In the baseline analysis, we always include the common intercept in the cross-sectional layer, that is, $\gamma_0=1$. Nevertheless, we also consider $\gamma_0=0$, i.e., no common intercept, in the robustness analysis.} In the presence of potentially weak factors and, hence, unidentified prices of risk, the posterior probabilities of models and factors are not well defined under flat priors. 

To solve this issue, BHJ introduce an (economically motivated) prior that, albeit not informative, restores the validity of posterior inference. In particular, the uncertainty underlying the estimation and model selection problem is encoded via a (continuous spike-and-slab) mixture prior, $\pi(\bm{\lambda},\sigma^2,\bm{\gamma},\bm{\omega})= \pi(\bm{\lambda} \mid \sigma^2, \bm{\gamma}) \pi(\sigma^2) \pi(\bm{\gamma} \mid \bm{\omega}) \pi(\bm{\omega})$, where
\begin{equation}
	\lambda_j \mid \gamma_j,\sigma^2 \sim \normal\big(0, r(\gamma_j) \psi_j\sigma^2 \big). \label{E:lamba_prior}
\end{equation}
Note the presence of three new elements, $r(\gamma_j)$, $\pi(\bm{\omega})$ and $\psi_j$, in the prior formulation.

First, $r(\gamma_j)$ captures the `spike-and-slab' nature of the prior formulation. When the factor should be included, we have $r(\gamma_j = 1)=1$, and the prior, the `slab,' is just a diffuse distribution centred at zero. When instead the factor should not be in the model, $r(\gamma_j =0)=r\ll 1$, the prior is extremely concentrated---a `spike' at zero. As $r\rightarrow 0$, the prior spike is just a Dirac distribution at zero, hence it removes the factor from the SDF.\footnote{We set $r=0.001$ in our empirical analysis.}

Second, the prior $\pi(\bm{\omega})$ not only gives us a way to sample from  the space of potential models, but also encodes belief about the sparsity of the true model using the prior distribution $\pi(\gamma_j=1 |\omega_j)=\omega_j$. Following the literature on predictor selection, we set
\begin{equation}\label{eq:omega_prior}
\pi(\gamma_j=1 | \omega_j)=\omega_j, \,\,\,\, \omega_j \sim Beta\left(a_\omega,b_\omega\right).
\end{equation}
Different hyperparameters $a_\omega$ and $b_\omega$ determine whether one a priori favors more parsimonious models or not. The prior expected probability of selecting a factor is  $\frac{a_\omega}{a_\omega+b_\omega}$ and we set $a_\omega=b_\omega=1$ in the benchmark case, that is, we have a uniform (flat) prior for the model dimensionality and each factor has an ex ante expected probability of being selected equal to 50\%.\footnote{However, we could set for instance, $a_\omega=1$ and $b_\omega>>1$ to favor sparser models.}

Third, the Bayesian solution to the weak factor problem in BHJ is to set
	\begin{equation}\label{eq:psi-pen2}
		\psi_j = \psi \times \widetilde{\bm{\rho}_j}^\top \widetilde{\bm{\rho}_j},
	\end{equation}
where $\widetilde{\bm{\rho}_j} \equiv \bm{\rho}_j - \left(\frac{1}{N}\sum_{i=1}^N \rho_{j,i} \right)\times \bm{1_N}$, $\bm{\rho_j}$ is an $N \times 1$ vector of correlation coefficients between factor $j$ and the test assets, and $\psi \in \mathbb{R}_{+}$ is a tuning parameter that controls the degree of shrinkage across all factors. That is, factors that have vanishing correlation with asset returns, or extremely low cross-sectional dispersion in their correlations (hence cannot help in explaining cross-sectional differences in returns), have a low value of $\psi_j$ and are therefore endogenously shrunk toward zero. Instead, such a prior has no effect on the estimation of strong factors since these have large and dispersed correlations with the test assets, yielding a large $\psi_j$ and consequently a diffuse prior.

Finally, for the cross-sectional variance scale parameter, $\sigma^2$, estimation and inference can be based on the canonical diffuse prior $\pi(\sigma^2) \propto \sigma^{-2}$. As per Proposition 1 of \citet{ChibZengZhao_2020}, since the parameter $\sigma$ is common across models and has the same support in each model, the marginal likelihoods obtained under this improper prior are valid and comparable.

The above hierarchical system yields a well-defined posterior distribution from which all the unknown parameters and quantities of interest can be sampled. Nevertheless, the prior formulation of BHJ might be overly restrictive when applied, as in our empirical analysis, to different asset classes jointly. To illustrate this, consider the case in which (as in our empirical application) all factors are standardized, and note that equations (\ref{E:lamba_prior}) to (\ref{eq:psi-pen2}) then yield the following (squared) prior Sharpe ratio (SR) for each factor $f_{k,t}$:
\[
\mathbb{E}_{\pi} [SR^2_{f_k} \mid \sigma^2] = \frac{a_\omega}{a_\omega+b_\omega} \psi \sigma^2 \bm{\tilde{\rho}}_k^\top  \bm{\tilde{\rho}}_k, \text{ as } r\rightarrow 0.
\] 
This implies that two factors with the same (sum of squared) demeaned correlations with asset returns will have exactly identical prior Sharpe ratios. This feature is unsatisfactory when considering factors proposed for pricing different asset classes, as the maximum Sharpe ratio achievable in different market segments might actually be quite different. We relax this constraint in the next subsection by introducing a new, more flexible prior formulation that preserves the robustness of the estimator to weak and spurious factors.

\subsection{A spike-and-slab prior for heterogeneous classes of factors} \label{subsec:kappa}

We now generalize the prior specification in equation (\ref{E:lamba_prior}). As in BHJ, we formalize a continuous spike-and-slab prior that, using the correlation between factors and asset returns, endogenously solves the problems arising from weak factor identification. However, unlike them, we introduce an additional hyperparameter that researchers can use to encode their prior belief about how much of the SDF Sharpe ratio in the data can be captured with factors coming from, respectively, the bond and stock factor zoos. Specifically, we formulate a spike-and-slab prior for the vector of all factors' market prices of risk as\footnote{More precisely, the first element of $\bm{\lambda}$ is the coefficient associated with the common cross-sectional intercept, while the remaining elements are the market prices of risks of the factors under consideration.} 
\begin{equation}\label{E:lamba_prior2}
	\bm{\lambda} | \sigma^2, \bm{\gamma} \sim \normal (\bm{0}, \sigma^2\bm{D}^{-1}).
\end{equation} 
For illustrative purposes, consider first the case in which we have only two types of factors under consideration: $K_1$ bond-market-based factors (ordered first) and $K-K_1$ stock-market-based factors (ordered last). In this case we can encode our prior beliefs about which factors are more likely drivers of observed risk premia by setting $\bm{D}$ as a diagonal matrix with elements $c$ (the prior precision for the intercept), $[(1+\kappa)r(\gamma_1) \psi_1)]^{-1}$, ..., $[(1+\kappa)r(\gamma_{K_1}) \psi_{K_1}]^{-1} $, $[(1-\kappa)r(\gamma_{K_1+1}) \psi_{K_1 +1}]^{-1} $, ...,  $[(1-\kappa)r(\gamma_{K}) \psi_{K}]^{-1} $. The $\psi_j$ elements are defined as in equation (\ref{eq:psi-pen2}) and endogenously solve the problems arising from weak factors. Similarly, $r(\gamma_j)$, as before, captures the spike-and-slab nature of the prior formulation. 

The new hyperparameter $\kappa \in (-1,1)$ encodes the prior belief about which class of factors is more likely to explain the Sharpe ratio of asset returns. To see this, consider the case in which both factors and returns are standardized (as in our empirical implementation). In this case:
\begin{equation*}
	\frac{\expec_\pi \left[ SR^2_{\bmf} | \bm{\gamma}, \sigma^2 \right]}{\expec_\pi \left[SR^2_{\bm{\alpha}} | \sigma^2 \right] } = \frac{\psi}{N}\left[(1+\kappa)\sum_{k = 1}^{K_1} r(\gamma_k)\bm{\tilde \rho}^\top_k \bm{\tilde \rho}_k  + (1-\kappa)\sum_{k = K_1 +1}^{K} r(\gamma_k)\bm{\tilde \rho}^\top_k \bm{\tilde \rho}_k  \right],
\end{equation*}
where $SR_{\bmf}$ and $SR^2_{\bm{\alpha}}$ denote, respectively, the Sharpe ratios achievable with all factors and the Sharpe ratio of the pricing errors. 

The above implies that the only free `tuning' parameters in our setting, $\psi$ and $\kappa$, have straightforward economic interpretations and can be transparently set. To see this, first consider $\kappa=0$ (the homogeneous prior specification). In this case (with a uniform prior of factor inclusion), the expected prior Sharpe ratio achievable with the factors is just $\mathbb{E}_\pi [SR^2_{\bm{f}} \mid \sigma^2] = \frac{1}{2} \psi \sigma^2 \sum_{k=1}^{K} \bm{\tilde{\rho}}_k^\top  \bm{\tilde{\rho}}_k $ as $r\rightarrow 0$. Hence, prior beliefs about the achievable Sharpe ratio with the factors fully pin down $\psi$.\footnote{Without a uniform prior for the SDF dimensionality, the prior Sharpe ratio value becomes $\mathbb{E}_\pi [SR^2_{\bm{f}} \mid \sigma^2] = \frac{a_\omega}{a_\omega+b_\omega} \psi \sigma^2 \sum_{k=1}^{K} \bm{\tilde{\rho}}_k^\top  \bm{\tilde{\rho}}_k $ as $r\rightarrow 0$. Hence, beliefs about the prior Sharpe ratio and model dimensionality fully pin down the hyperparameters.} When instead $\kappa \neq 0$, the prior is heterogeneous across types of factors, and this parameter encodes our prior expectation about which type of factors explains a larger share of the Sharpe ratio of the asset returns. As $\kappa \rightarrow 1^-$ ($\kappa \rightarrow -1^+$), the prior becomes concentrated on only bond (stock) factors being able to explain the Sharpe ratio of asset returns. For example, setting $\kappa = 0.5$ encodes the prior belief that, ceteris paribus, bond factors explain a $\frac{1+\kappa}{1-\kappa} =$ 3 times as large a share of the squared Sharpe ratio than equity factors.

More generally, we can flexibly encode prior beliefs about the saliency of more than two categories of factors by setting $\bm{D} = \bm{\tilde{D}} \times \bm{\kappa}$, where $ \bm{\tilde{D}} $ is a diagonal matrix with elements $c$, $(r(\gamma_1) \psi_1)^{-1}$, ..., $(r(\gamma_{K}) \psi_{K})^{-1} $ and $\bm{\kappa}$ is a conformable column vector with elements $1, \, 1+\kappa_1, \, \dots, 1+\kappa_K$ such that $\sum_{k=1}^{K} \kappa_j =0$ and $0<|\kappa_j|<1$ $\forall j$. 

Note that this general prior encoding maintains the same assumption of \textit{exponential tails} for all factors (given the Gaussian formulation in equation (\ref{E:lamba_prior2})). And there is a very good reason for this: useless factors generate heavy-tailed cross-sectional likelihoods (in the limit, the likelihood is an improper ``uniform'' on $\mathbb{R}$), with peaks for the market prices of risk that deviate toward infinity. But, as first pointed out by \citet{Jeffreys_1961}, as the peak of a thick-tailed likelihood moves away from the exponential-tail prior, the posterior distribution eventually \emph{reverts back to the prior}. Hence, in our setting, the exponential tails of the prior play an important role: they shrink the price of risk of useless factors toward zero.

The transparency and interpretability of our prior formulation allows us, in the empirical analysis, to report results for various prior expectations of the Sharpe ratio achievable in the economy,\footnote{More precisely, we report results for different prior values of $\sqrt{\mathbb{E}_\pi [SR^2_{\bm{f}} \mid \sigma^2]}$.} prior probability of factor inclusion, shares of the prior Sharpe ratio achievable with the different types of factors that we consider, and account for a potential ``mismeasurement alpha'' in the corporate bond data.

Furthermore, note that pure `level' factors---i.e., factors that have no explanatory power for cross-sectional differences in asset returns but capture the average level of risk premia across assets---can be accommodated by removing the free intercept in the SDF (since it would be collinear with a pure level factor) and using simple correlations (instead of cross-sectionally demeaned ones) in equation (\ref{eq:psi-pen2}), i.e. setting $\psi_j = \psi \times \bm{\rho}_j^\top \bm{\rho}_j$. We consider this particular case among our robustness exercises, and it leaves our main findings virtually unchanged.

\subsection{Model and factor selection and aggregation}\label{sec:ModelSel}

Our Bayesian hierarchical system defined in the previous subsections yields a well-defined posterior distribution from which all the unknown parameters and quantities of interest (e.g., $R^2$, SDF-implied Sharpe ratio, and model dimensionality) can be sampled to compute posterior means and credible intervals via the Gibbs sampling algorithm described in \ref{post_samp}. Most importantly, these posterior draws can be used to compute posterior model and factor probabilities, and, hence, identify robust sources of priced risk and---\emph{if} such a model exists---a dominant model for pricing assets.

Model and factor probabilities can also be used for aggregating optimally, rather than selecting, the pricing information in the factor zoo. For each possible model $\bm{\gamma}^m$ that one could construct with the universe of factors, we have the corresponding SDF: $M_{t,\bm{\gamma}^m} =1- \left(\bm{f}_{t,\bm{\gamma}^m} -\mathbb{E}[\bm{f}_{t,\bm{\gamma}^m}]\right)^\top \bm{\lambda_{\bm{\gamma}^m}}$. Therefore, we construct a BMA-SDF by averaging all possible SDFs using the posterior probability of each model as weights:
\begin{equation}\label{E:BMA}
M^{BMA}_t  =\sum_{m=1}^{\bar{m}}M_{t, \bm{\gamma}^m} 
	\Pr\left( \bm{\gamma}^m|\text{data}\right),
\end{equation}
where $\bar{m}$ is the total number of possible models.\footnote{See, e.g., \citet{HoetingMadiganRaftery_1997} and \citet{HoetingMadiganRafteryVolinsky_1999}.}

The BMA aggregates information about the true latent SDF over the space of all possible models, rather than conditioning on a particular model. At the same time, if a dominant model exists (a model for which $\Pr\left( \bm{\gamma}^m|\text{data}\right)\approx 1$), the BMA will use that model alone. Pricing with the BMA-SDF is also robust to the problems arising from collinear loadings of assets on the factors, since any convex linear combination of factors with collinear loadings has exactly the same pricing implications. Moreover, the BMA-SDF can be microfounded, as in \citet{HeyerdahlLarsenIlleditschWalden_2023}, thanks to the equivalence of a log utilities and heterogeneous beliefs economy with a representative agent using the Bayes rule. Furthermore, BMA aggregation is optimal under a wide range of criteria, but in particular, it is \emph{optimal on average}: no alternative estimator can outperform it for all possible values of the true unknown parameters.\footnote{See, e.g., \cite{Raftery_2003} and \cite{Schervish_1995}.}  Finally, since its predictive distribution minimizes the Kullback-Leibler information divergence relative to the true unknown data-generating process, the BMA aggregation delivers the most likely SDF given the data, and the estimated density is as close as possible to the true unknown one, even if all of the models considered are misspecified.

The BMA has further useful properties when applied to the construction of the SDF. To see this, note that the BMA-SDF defined in equation (\ref{E:BMA})---thanks to the linearity of the models considered---can be rewritten as a weighted sum over the space of factors, rather than over the space of models. That is:
\begin{equation} 
M^{BMA}_t  =  1-\sum_{j=1}^{K} \underset{\equiv \; \expec [\lambda_j | \text{data}] }{\underbrace{\expec [\lambda_j | \text{data}, \gamma_j =1] \Pr(\gamma_j =1 | \text{data})}} \left(f_{j,t} -\expec [f_{j,t}]\right), \;\;\;\; \text{as} \; r\rightarrow 0.  \label{E:BMA2a}
\end{equation}
This expression makes clear that the weight attached to each factor in the BMA-SDF is driven by two elements. First, the probability of the factor being a ``true'' source of priced risk,  $\Pr(\gamma_j =1 | \text{data})$. Hence, naturally, when a factor is more likely (given the data) to drive asset risk premia, it features more prominently in the BMA-SDF. Second, when a factor commands a large market price of risk in the models that include it, i.e. when  $\expec [\lambda_j | \text{data}, \gamma_j =1]$ is large, it will, ceteris paribus, have a larger role in the BMA-SDF. These two forces are jointly captured in $\expec [\lambda_j | \text{data}]$, the posterior expectation of the market price of risk given the data \emph{only}, i.e., independently of the individual models.

This property of the BMA-SDF implies that, when parsing the factor zoo, there are two quantities of key interest. First, $\Pr(\gamma_j =1 | \text{data})$, as we want to discern which variables are more likely, given the data, to be fundamental sources of risk and, hence, should be included in our theoretical models for explaining asset returns. Second, and arguably as important,  $\expec [\lambda_j | \text{data}]$, as this quantity pins down how salient the given factor is in the BMA approximation of the SDF. Furthermore, $\expec [\lambda_j | \text{data}]$ yields the weights that should be assigned to the factors in a portfolio that best approximates the true latent SDF. For these reasons, we track both quantities in our empirical analysis.

Furthermore, this implies that posterior probabilities of factors that are not true sources of fundamental risk will not necessarily tend to zero if they nevertheless help span the true latent risks driving asset returns. That is, it might well be the case that, for a given factor, the posterior probability of being part of the SDF ($\Pr(\gamma_j =1 | \text{data})$) is smaller than the prior one---hence indicating that the data do not support the factor being a fundamental risk---while at the same time its estimated posterior market price of risk ($\expec [\lambda_j | \text{data}]$) is substantial, since the factor helps the BMA-SDF span the risks in asset returns. This is not a contradiction, but rather an important element of strength of our method.

To illustrate these properties, consider the case in which the ``true'' SDF contains only one factor. That is, $M^{true}_t = 1 - \lambda_f f_{t,true}$, where $f_{true}$ is the true source of fundamental risk and to simplify exposition, we employ the normalizations $\expec [f_{t,true}] = 0$ and $\var (f_{t,true}) =1$. Note that under this innocuous normalization the risk premium and market price of risk of the factor coincide, i.e. $\lambda_{true} = \sqrt{\var(M^{true}_t)} = - \cov(M^{true}_t, f_{t,true}) = \mu_{true}$. Consistent with the postulated one factor structure, the vector of test assets' excess returns $\bm{R}_t$ follows the process
$$
\bm{R}_t = \bm{\mu_R} + \bm{C} f_{t,true} + \bm{w}_{\bm{R}, t},
$$
where $\bm{w}_{\bm{R}, t} \perp f_{t,true}$ and $\expec [\bm{w}_{\bm{R}, t}] = \bm{0}$. Hence, it follows that the true factor prices perfectly (in population) the asset returns, as $\bm{\mu_R} = -\cov (M^{true}_t, \bm{R}_t) =\bm{C} \lambda_{true}$.

Suppose further that there are a set of factors, ``noisy proxies'' of the true factor $f_{true}$, that the researcher considers as potential sources of fundamental risk,
$$
f_{j,t} = \delta_j f_{t,true} + \sqrt{1-\delta_j^2} \,  w_{j,t}, \;\;\; | \delta_j | <1,      
$$
for each noisy proxy $j$, with $w_{j,t} \perp f_{t,true}$ and $w_{j,t} \iid (0,1)$. Note that in this handy encoding $\delta_j$ captures both the correlation between the true source of risk and the $j$-th noisy proxy and the latter's signal-to-noise ratio (as $\sqrt{\var(f_{j,t})} =1$ by construction).

Suppose that a researcher tests the pricing ability of the $j$-th noisy proxy by considering the misspecified SDF $\widetilde M_{j,t} = 1 - \tilde \lambda_j f_{j,t}$. We then have that the misspecified SDF prices the test assets perfectly in population (as long as the noise in the factor is ``classical,'' i.e. $w_{j,t} \perp \bm{w}_{\bm{R}, t} $):
\begin{equation} \label{E:misspecifiedlambda}
\bm{\mu_R} = - \cov(\widetilde M_{j,t}, \bm{R}_t) = \bm{C}\delta_j \tilde \lambda_j \;\;\; \text{with} \; \; \tilde \lambda_j = \lambda_{true}/\delta_j.
\end{equation}
That is, the noisy proxy seems indistinguishable from the true factor in its pricing ability for the test assets, and it yields an estimated  market price of risk (in population) that is larger (in absolute terms) than that of the true factor.\footnote{Furthermore, $|\tilde \lambda_j| \rightarrow \infty$ as $|\delta_j| \rightarrow 0$, in yet another manifestation of the weak factor problem.} 

Nevertheless, our method will detect such factor as a noisy proxy  since our hierarchical Bayesian framework requires factors to self-price. To see this, note that the true risk premium of the noisy proxy is $\mu_{j} \equiv - \cov(M^{true}_t, f_{j,t}) \equiv \delta_j \lambda_{true}$, while instead the misspecified SDF that prices the cross-section of test assets yields an implied risk premium for the factor given by $\tilde \mu_{j} : = -\cov(\widetilde M_{j,t}, f_{j,t}) = \tilde \lambda _j = \lambda_{true}/\delta_j$. Thus, the noisy proxy will fail to self-price, since $|\tilde \mu_{j}| > |\mu_{j}| \,\,\, \forall |\delta_j| <1$, and its self-mispricing will be proportional to $|\frac{1}{\delta_j^2} -1|$. 

This implies that, once the candidate factors are added to the set of test assets, factors that have a higher correlation ($\delta_j$) with the true source of risk will have overall better performance in the cross-sectional likelihood in equation (\ref{eq:xs-lf}). Moreover, since $\tilde \mu_{j} \underset{|\delta_j|\rightarrow 1}{\longrightarrow} \mu_{j}$, noisy proxies with a higher signal-to-noise ratio will tend to have higher posterior probabilities.
As a result, the BMA-SDF is more robust in recovering the pricing of risk than other canonical estimators. The reason being that, as per equation (\ref{E:misspecifiedlambda}), simple cross-sectional estimation with the noisy proxy included in the SDF yields an upward biased market price of risk for this factor, $\expec [\lambda_j | \text{data}, \gamma_j =1]$. Nevertheless, due to the self-pricing restriction that the noisy proxy will not satisfy, the posterior probability of such factors, $\Pr(\gamma_j =1 | \text{data})$, will be strictly smaller than one. This, in turn, will counteract the upward bias in the market price of risk since the factor enters the BMA in equation (\ref{E:BMA2a}) with a weight equal to $\expec [\lambda_j | \text{data}, \gamma_j =1] \Pr(\gamma_j =1 | \text{data})$ (as $r\rightarrow 0$).

Note that this analytical example of the properties of our estimator is without loss of generality. For instance, a misspecified SDF with multiple noisy proxies will also yield an upward-biased measure of the market price of risk. Consequently, the misspecified SDF will not be able to satisfy the self-pricing restriction of the factors; hence, it will achieve a posterior probability strictly smaller than one. Therefore, this upward biased measure of the market price of risk implied by the misspecified SDF will be counteracted in the BMA in equation (\ref{E:BMA2a}) by a $\Pr\left( \bm{\gamma}^m|\text{data}\right) <<1$.

But are these population (hence asymptotic) properties of our method likely to hold in a finite sample? We address this question with a realistic simulation exercise.

\subsubsection{Simulation}\label{sec:Simulation}
We calibrate a single (pseudo-true) useful factor ($f_{true}$) that mimics the pricing ability of the HML factor in the cross-section of the 25 Fama-French size and book-to-market portfolios. That is,  we consider a setting with a partially misspecified pricing kernel (as HML yields sizable pricing errors in the cross-section used for calibration). To make the estimation challenging, we always include a useless factor (as this breaks the validity of canonical estimation methods), and consider noisy proxies with different correlations with the useful factor. In each experiment we include a variable number of noisy proxies $f_j$, $j=1,..., 4$ with correlations with the pseudo-true factor equal to 0.4, 0.3, 0.2, and  0.1, respectively. Further details of the simulation design are reported in Internet Appendix~IA.2. 

\begin{figure}[tbp]
\begin{center}
\includegraphics[scale=.16, trim = 0cm 0cm 0cm 0cm, clip]{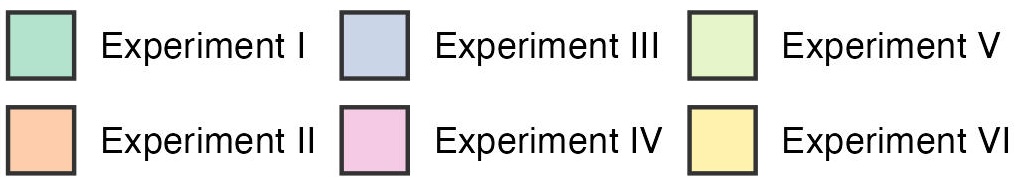}%
\vspace{.15cm}

\begin{subfigure}[b]{0.45\textwidth}
 \includegraphics[scale=.1,trim = 0cm 0cm 0cm 16cm, clip]{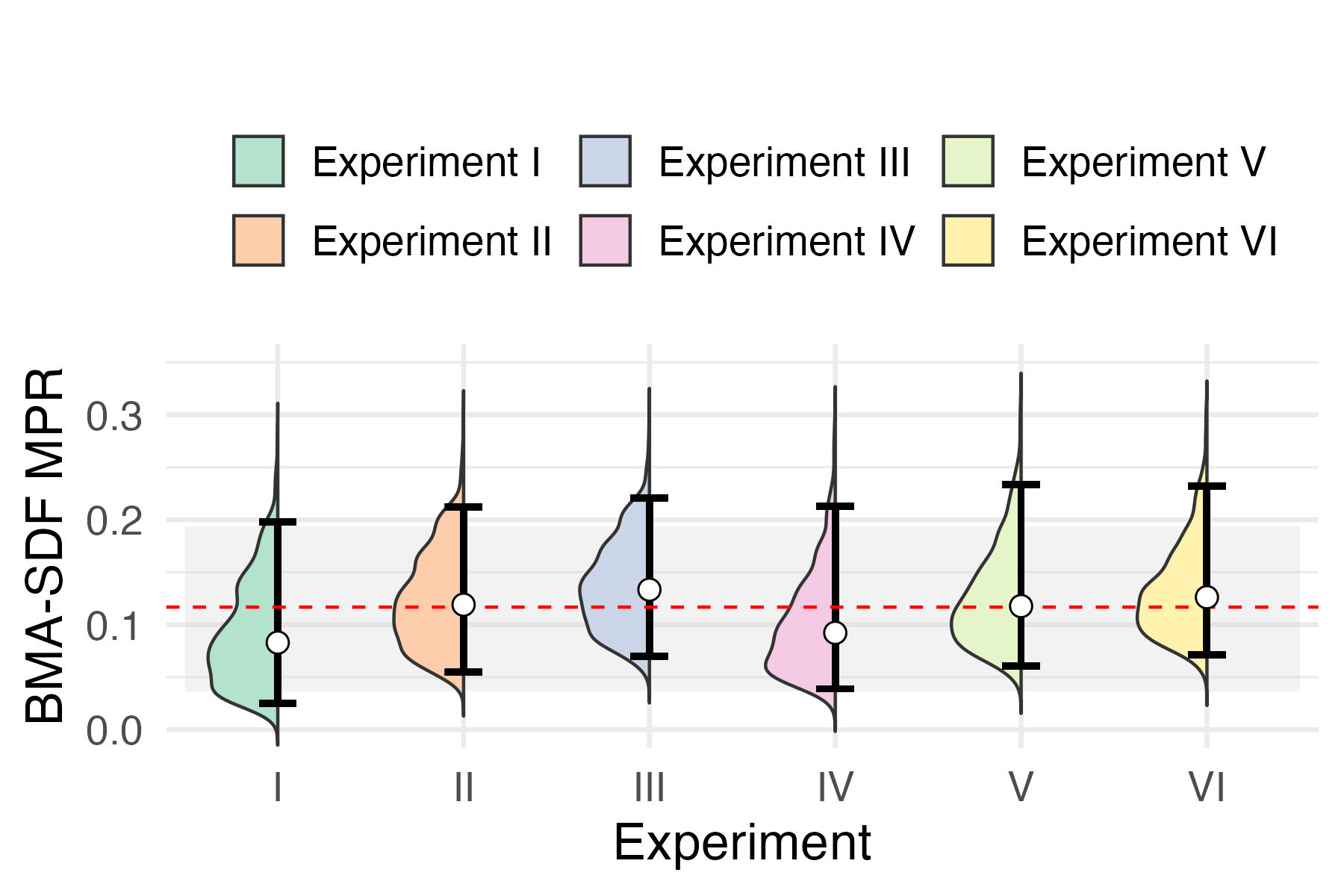}\caption{BMA-SDF market price of risk, $T=400$}
\end{subfigure}
\hspace{.2cm}
\begin{subfigure}[b]{0.45\textwidth}
 \includegraphics[scale=.1,trim = 0cm 0cm 0cm 16cm, clip]{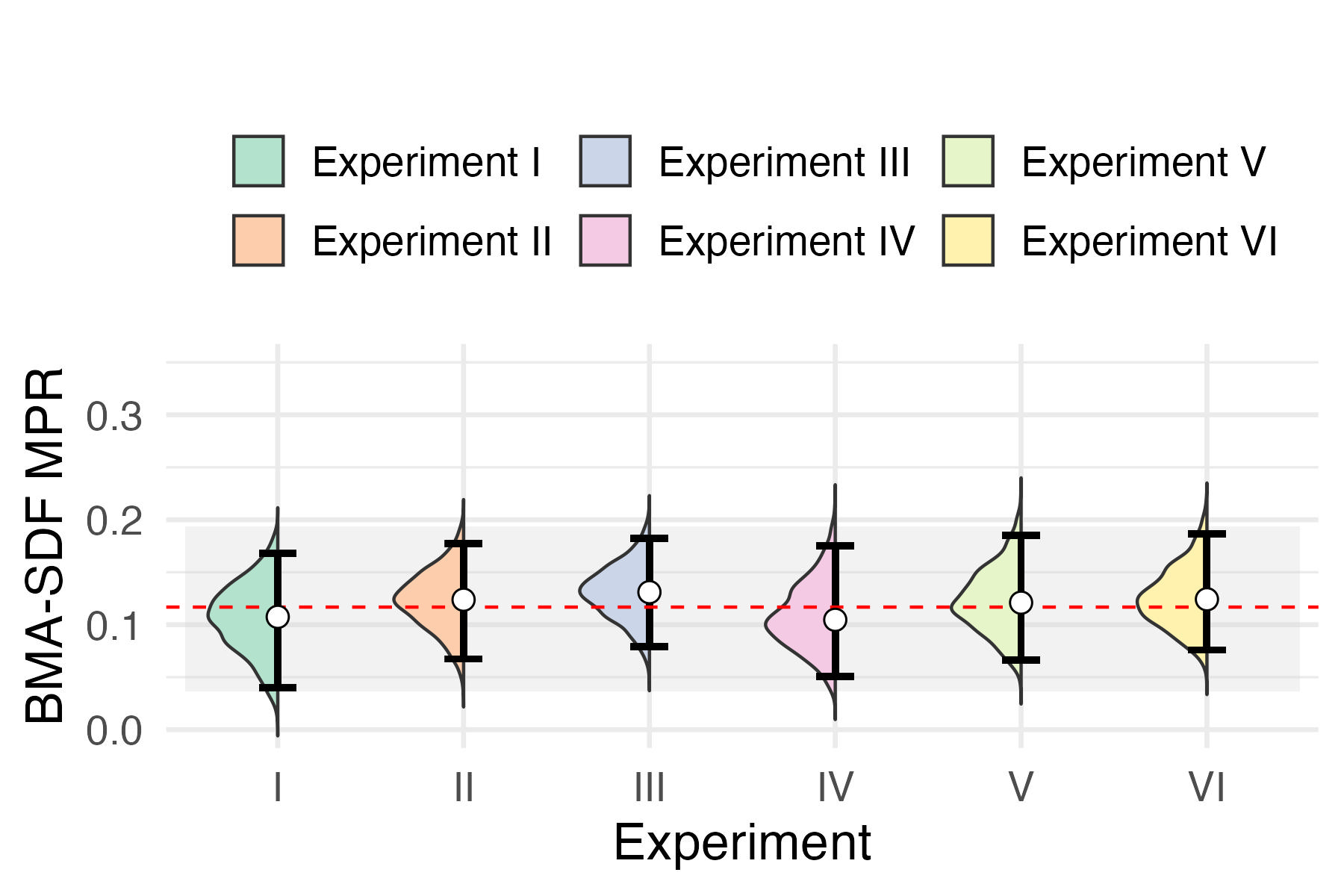}\caption{BMA-SDF market price of risk, $T=1600$}
\end{subfigure}

\begin{subfigure}[b]{0.45\textwidth}
 \includegraphics[scale=.1,trim = 0cm 0cm 0cm 0cm, clip]{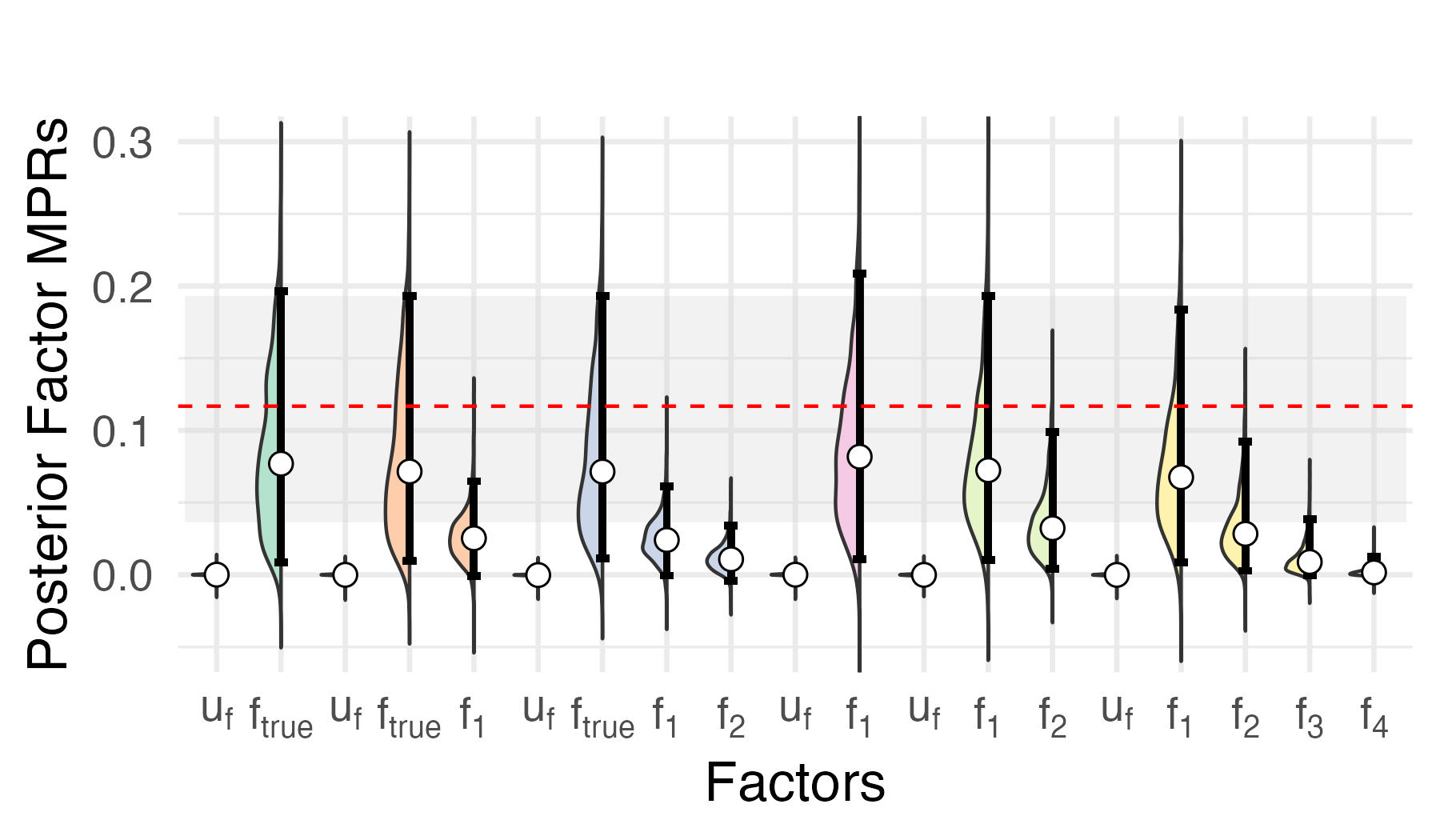}\caption{Factors' market price of risk, $T=400$}
\end{subfigure}
\hspace{.2cm}
\begin{subfigure}[b]{0.45\textwidth}
 \includegraphics[scale=.1,trim = 0cm 0cm 0cm 0cm, clip]{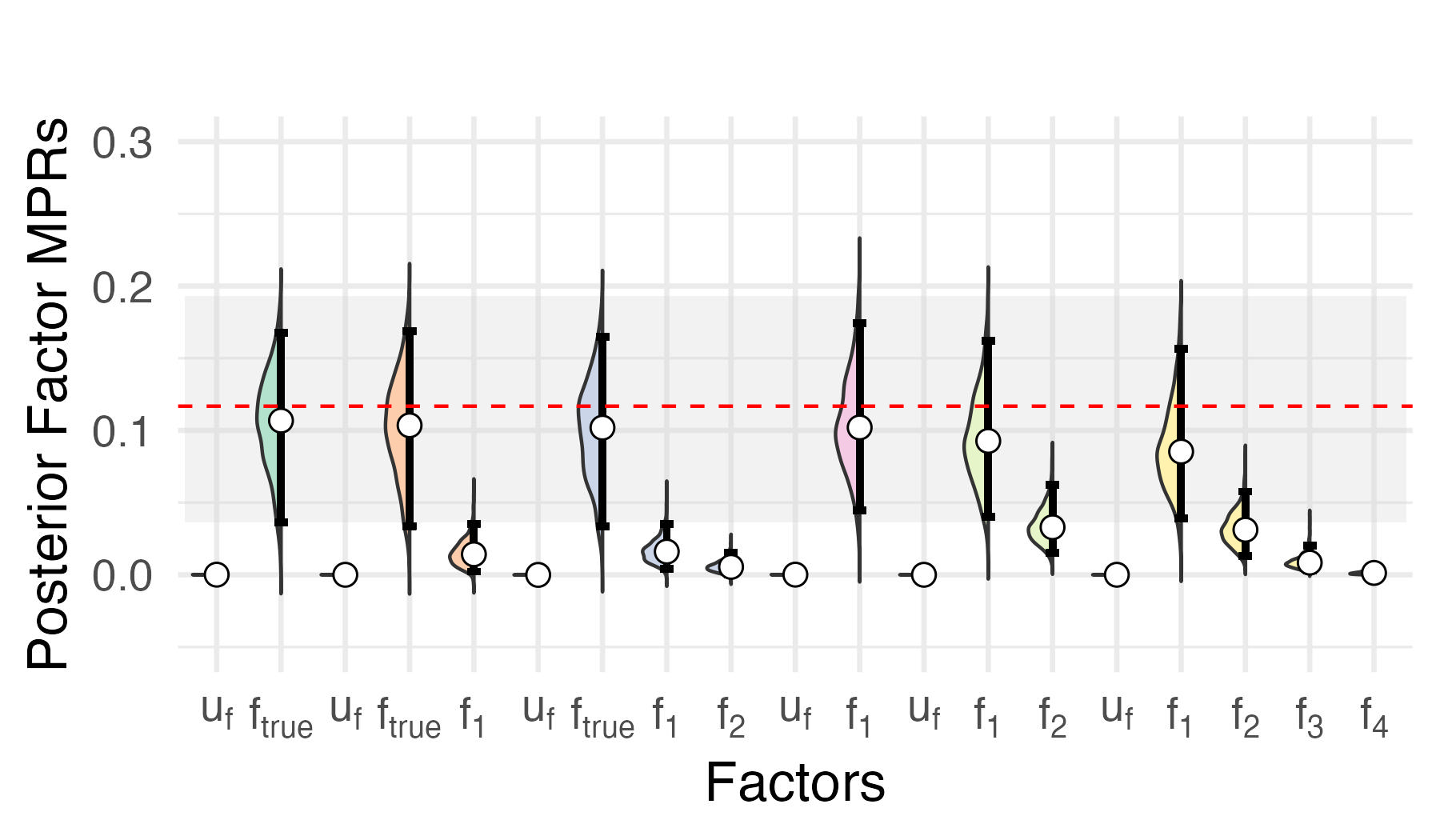}\caption{Factors' market price of risk, $T=1600$}
\end{subfigure}

\begin{subfigure}[b]{0.45\textwidth}
 \includegraphics[scale=.12,trim = 0cm 0cm 0cm 0cm, clip]{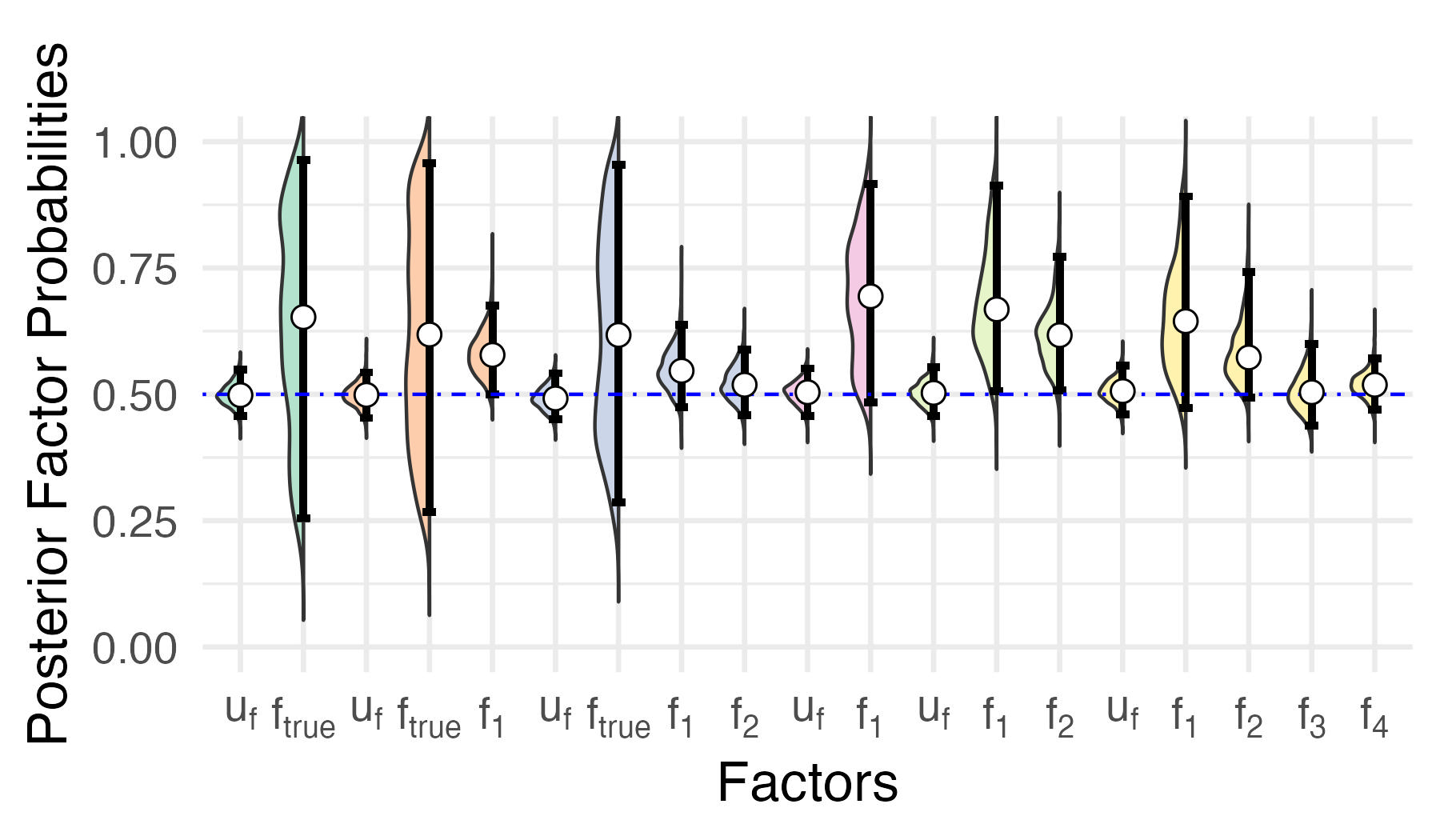}\caption{Factors' posterior probabilities, $T=400$}
\end{subfigure}
\hspace{.2cm}
\begin{subfigure}[b]{0.45\textwidth}
 \includegraphics[scale=.1,trim = 0cm 0cm 0cm 0cm, clip]{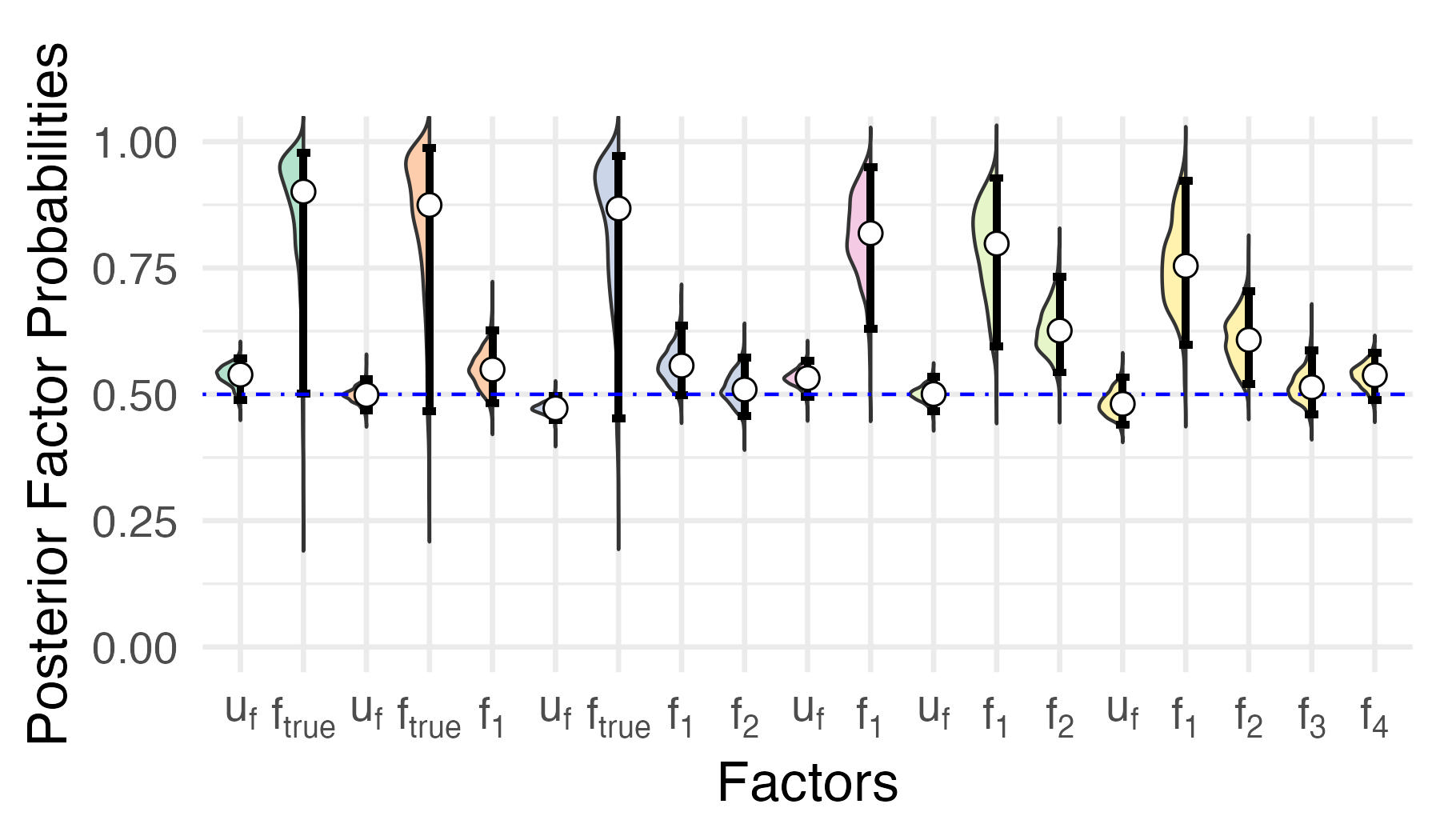}\caption{Factors' posterior probabilities, $T=1600$}
\end{subfigure}
\end{center}
\vspace{-4mm}
\caption{Simulation evidence with useless factors and noisy proxies.}
\vspace{-2mm}
\begin{justify}
\begin{spacing}{1}
\footnotesize{
Simulation results from applying our Bayesian methods to different sets of factors. Each experiment is repeated 1,000 times with the specified sample size ($T$). The data-generating process is calibrated to match the pricing ability of the HML factor (as a pseudo-true factor) for the Fama-French 25 size and book-to-market portfolios. Horizontal red dashed lines denote the market price of risk of HML, and the grey shaded area the frequentist 95\% confidence region of its GMM estimate in the historical sample of 665 monthly observations. The prior is set to 60\% of the ex post maximum Sharpe ratio. Simulation details are in Internet Appendix~IA.2. Half-violin plots depict the distribution of the estimated quantities across the simulations, with black error bars denoting centered 95\% coverage, and white circles denoting median values, across repeated samples. In all experiments we include a useless factor ($u_f$), while the pseudo-true factor ($f_{true}$) is included only in experiments I to III. In each experiment we include a variable number of noisy proxies $f_j$, $j=1,..., 4$ with correlations with the pseudo-true factor equal to, respectively, 0.4, 0.3, 0.2, and  0.1. The factors considered in the various experiments are: \\
\noindent \begin{tabular}{@{}p{0.49\linewidth}@{}p{0.49\linewidth}@{}}
\textbf{Experiment I}:  $u_f$ and $f_{true}$.& 
\textbf{Experiment IV}: $u_f$, and $f_1$. \\
\textbf{Experiment II}: $u_f$, $f_{true}$ and $f_1$. & 
\textbf{Experiment V}: $u_f$, $f_1$ and $f_2$.  \\
\textbf{Experiment III}: $u_f$, $f_{true}$, $f_1$ and $f_2$. &  
\textbf{Experiment VI}: $u_f$, $f_1$, $f_2$, $f_3$ and $f_4$.\\
\end{tabular}}
\end{spacing}
\end{justify}
\label{fig:simulation}
\end{figure}

Simulation results are reported in Figure~\ref{fig:simulation} for different sample sizes and a prior Sharpe ratio of 60\% of the ex post maximum Sharpe ratio in the simulated samples. Results for different priors and sample sizes are reported in the Internet Appendix. We conduct six experiments. In the first three (experiments I to III), the pseudo-true factor is included among the candidate factors, while in the latter three (experiments IV to VI) only its noisy proxies are included. 

Panel A of Figure~\ref{fig:simulation} reports the BMA-SDF-implied market price of risk for several simulation designs in time series samples with only 400 monthly observations. The horizontal red dashed line denotes the Sharpe ratio of the pseudo-true factor, while the shaded grey area denotes the frequentist 95\% confidence region for the market price of risk of the HML factor estimated via GMM in the (true) cross-section of 25 size and book-to-market portfolios with 665 monthly observations. Remarkably, the BMA-SDF estimator accurately recovers the market price of risk of the SDF not only when the pseudo-true factor is included among the candidate pricing factors (experiments I to III), but also when \emph{only} noisy proxies of the true source of risk  are included (experiments IV to VI). Moreover, the estimates are sharp---the distributions of the BMA-MPRs across simulation runs have 95\% coverage areas very similar to the ones obtained (without accounting for model uncertainty) in the much longer true sample. Furthermore, as the time series sample size increases, Panel B of Figure~\ref{fig:simulation} illustrates that the BMA estimates of the MPRs of the SDF become progressively more concentrated on the pseudo-true value, and converge to it in the large sample (see Panel B of Figure~IA.9 of the Internet Appendix), even if only noisy proxies of the true source of risk are among the factors considered. 

That is, our method can correctly recover the pricing of risk in the economy even when the true source of risk is not among the set of tested factors. Nevertheless, as illustrated in Panels C to F of Figure~\ref{fig:simulation}, this goal is achieved by the BMA in two different ways, depending on whether the pseudo-true factor is included among the tested ones or not.

First, when the pseudo-true factor is among the tested ones (experiments I to III), its estimated MPR (Panels C and D) is concentrated on the pseudo-true value, and converges to it as the time series sample size increases (as per Figure~IA.9 of the Internet Appendix), and its posterior probability of being part of the SDF becomes progressively closer to one. On the contrary, the estimated MPRs of the noisy proxies are small and tend to zero as the sample size increases. Similarly, the market price of risk of the useless factor is effectively shrunk to zero. Note that while the posterior probability of the pseudo-true factor goes to one as the sample size increases, the probabilities of the useless factor and noisy proxies do revert to their prior value (Panels E and F). This might seem counterintuitive at first, but it is exactly what should be expected: as the posterior MPR of a given factor goes to zero, the fit of a model that includes that factor becomes indistinguishable from the one of a model that does not include said factor. Hence, the posterior probability of a factor whose MPR is sharply estimated to be close to zero should revert to its prior value---exactly what our method delivers. Note also that such factors, as shown in equation (\ref{E:BMA2a}), will have zero weight in the BMA-SDF (as $\expec [\lambda_j | \text{data}]\rightarrow 0$).

Second, when the pseudo-true factor is \emph{not} among the tested factors (experiments IV to VI), the BMA-SDF still correctly recovers the overall price of risk (Panels A and B), but does so by assigning non-zero MPRs (Panels C and D), and posterior probabilities above their prior values, to the noisy proxies. Furthermore, as in the above-derived analytical results, noisy proxies more correlated with the pseudo-true factor have higher posterior probabilities and MPRs. Nevertheless, even asymptotically (Panel F of Figure~IA.9 of the Internet Appendix), the posterior probability of the noisy proxies will not tend to one---as discussed above, thanks to the self-pricing restriction imposed by our estimator. This also implies that the BMA will not simply select the ``best'' noisy proxy. Instead, it will use multiple proxies in order to maximize the signal, and minimize the noise, that noisy proxies bring to the table.

The robustness of this last result should not be overstated. In the presence of the true factor among the tested ones, the data will always overcome the prior and converge to the truth under standard conditions (see, e.g., \citet[Thm. 7.78]{Schervish_1995}). Nevertheless, when  the true factor is \emph{not} among the tested ones \emph{and} the prior encodes a very high degree of shrinkage (via a very small prior Sharpe ratio), we should expect an attenuation bias in the BMA-SDF-implied MPR in the economy. This is due to the fact that, in the presence of only noisy proxies, no linear combination of them will be able to perfectly price (even asymptotically) both test assets and the factors themselves. Hence, the data will always provide some support for the case in which none of the factors should be included in the SDF, in turn reducing the BMA estimation of the overall MPR achievable with the factors (see, e.g., Panel B of Figure~IA.10 of the Internet Appendix). This does not imply that one should prefer very little or no shrinkage at all, as this is crucial to preempt weak and useless factors from invalidating inference. Hence, exactly as we do in our empirical exercises, one should analyze the sensitivity of the results to the prior degree of shrinkage. 

The above theoretical and simulation-based results stress the robustness of our method in both a large and small sample. Furthermore, they highlight that factor posterior probabilities and market prices of risk carry different, yet salient, information. Hence, both quantities should be tracked and analyzed (as we do in our empirical exploration). For instance, one might find that a given factor has both a posterior probability below its prior value---hence, it is unlikely to be a source of fundamental risk---and a large posterior MPR---since it is highly correlated with the true sources of priced risk, and it will consequently have a large weight in the BMA approximation of the true latent SDF in equation (\ref{E:BMA2a}).  In a nutshell, posterior probabilities tell us which factors should be included in a theoretical model given the data, since they identify the most likely sources of priced risk, while instead posterior market prices of risk tell us which factors should be included (and with what weight) in a portfolio that best approximates the true latent SDF and delivers the maximum achievable Sharpe ratio with the factors at hand.

\section{Estimation results}\label{sec:empirics}

In this section, we apply the hierarchical Bayesian method to a large set of factors proposed in the previous bond and equity literature. Overall, we consider 40 tradable and 14 nontradable factors, yielding  $2^{54} \approx 18$ quadrillion possible models for the combined bond and stock factor zoo. In Sections~\ref{sec:Copricing} and \ref{sec:Properties} we only consider returns for the bond portfolios in excess of the short-term risk-free rate (calculated as outlined in Section~\ref{sec:bond_returns}). In Section~\ref{sec:Information}, we also use duration-adjusted excess returns, as well as U.S. Treasury portfolios, to disentangle the credit and Treasury term structure components of corporate bond returns. 

\subsection{Co-pricing bonds and stocks}\label{sec:Copricing}

We now consider the pricing power of the 54 factors to gauge the extent to which the cross-section of corporate bond and stock returns is priced by the joint factor zoo. The IS test assets include the 50 bond and 33 stock portfolios described in Section~\ref{sec:test_portfolios_IS} in addition to the 40 tradable factor portfolios (for a total $N=123$).  Throughout, we use the continuous spike-and-slab approach described in Section~\ref{sec:econ-method}.  To report the results, we refer to the priors as a fraction of the ex post maximum Sharpe ratio in the data, which is equal to 5.4 annualized for the joint cross-section of portfolios, from a very strong degree of shrinkage (20\%, i.e., a prior annualized Sharpe ratio of 1.0), to a very moderate one (80\% or a prior annualized Sharpe ratio of 4.2). Given that the results demonstrate considerable stability across a wide range of prior Sharpe ratio values, we present selected findings for a prior set at 80\% of the ex post maximum, as this choice tends to yield the best out-of-sample performance.\footnote{Additional results for different values of the prior Sharpe ratio are reported in Table~\ref{tab:table-app-probs} of  \ref{sec:all_Sharpe_priors}.} 

\subsubsection{The co-pricing SDF}\label{sec:Copricing_SDF}

\begin{figure}[tbh!]
\begin{center}
     \includegraphics[scale=.45, trim = 0cm 0.5cm 0cm 0cm,clip]{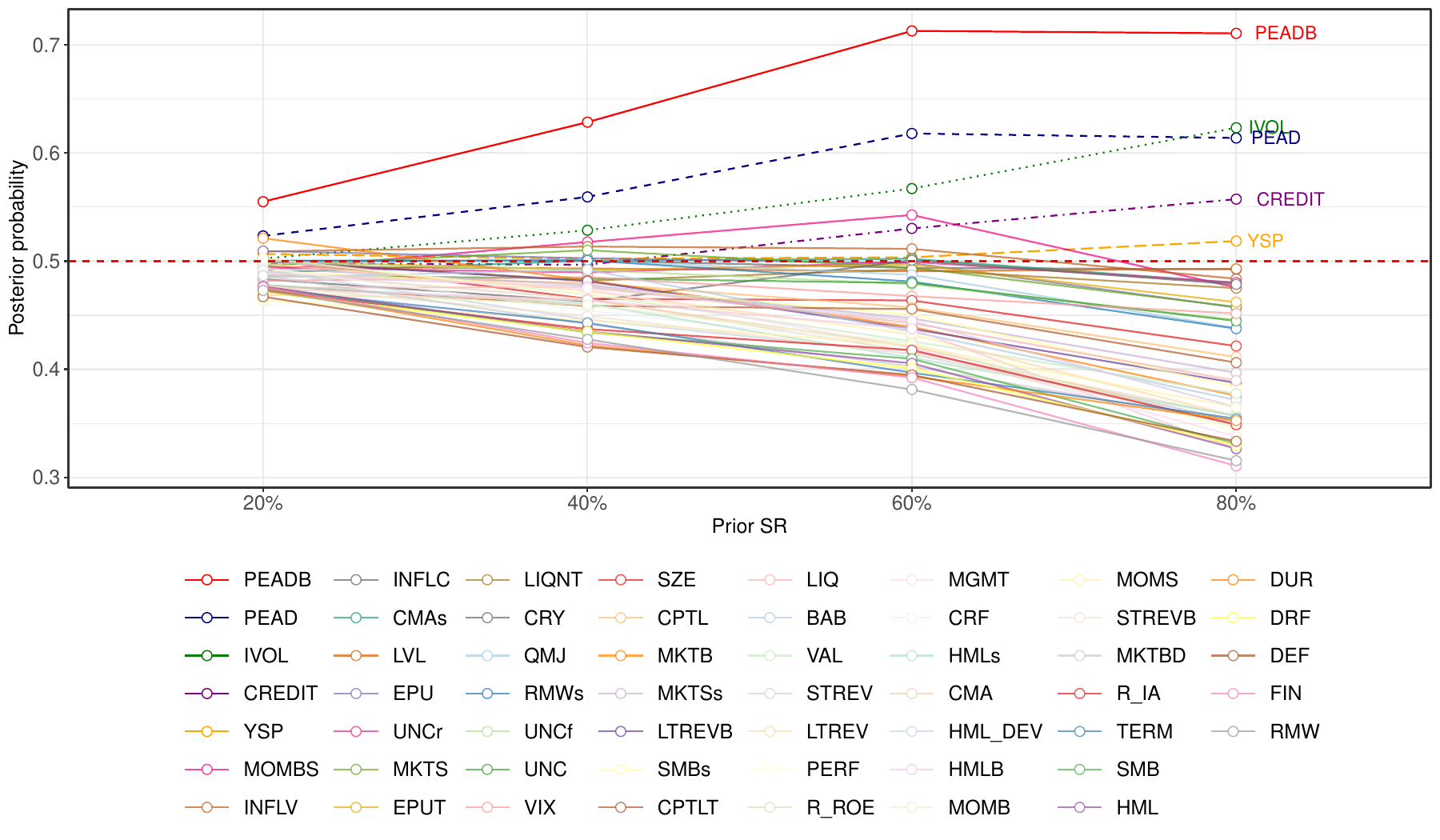} \\
\end{center}     
\vspace{-4mm}
\caption{Posterior factor probabilities: Co-pricing factor zoo.}
\vspace{-2mm}
\begin{justify}
\begin{spacing}{1}
\footnotesize{
Posterior probabilities, $\mathbb{E}[\gamma_j|\text{data}]$, of the 54 bond and stock factors described in \ref{sec:factor_zoo}. The prior for each factor inclusion is a Beta(1, 1), yielding a prior expectation for $\gamma_j$ of 50\%. Results are shown for different values of the prior Sharpe ratio, $\sqrt{\mathbb{E}_\pi [SR^2_{\bm{f}} \mid \sigma^2]}$, with values set to 20\%, 40\%, 60\% and 80\% of the ex post maximum Sharpe ratio of the test assets. Labels are ordered by the average posterior probability across the four levels of shrinkage. Test assets are the 83 bond and stock portfolios and 40 tradable bond and stock factors described in Section~\ref{sec:data}. The sample period is 1986:01 to 2022:12 ($T = 444$).
}
\end{spacing}
\end{justify}
\vspace{-4mm}
\label{Fig:post_probs}
\end{figure}

We start by asking which factors are likely components of the latent SDF in the economy. Figure~\ref{Fig:post_probs} reports the posterior probabilities (given the data) of each factor (i.e., $\mathbb{E}[\gamma_j|\text{data}], \forall j$) for different values of the prior Sharpe ratio achievable with the linear SDF (expressed as a percentage of the ex post maximum Sharpe ratio). See Table~\ref{tab:table-001-appendix} of \ref{sec:factor_zoo} for a detailed description of the factors.

Recall that we have a uniform (hence flat) prior for the model dimensionality and each factor has an ex ante expected probability of being selected equal to 50\%, depicted by the dashed horizontal line in Figure~\ref{Fig:post_probs}. Several observations are in order. First, with some notable exceptions, most factors proposed in the corporate bond and equity literatures have (individually) a posterior probability of being part of the SDF that is below its prior value of 50\%. That is, given the data, they are unlikely sources of fundamental risks.

Second, given that their posterior probabilities are above the prior 50\% value for the entire range of prior Sharpe ratios considered, five factors are identified as likely sources of fundamental risk in the bond and equity markets. In particular, there is strong evidence for including two tradable factors, PEADB and PEAD (i.e., respectively, the bond and stock post-earnings announcement drift factors), as a source of priced risk in the SDF. Partially, this is a surprising result, as PEADB has not specifically been proposed as a priced risk factor in the previous literature. \citet{NozawaQiuXiong_2025} are the first to document a post-earnings announcement drift in corporate bond prices, and they rationalize their finding with a stylized model of disagreement. They also show that a strategy that purchases bonds issued by firms with high earnings surprises and sells bonds of firms with low earnings surprises generates sizable Sharpe ratios and large risk-adjusted returns. On the other hand, \citet{BryzgalovaHuangJulliard_2023} and \citet{AvramovChengMetzkerVoigt_2023} find strong evidence that the \textit{stock market} post-earnings announcement drift (PEAD) factor of \citet{DanielHirshleiferSun_2020} exhibits a particularly high posterior probability of being part of the SDF for stock returns. In fact, PEAD is the only other tradable factor with a posterior probability of being part of the SDF that prices the joint cross-section of corporate bond and stock returns that is above 50\%. That is, the only two tradable factors with high posterior probabilities are the bond and stock versions of the post-earnings announcement drift. 
Note that, in equilibrium models in which rational agents with limited risk-bearing capacity face behavioral asset demand, the drivers of the latter become part of the pricing measure---exactly as we find (see, e.g., \citet{DeLongShleiferSummersWaldmann_1990}). Note also that, as shown in Table~IA.III of the Internet Appendix, these are the tradable factors with the highest Sharpe ratio in our full sample. Moreover, PEADB has the highest Sharpe ratio among bond factors when the sample is split in half, while PEAD has the highest Sharpe ratio among stock factors in the first half, and one of the highest in the second half of the sample (see Table~IA.IV of the Internet Appendix).\footnote{Despite its reduced \emph{time series} predictability in most recent data (see, e.g., \citet{Martineau_2022}), we document remarkable stability of the post-earnings announcement drift for forming long-short corporate bond and stock portfolios across subsamples in Internet Appendix~IA.4. That is, the \emph{cross-sectional} predictability of the post-earnings announcement drift within a portfolio context remains robust and does not appear to be driven by micro-cap stocks.}

Furthermore, the \textit{nontradable} idiosyncratic equity volatility factor (IVOL) of \cite{CampbellTaksler_2003} is supported by the data as a fundamental source of priced risk. Interestingly, the rationale behind this factor closely connects bond and stock markets: as per the seminal insight of \cite{Merton_1974}, equity claims are akin to a call option on the value of the assets of the firm, while the debt claim contains a short put option on the same. Consequently, \cite{CampbellTaksler_2003} suggest that changes in the firm's volatility should be expected to affect bond and stock prices.\footnote{See, e.g., \citet{DickersonFournierJeanneretMueller_2025} for a model of the correlation of bonds and stocks of the same firm.}

Additionally, two more \textit{nontradable} factors have posterior probabilities of being part of the SDF above 50\% for all values of the prior Sharpe ratio: the slope of the Treasury yield term structure (YSP, \citet{KoijenLustigVanNieuwerburgh_2017}), a well-known predictor of business cycle variation, and the AAA/BAA yield spread (CREDIT, \citet{FamaFrench_1993}), a common metric of the risk compensation differential between safer and riskier securities. Interestingly, the term premium and default risk factors are originally suggested in \cite{FamaFrench_1993} exactly for the purpose of co-pricing bonds and stocks.

\begin{figure}[tb]%
\begin{center}
    \includegraphics[scale=.45, trim = 0cm 0.5cm 0cm 0cm,clip]{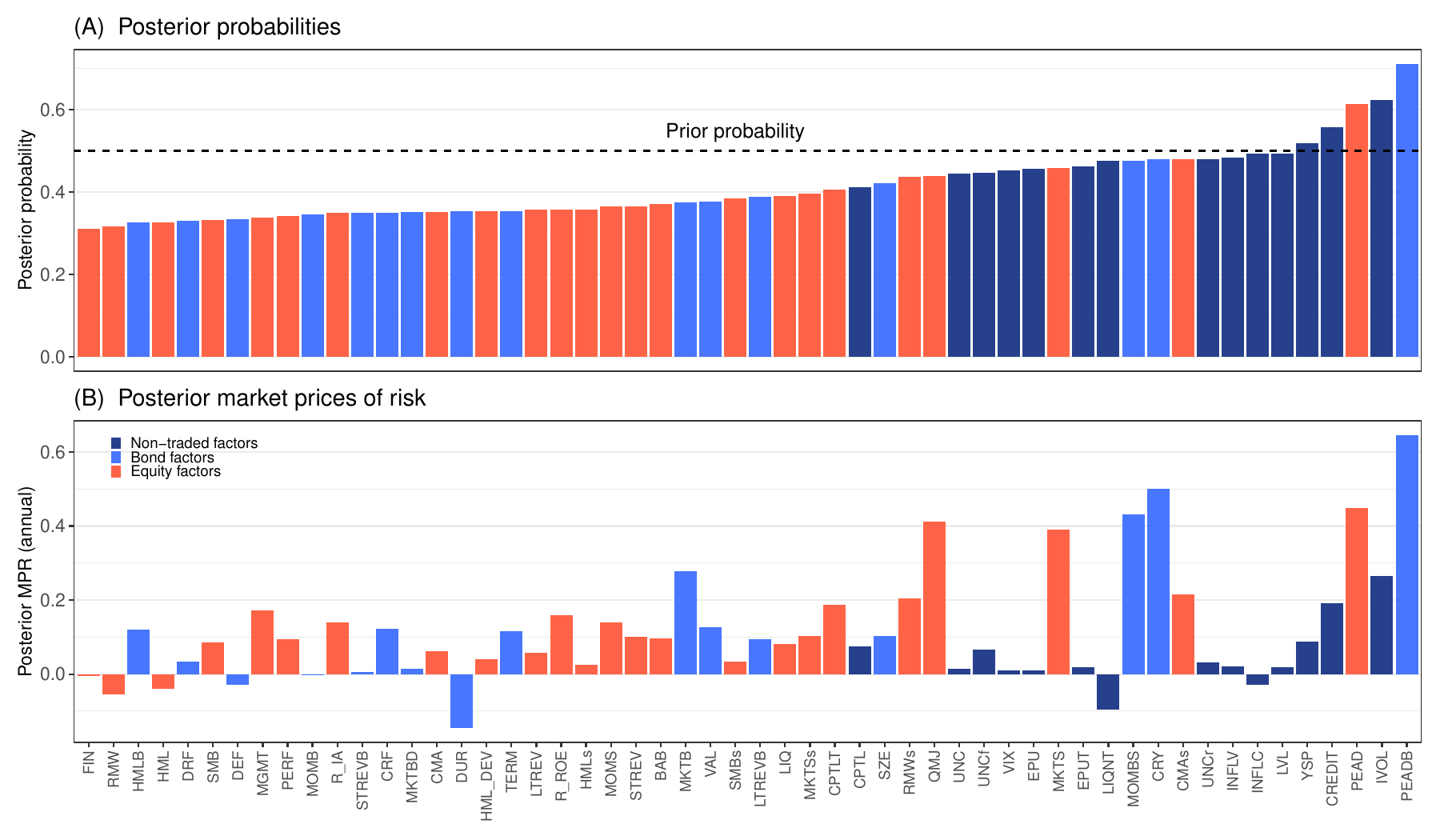} \\
\end{center} 
\vspace{-4mm}
\caption{Posterior factor probabilities and risk prices: Joint factor zoo (excess bond returns).}
\vspace{-2mm}
\begin{justify}
\begin{spacing}{1}
\footnotesize{
The figure reports posterior probabilities, $\mathbb{E}[\gamma_j|\text{data}]$, and posterior means of annualized market prices of risk, $\mathbb{E}[\lambda_j|\text{data}]$, of the 54 bond and stock factors described in \ref{sec:factor_zoo}. The prior for each factor inclusion is a Beta(1, 1), yielding a prior expectation for $\gamma_j$ of 50\%. The prior Sharpe ratio is set to 80\% of the ex post maximum Sharpe ratio of the 83 stock and bond portfolios and the 40 tradable factors described in Section~\ref{sec:data}. 
The sample period is 1986:01 to 2022:12 ($T = 444$).
}
\end{spacing}
\end{justify}
\vspace{-4mm}
\label{Fig:post_probs_mprs}
\end{figure}

Third, there are a few factors for which the posterior probability is roughly equal to the prior (implying that at least some of these factors are likely to be weakly identified at best), and there is a large set of factors that are \textit{individually} unlikely to be sources of fundamental risk in the SDF pricing the joint cross-section of bond and stock returns.  In particular, besides PEADB and PEAD, \textit{all} tradable bond and stock market factors are individually unlikely to capture fundamental risk in the SDF. For instance, with a prior Sharpe ratio set to 80\% of the ex post maximum, the posterior probabilities for 30 of the 40 tradable bond and stock factors are below 40\% (see Figure~\ref{Fig:post_probs} as well as the top panel of Figure~\ref{Fig:post_probs_mprs}). Nevertheless, as shown theoretically and in the simulation in Section~\ref{sec:ModelSel}, and discussed extensively below, this does \textit{not} imply that these factors, \textit{jointly}, do not carry relevant information to characterize the true latent SDF.

Notably, the stock as well as the bond market factors (MKTS and MKTB, respectively) both exhibit posterior probabilities below 50\% for almost the full range of prior Sharpe ratios for the joint cross-section of returns. Nevertheless, when separately pricing the cross-sections of stock and bond returns with only the factors in their respective zoos, both market indices become likely components of the SDF: for all prior levels in the MKTS case, and for all but one in the MKTB case
(see Tables IA.V and IA.VI of the Internet Appendix). This confirms the finding that the equity market index contains valuable information for pricing stocks in an unconstrained SDF based on stock factors only (as in \citet{BryzgalovaHuangJulliard_2023}). However, when the space of potential factors is expanded to include both stock and bond factors, without dimensionality restrictions on the SDF as we do in our baseline co-pricing exercise, models with MKTS (and more so in the MKTB case) overall perform worse than denser models containing factors from both zoos. That is, the information in the market indices appears to be spanned by the other factors in the zoo. This finding is unlikely to be driven by the market indices acting as ``level'' or ``weak'' factors since asset returns display large and well-dispersed loadings on these factors, and the market prices of risk they command are substantial when included in the SDF (see Table~\ref{tab:table-app-probs} of \ref{sec:all_Sharpe_priors} and the bottom panel of Figure~\ref{Fig:post_probs_mprs}). Moreover, we show in Internet Appendix~IA.3.1 that removing the free intercept, and the prior penalization of pure level factors, leaves \textit{all} of the above results virtually unchanged.

Given the focus of most (yet not all) of the previous literature on selecting models characterized by a small number of factors, the above findings raise the question of whether the handful of most likely factors that we have identified are enough to capture the span of the true, latent, SDF that jointly prices bonds and stocks. Moreover, are factors less likely to be sources of fundamental risk really devoid of useful pricing information? Since our Bayesian method does not ex ante impose the existence of a unique, low-dimensional, and correctly specified model---all assumptions that are needed with conventional frequentist asset pricing methods---we can formally answer these questions. 

\begin{figure}[tbh!]
\begin{center}
    	\includegraphics[scale=.45, trim = 0cm 0.5cm 0cm 0cm,clip]{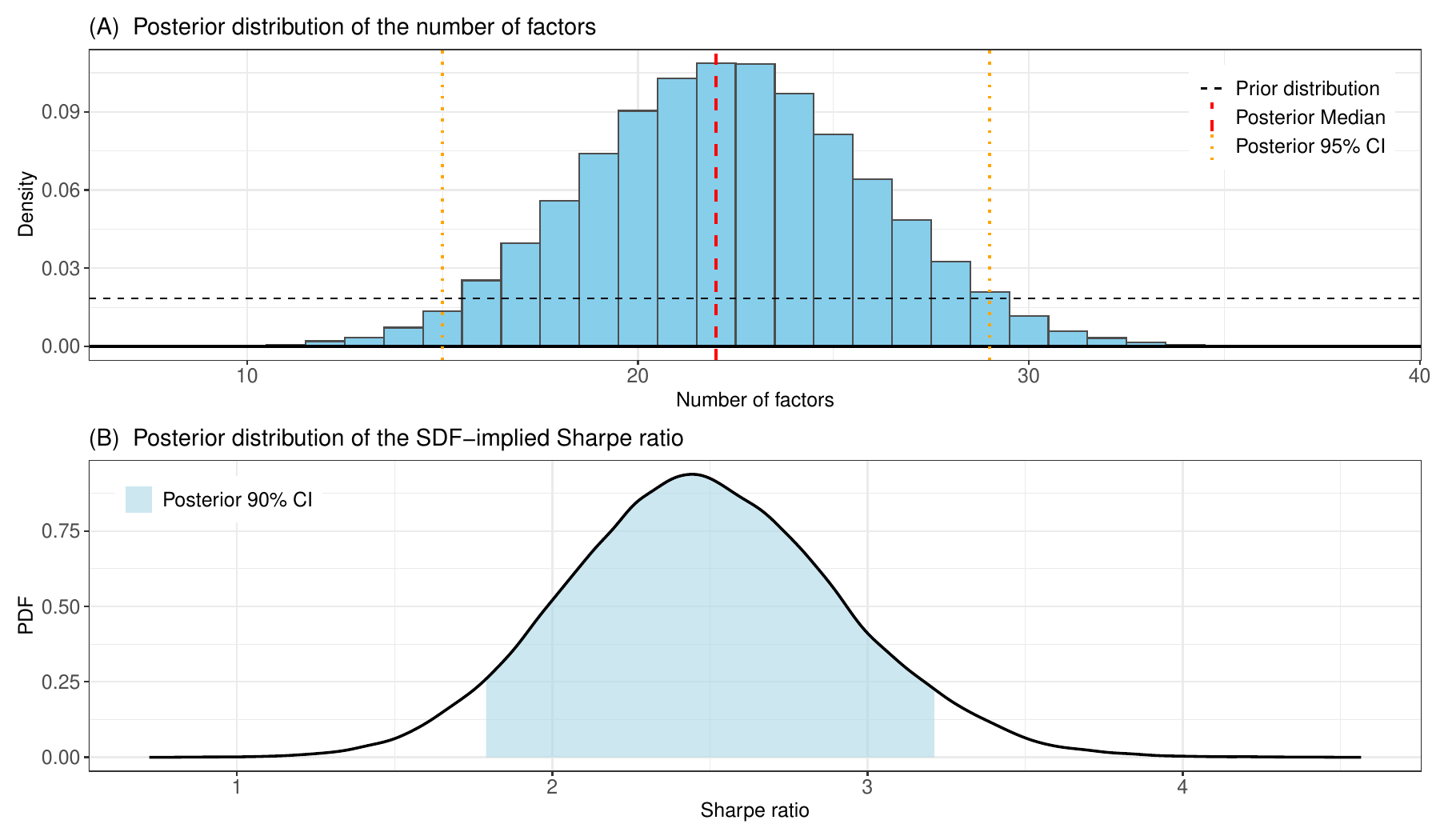} \\
\end{center}
\vspace{-4mm}
\caption{Posterior SDF dimensionality and Sharpe ratios: Co-pricing factor zoo.}
\vspace{-2mm}
\begin{justify}
\begin{spacing}{1}
\footnotesize{
Posterior distributions of the number of factors to be included in the co-pricing SDF (top panel) and of the SDF-implied Sharpe ratio (bottom panel), computed using the 54 bond and stock factors described in \ref{sec:factor_zoo}. The prior distribution for the $j^{\text{th}}$ factor inclusion is a Beta(1, 1), yielding a flat prior for the SDF dimensionality depicted in the top panel. The prior Sharpe ratio is set to 80\% of the ex post maximum Sharpe ratio of the 83 bond and stock portfolios and 40 tradable factors described in Section~\ref{sec:data}. The sample period is 1986:01 to 2022:12 ($T = 444$). }
\end{spacing}
\end{justify}
\vspace{-4mm}
\label{Fig:post_dim}
\end{figure}

The top panel of Figure~\ref{Fig:post_dim} reports the posterior dimensionality of the SDF  in terms of observable factors to be included in it, and the bottom panel shows the posterior distribution of the Sharpe ratios achievable with such an SDF. It is evident that the \textit{sparse} models suggested in the previous literature have very weak support in the data, and are misspecified with very high probability, as a substantial number of factors is needed to capture the span of the true latent SDF: the posterior median number of factors is 22 with a centered 95\% coverage of 15 to 29 factors. In fact, the posterior probability of a model with less than 10 factors is virtually zero, indicating that the quest for a sparse, unique, SDF model among the observable factors in the joint bond and stock factor zoo is misguided at best.

But, as often argued, wouldn't a \textit{dense} SDF imply an unrealistically high Sharpe ratio achievable in the market? The bottom panel of Figure~\ref{Fig:post_dim} highlights that the SDF-implied Sharpe ratio is not unrealistically large (recall that the ex post maximum Sharpe ratio in the data is 5.4), suggesting that many factors are likely to span a lot of common risks. Furthermore, Table \nolinebreak \ref{tab:table-model-dim2-top-non-top} shows that albeit the most likely (top five) factors to be included in the SDF for pricing bonds and stocks (jointly or separately) are responsible for a substantial share of the Sharpe ratio (e.g., $\mathbb{E}[SR_f|\text{data}]$ ranges from $0.78$ to $1.46$ for a $60\%$ to $80\%$ prior), the share of the SDF squared Sharpe ratio generated by these factors alone ($\mathbb{E}\big[SR^2_f/SR^2_m|\text{data}\big]$) is quite limited. This means that there is substantial additional priced risk in the factor zoo that is \emph{not} captured by the most likely factors. That is, the \textit{less} likely factors are noisy proxies for latent fundamental risks and are needed, \textit{jointly}, to provide an accurate characterization of the risks priced by the true latent SDF. This feature of the data arises not only when jointly pricing  bonds and stocks (Panel A), but also when separately focusing  on the pricing of the two asset classes using their respective factor zoos (Panels B and C).

\begin{table}[tb!]
\begin{center}
\caption{Most likely (top five) factor contribution to the SDF}\label{tab:table-model-dim2-top-non-top}\vspace{-2mm}
\scalebox{.8}{
\begin{tabular}{lcccccccccccccc}
\toprule
& \multicolumn{4}{c}{\textbf{Panel A}: Co-pricing SDF} & & \multicolumn{4}{c}{\textbf{Panel B}: Bond SDF} & & \multicolumn{4}{c}{\textbf{Panel C}: Stock SDF} \\ \cmidrule{2-5}\cmidrule{7-10}\cmidrule{12-15}
\text{Total prior SR: }& 20\% & 40\% & 60\% & 80\% & & 20\% & 40\% & 60\% & 80\% & & 20\% & 40\% & 60\% & 80\% \\ \midrule
$\mathbb{E}[SR_f|\text{data}]$ & 0.26 & 0.57 & 1.06 & 1.24 &  & 0.28 & 0.71 & 1.10 & 1.46 &  & 0.17 & 0.42 & 0.78 & 1.10 \\
$\mathbb{E}\left[\frac{SR^2_f}{SR^2_m}|\text{data}\right]$ & 0.13 & 0.20 & 0.32 & 0.28 &  & 0.34 & 0.57 & 0.65 & 0.70 &  & 0.12 & 0.22 & 0.35 & 0.42 \\
\bottomrule
\end{tabular}
}
\end{center}
\begin{spacing}{1}
	{\footnotesize Posterior mean of implied Sharpe ratios achievable with the most likely (top five) factors, $\mathbb{E}[SR_f|\text{data}]$, and their share of the SDF squared Sharpe ratio, $\mathbb{E}\big[SR^2_f/SR^2_m|\text{data}\big]$. 
    Panels A, B and C report results using the corresponding factor zoos, for the co-pricing, bond-only, and stock-only BMA-SDFs, respectively. 
    Top five co-pricing factors are PEADB, IVOL, PEAD, CREDIT and YSP.
    Top five bond factors are PEADB, CREDIT, MOMBS, YSP and IVOL.
    Top five stock factors are PEAD, IVOL, MKTS, CMAs and LVL.
    The total prior Sharpe ratio  is expressed as a share of the ex post maximum Sharpe ratio of the test assets. 
}
\end{spacing}
\vspace{-4mm}
\end{table}

As shown in Section~\ref{sec:ModelSel}, if a dominant, low-dimensional, model is not supported by the data---as the above evidence implies---we can optimally aggregate the pricing information in the factor zoo by constructing a Bayesian model averaging of all possible models. Moreover, the \textit{model averaging} is equivalent to a \textit{factor averaging}, where the weights of the individual factors are simply the factors' posterior market prices of risk ($\mathbb{E}[\lambda_j|\text{data}]$). Hence, large posterior market prices of risk reveal which factors (true sources of risk or noisy proxies) are useful in approximating the true, latent, SDF.

In Figure~\ref{Fig:post_probs_mprs} we list all 54 factors in increasing order of posterior probabilities (i.e., $\Pr(\gamma_j =1 | \text{data})$, top panel), for a prior Sharpe ratio of 80\% of the maximum ex post Sharpe ratio, along with the corresponding annualized posterior means of the price of risk of the factors ($\mathbb{E}[\lambda_j|\text{data}]$, bottom panel). Posterior probabilities and market prices of risk for different priors are tabulated in Table~\ref{tab:table-app-probs} of  \ref{sec:all_Sharpe_priors}. 

All five factors with posterior probabilities higher than their prior values (i.e., PEADB, IVOL, PEAD, CREDIT and YSP) command substantial market prices of risk, implying a considerable weight in a portfolio that best approximates the true latent SDF. Hence, not only does the data support their inclusion in the SDF, but they also play an important role in its BMA estimate.

Out of the next fifteen factors with the highest (individual) posterior probabilities, ten are  nontradable in nature. Nevertheless, the risk prices of several of these nontradable factors are small and, in some cases, effectively shrunk toward zero. This is due to the fact that these are likely \textit{weak factors} in the joint cross-section of corporate bond and stock returns and, consequently, carry a near-zero weight in the portfolio that approximates the SDF.\footnote{That is, their correlations with the test assets are small and have little cross-sectional dispersion. See, e.g., \citet{GospodinovKanRobotti_2019} and \cite{Kleibergen_2009} for a formal definition of weak and level factors.} The occurrence of weak factors, which, in fact, is most common among the nontradable ones, causes identification failure and invalidates canonical estimation approaches (e.g., GMM, MLE, and two-pass regressions). This is \textit{not} an issue for our Bayesian method, which restores inference by design, by regularizing the marginal likelihood. Furthermore, for these factors, both shown theoretically and in the simulation in Section~\ref{sec:ModelSel}, the posterior probabilities revert to their prior value as the market prices of risk tend to zero.

Interestingly, several factors with posterior probabilities below their prior values---hence unlikely sources of fundamental risk---do carry very sizeable posterior market prices of risk. For example, the equity market index factor carries the third largest MPR among equity factors and the sixth largest among the tradable ones. Section~\ref{sec:ModelSel} informs us exactly how to interpret such findings: these are factors that the data do not support as being fundamental sources of risk (hence the posterior probability being below the prior value), but that nevertheless have a high correlation with the true latent priced risk and, hence, feature prominently in the BMA-SDF to provide an accurate approximation of the true latent SDF.

This aggregation property of the BMA-SDF is clearly displayed in Figure~IA.12 of the Internet Appendix, where we plot the cumulative SDF-implied Sharpe ratio when subsequently adding factors ordered by their (individual) posterior probability. While the Sharpe ratio increases with the number of factors, some factors seem to contribute more to the implied Sharpe ratio than others. For example, the factors ranked 6 to 9 (LVL, INFLC, INFLV, UNCr) do not appear to add much individually, while the Sharpe ratio increases markedly once factors 10 (CMAs) and 11 (CRY) are included. This is because many factors are potentially noisy proxies for the same fundamental sources of risk that are important for the SDF. As shown in Section~\ref{sec:ModelSel}, factors that are useful noisy proxies for a particular fundamental source of risk not fully spanned by individual factors will exhibit nonzero market prices of risk (or portfolio weights). However, the Sharpe ratio only substantially increases once the first of the factors spanning (at least partially) a common risk is included in the analysis. In contrast, subsequent factors spanning the same risk generate a much smaller increase in the Sharpe ratio due to the enhanced signal extraction of the common risk. Further examining the four factors in positions 8 to 11, these are all nontradable in nature and related to inflation, interest rates, and uncertainty. Similarly, factors in positions 16 to 19 are all related to different measures of macroeconomic uncertainty. While it is important to include all of these factors in the SDF to increase the signal to noise ratio of latent fundamental risk, their individual marginal contribution to the Sharpe ratio may be minimal as they share common spanning. This is highlighted by the posterior confidence interval in the figure. As more factors are added sequentially, one might expect the posterior uncertainty to increase, as the uncertainty about the individual market prices of risk is compounded in the SDF. Nevertheless, the opposite occurs in Figure~IA.12---overall, the posterior confidence region \emph{shrinks} as factors are added. Moreover, the last few factors have virtually no effect on the posterior mean of the Sharpe ratio, but they do reduce the confidence region significantly, as the BMA aggregation increases the signal to noise ratio.

\subsubsection{Cross-sectional asset pricing}\label{sec:xs-pricing}

We now turn to the asset pricing performance of the BMA-SDF based on the joint cross-section and factor zoos, as well as based on bond and stock portfolios separately. In Table~\ref{tab:tab-is-pricing-excess} we report results for in-sample cross-sectional pricing using various performance measures, while out-of-sample results are summarized in Table~\ref{tab:tab-os-pricing-excess}. The in-sample assets for the joint cross-section in Panel A of Table~\ref{tab:tab-is-pricing-excess} are the 83 portfolios of bonds and stocks (described in Section~\ref{sec:test_portfolios_OOS}) plus 40 tradable factors. Panels B and C focus only on bonds (50 portfolios and 16 bond tradable factors) and stocks (33 anomaly portfolios and 24 stock tradable factors), respectively. The out-of-sample test assets in Table~\ref{tab:tab-os-pricing-excess} comprise 77 bond portfolios and 77 stock portfolios (described in Section~\ref{sec:test_portfolios_OOS}), which are considered jointly in Panel \nolinebreak A and separately in Panels B and C.

When assessing the pricing performance, we compare our BMA-SDF for different levels of prior Sharpe ratio shrinkage with the performance of a number of benchmark models. In particular, we consider the bond CAPM (CAPMB), the stock CAPM, the  \cite{FamaFrench_1993} five-factor model (FF5), the intermediary asset pricing model of \citet{HeKellyManela_2017} (HKM), the PCA-based SDF of \citet{KozakNagelSantosh_2020} (KNS) and the risk premia PCA approach of \cite{LettauPelger_2020} (RPPCA).\footnote{The SDFs of both KNS and RPPCA are re-estimated using our data and the methods proposed in the original papers. Details of the estimation for all benchmark models are reported in \ref{sec:benchmark_models}.} In addition, since most of the previous literature focuses on selection (rather than aggregation) of pricing factors, we also include the respective `top' factor models (TOP) from our Bayesian analysis that comprise only the five factors with the highest posterior probabilities
(for the joint cross-section for example, this is a five-factor model with PEADB, IVOL, PEAD, CREDIT, and YSP). All the benchmark model SDFs are estimated via a GLS version of GMM.\footnote{See \ref{sec:benchmark_models} for further details.} Note that for the cross-sectional OS pricing, we do not refit the BMA-SDF or the other benchmark models to the new data. Instead, we use the estimated parameters from the respective IS pricing exercises.

For the in-sample pricing in Table~\ref{tab:tab-is-pricing-excess}, a few observations are in order. First and foremost, the BMA-SDF using moderate shrinkage (80\% of the prior Sharpe ratio) outperforms virtually all benchmark models on almost all dimensions considered, with the best alternative model being KNS. Second, no low dimensional model performs well. This should not come as a surprise given the discussion in Section~\ref{sec:Dimensionality}, which implies that all low-dimensional models are both misspecified with a very high probability and are strongly rejected by the data. In fact, the performance of both the bond and stock CAPM is rather disappointing compared to the BMA-SDF. Moreover, popular models such as FF5 and HKM do not perform particularly well either. Third, the low dimensional TOP factor model, albeit better performing than the low dimensional models from the literature, delivers inferior pricing compared to the BMA-SDF with moderate shrinkage, once again pointing out that aggregation of factors, rather than selection, is preferred by the data. This highlights
that just the most likely factors are not sufficient to provide an accurate characterization of the risks spanned by the true latent SDF. Fourth, the results are fairly consistent across the three panels. Apart from the BMA-SDF, KNS, and
RPPCA deliver consistently better IS pricing performance than the low dimensional models.

\begin{table}[tbh!]
\begin{center}
\caption{In-sample cross-sectional asset pricing performance}\label{tab:tab-is-pricing-excess}\vspace{-2mm}
\scalebox{.8}{
\begin{tabular}{lcccc|ccccccc}\toprule
 & \multicolumn{4}{c}{BMA-SDF prior Sharpe ratio} & CAPM & CAPMB & FF5 & HKM & TOP & KNS & RPPCA \\ \cmidrule(lr){2-5}
 & 20\% & 40\% & 60\% & \multicolumn{1}{c}{80\%} &  &  &  &  &  &  &  \\ \midrule
\multicolumn{12}{c}{\textbf{Panel A:} Co-pricing bonds and stocks} \\ \midrule
RMSE & 0.214 & 0.203 & 0.185 & 0.167 & 0.260 & 0.278 & 0.258 & 0.259 & 0.230 & 0.166 & 0.197 \\
MAPE & 0.167 & 0.154 & 0.139 & 0.125 & 0.194 & 0.221 & 0.198 & 0.192 & 0.171 & 0.126 & 0.132 \\
$R^2_{\text{OLS}}$ & 0.155 & 0.240 & 0.367 & 0.487 & $-$0.244 & $-$0.426 & $-$0.233 & $-$0.238 & 0.023 & 0.489 & 0.282 \\
$R^2_{\text{GLS}}$ & 0.106 & 0.168 & 0.232 & 0.285 & 0.078 & 0.083 & 0.087 & 0.078 & 0.263 & 0.176 & 0.267 \\
 \midrule
\multicolumn{12}{c}{\textbf{Panel B}: Pricing bonds} \\ \midrule
RMSE & 0.180 & 0.148 & 0.121 & 0.104 & 0.209 & 0.214 & 0.201 & 0.206 & 0.162 & 0.192 & 0.091 \\
MAPE & 0.129 & 0.109 & 0.091 & 0.079 & 0.146 & 0.135 & 0.143 & 0.146 & 0.128 & 0.111 & 0.067 \\
$R^2_{\text{OLS}}$ & 0.196 & 0.455 & 0.638 & 0.733 & $-$0.083 & $-$0.134 & $-$0.006 & $-$0.049 & 0.347 & 0.088 & 0.794 \\
$R^2_{\text{GLS}}$ & 0.211 & 0.299 & 0.381 & 0.444 & 0.172 & 0.195 & 0.238 & 0.175 & 0.549 & 0.071 & 0.419 \\
 \midrule
\multicolumn{12}{c}{\textbf{Panel C}: Pricing stocks} \\ \midrule
RMSE & 0.230 & 0.241 & 0.236 & 0.220 & 0.292 & 0.264 & 0.275 & 0.292 & 0.352 & 0.162 & 0.175 \\
MAPE & 0.186 & 0.189 & 0.181 & 0.166 & 0.229 & 0.211 & 0.221 & 0.226 & 0.294 & 0.133 & 0.141 \\
$R^2_{\text{OLS}}$ & 0.023 & $-$0.075 & $-$0.029 & 0.103 & $-$0.570 & $-$0.282 & $-$0.392 & $-$0.574 & $-$1.288 & 0.515 & 0.433 \\
$R^2_{\text{GLS}}$ & 0.145 & 0.213 & 0.287 & 0.353 & 0.120 & 0.118 & 0.130 & 0.121 & 0.330 & 0.311 & 0.493 \\
 \bottomrule
\end{tabular}
}
\end{center}
\begin{spacing}{1}
	{\footnotesize 
The table presents the cross-sectional in-sample asset pricing performance of different models pricing bonds and stocks jointly (Panel A), bonds only (Panel B) and stocks only (Panel C), respectively. For the BMA-SDF, we provide results for prior Sharpe ratio  values set to 20\%, 40\%, 60\% and 80\% of the ex post maximum Sharpe ratio of the test assets. TOP includes the top five factors with an average posterior probability greater than 50\%. CAPM is the standard single-factor model using MKTS, and CAPMB is the bond version using MKTB. FF5 is the five-factor model of \cite{FamaFrench_1993}, HKM is the two-factor model of \citet{HeKellyManela_2017}. KNS stands for the SDF estimation of \citet{KozakNagelSantosh_2020} and RPPCA is the  risk premia PCA of \cite{LettauPelger_2020}. Estimation details for the benchmark models are given in \ref{sec:benchmark_models}. Bond returns are computed in excess of the one-month risk-free rate of return.
By panel the models are estimated with the respective factor zoos and test assets. Test assets are the 83 bond and stock portfolios and the 40 tradable bond and stock factors (Panel A), the 50 bond portfolios and  16 tradable bond factors (Panel B), and the 33 stock portfolios and 24 tradable stock factors (Panel C), respectively. All are described in Section~\ref{sec:data}. All data are standardized, that is, pricing errors are in Sharpe ratio units. The sample period is 1986:01 to 2022:12 ($T=444$).
}
\end{spacing}
\vspace{-4mm}
\end{table}

\begin{table}[tbh!]
\begin{center}
\caption{Out-of-sample cross-sectional asset pricing performance}\label{tab:tab-os-pricing-excess}\vspace{-2mm}
\scalebox{.8}{
\begin{tabular}{lcccc|ccccccc} \toprule
 & \multicolumn{4}{c}{BMA-SDF prior Sharpe ratio} & CAPM & CAPMB & FF5 & HKM & TOP & KNS & RPPCA \\ \cmidrule(lr){2-5}
 & 20\% & 40\% & 60\% & \multicolumn{1}{c}{80\%} &  &  &  &  &  &  &  \\ \midrule
\multicolumn{12}{c}{\textbf{Panel A}: Co-pricing bonds and stocks} \\ \midrule
RMSE & 0.114 & 0.102 & 0.095 & 0.090 & 0.224 & 0.154 & 0.139 & 0.223 & 0.171 & 0.160 & 0.153 \\
MAPE & 0.081 & 0.074 & 0.069 & 0.065 & 0.192 & 0.129 & 0.102 & 0.190 & 0.135 & 0.143 & 0.130 \\
$R^2_{\text{OLS}}$ & 0.357 & 0.489 & 0.557 & 0.603 & $-$1.478 & $-$0.161 & 0.053 & $-$1.444 & $-$0.442 & $-$0.268 & $-$0.159 \\
$R^2_{\text{GLS}}$ & 0.038 & 0.070 & 0.098 & 0.124 & 0.028 & 0.034 & 0.036 & 0.028 & 0.090 & 0.065 & 0.028 \\
 \midrule
\multicolumn{12}{c}{\textbf{Panel B}: Pricing bonds} \\ \midrule
RMSE & 0.123 & 0.116 & 0.110 & 0.106 & 0.129 & 0.128 & 0.140 & 0.133 & 0.102 & 0.114 & 0.100 \\
MAPE & 0.090 & 0.085 & 0.081 & 0.079 & 0.094 & 0.092 & 0.104 & 0.098 & 0.084 & 0.083 & 0.073 \\
$R^2_{\text{OLS}}$ & 0.051 & 0.156 & 0.237 & 0.296 & $-$0.051 & $-$0.029 & $-$0.231 & $-$0.112 & 0.342 & 0.180 & 0.375 \\
$R^2_{\text{GLS}}$ & 0.019 & 0.056 & 0.081 & 0.102 & $-$0.004 & 0.024 & $-$0.032 & $-$0.007 & 0.101 & 0.066 & 0.045 \\
 \midrule
\multicolumn{12}{c}{\textbf{Panel C}: Pricing stocks} \\ \midrule
RMSE & 0.105 & 0.088 & 0.077 & 0.070 & 0.123 & 0.119 & 0.116 & 0.124 & 0.149 & 0.078 & 0.104 \\
MAPE & 0.078 & 0.067 & 0.062 & 0.057 & 0.089 & 0.085 & 0.082 & 0.091 & 0.115 & 0.060 & 0.082 \\
$R^2_{\text{OLS}}$ & 0.298 & 0.508 & 0.620 & 0.683 & 0.032 & 0.099 & 0.136 & 0.019 & $-$0.422 & 0.613 & 0.305 \\
$R^2_{\text{GLS}}$ & 0.090 & 0.160 & 0.227 & 0.280 & 0.103 & 0.065 & 0.099 & 0.107 & 0.079 & 0.207 & 0.072 \\
 \bottomrule
\end{tabular}
}
\end{center}
\begin{spacing}{1}
    {\footnotesize 
The table presents the cross-sectional out-of-sample asset pricing performance of different models pricing bonds and stocks jointly (Panel A), bonds only (Panel B) and stocks only (Panel C), respectively. For the BMA-SDF, we provide results for prior Sharpe ratio values set to 20\%, 40\%, 60\% and 80\% of the ex post maximum Sharpe ratio of the test assets. TOP includes the top five factors with an average posterior probability greater than 50\%. CAPM is the standard single-factor model using MKTS, and CAPMB is the bond version using MKTB. FF5 is the five-factor model of \cite{FamaFrench_1993}, HKM is the two-factor model of \citet{HeKellyManela_2017}. KNS stands for the SDF estimation of \citet{KozakNagelSantosh_2020} and RPPCA is the  risk premia PCA of \cite{LettauPelger_2020}. Estimation details for the benchmark models are given in \ref{sec:benchmark_models}. Bond returns are computed in excess of the one-month risk-free rate of return. The models are first estimated using the baseline IS test assets. The resulting SDF is then used to price (with no additional parameter estimation) each set of the OS assets. The IS test assets are the same as in Table~\ref{tab:tab-is-pricing-excess}. OS test assets are the combined 154 bond and stock portfolios (Panel A), as well as the separate 77 bond and stock portfolios (Panels B and C). All are described in Section~\ref{sec:data}. All data are standardized, that is, pricing errors are in Sharpe ratio units. The sample period is 1986:01 to 2022:12 ($T=444$).
}
\end{spacing}
\vspace{-4mm}
\end{table}

The co-pricing BMA-SDF performs exceptionally well out-of-sample (see Panel A of Table \nolinebreak \ref{tab:tab-os-pricing-excess}). While KNS is a close contender regarding in-sample performance, the BMA-SDF strongly dominates KNS out-of-sample. In Internet Appendix~IA.3.2 we show that the strong OS performance of the co-pricing BMA-SDF is not driven by the specific, yet rich, selection of test assets in our baseline analysis presented here. In particular, we compare the performance of the BMA-SDF vis-\`{a}-vis the closest competitor, KNS, across $2^{14}-1 = 16,383$ OS cross-sections. Depending on the measure of fit (i.e., $R^2_{GLS}$, $R^2_{OLS}$, RMSE, and MAPE), the BMA-SDF outperforms KNS in 96.6\% to 99.9\% of all OS cross-sections we consider. 

Additionally, note that, as shown in Internet Appendix~IA.3.2, the pricing ability of the BMA-SDF significantly outperforms, in- and out-of-sample, not only the benchmark models in Tables \ref{tab:tab-is-pricing-excess} and \ref{tab:tab-os-pricing-excess}, but also a much broader set of additional benchmark models designed specifically to price the bond and stock cross-sections individually.\footnote{In Table~IA.XII of the Internet Appendix we consider an  expanded set of benchmarks that includes the models of \citet{BaiBaliWen_2019}, \citet{vanBinsbergenNozawaSchwert_2025}, \citet{BaliSubrahmanyamWen_2021_JFQA}, \citet{ChungWangWu_2019}, \citet{Carhart_1997}, \citet{HouXueZhang_2015}, \cite{FamaFrench_2015} (with and without the addition of the momentum factor), \citet{DanielMotaRottkeSantos_2020}, and the DEFTERM specification of \citet{FamaFrench_1993}. 
In addition, in Figures IA.14 and IA.15 of the Internet Appendix, we report an extensive comparison of the BMA-SDF performance relative to the \cite{DickNielsenFeldhuetterPedersenStolborg_2024} five-factor corporate bond model.}  

\begin{figure}[tbp]
\begin{center}
 \includegraphics[scale=.8, trim = 0cm 0cm 0cm 0cm, clip]{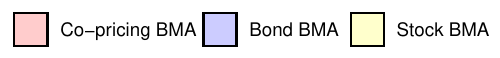}
 
\vspace{.25cm}
\begin{subfigure}[b]{0.46\textwidth}
  \includegraphics[scale=.8]{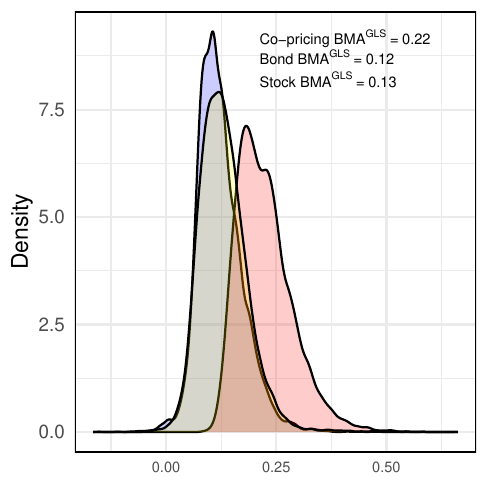}\caption{$R^2_{GLS}$}
\end{subfigure}
\hspace{.2cm}
\begin{subfigure}[b]{0.46\textwidth}
\includegraphics[scale=.8]{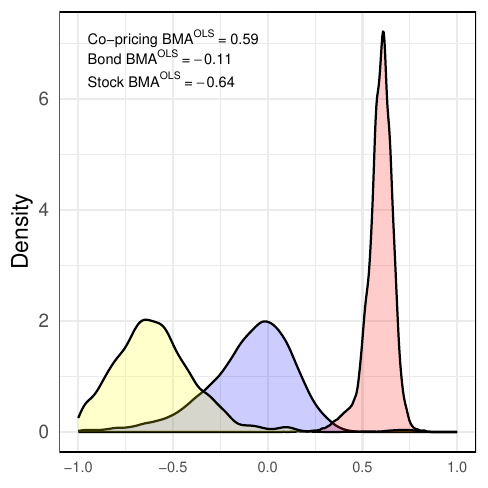}\caption{$R^2_{OLS}$}
\end{subfigure}

\vspace{.5cm} 
\begin{subfigure}[b]{0.46\textwidth}
 \includegraphics[scale=.8]{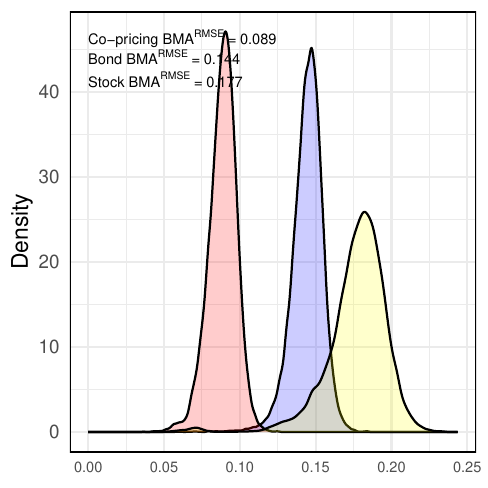}\caption{$R^2_{RMSE}$}
\end{subfigure}
\hspace{.2cm}
\begin{subfigure}[b]{0.46\textwidth}
\includegraphics[scale=.8]{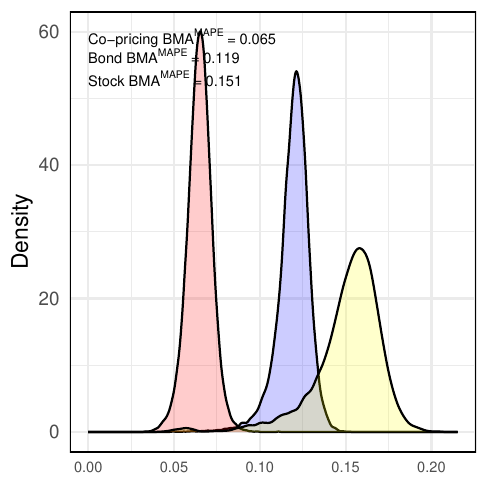}\caption{$R^2_{MAPE}$}
\end{subfigure}
\end{center}
\vspace{-4mm}
\caption{Pricing out-of-sample stocks and bonds with different BMA-SDFs.}
\vspace{-2mm}
\begin{justify}
\begin{spacing}{1}
\footnotesize{
This figure plots the distributions of $R^2_{GLS}$, $R^2_{OLS}$, RMSE and MAPE in Panels A, B, C and D respectively across 16,383 possible bond and stock cross-sections using the 14 sets of stock and bond test assets ($2^{14}-1 = 16,383$) priced using the respective BMA-SDF (the empty set is excluded). The models are first estimated using the baseline set of IS test assets and then used to price (with no additional parameter estimation) each set of the 16,383 OS combinations of test assets.  The red distributions correspond to the pricing performance of the co-pricing BMA-SDF. The blue (yellow) distributions correspond to the pricing performance of the bond (stock) only BMA-SDF. The BMA-SDFs are computed with a prior Sharpe ratio value set to 80\% of the ex post maximum Sharpe ratio of the IS test assets. All data are standardized, that is, pricing errors are in Sharpe ratio units. The sample period is  1986:01 to 2022:12 ($T=444$).   
}
\end{spacing}
\end{justify}
\vspace{-4mm}
\label{fig:bma_comparison_joint_excess}
\end{figure}

Given the findings in Tables \ref{tab:tab-is-pricing-excess} and \ref{tab:tab-os-pricing-excess} that bonds and stocks can be accurately priced \emph{separately} with BMA-SDFs constructed based only on their respective factor zoos, a natural question is whether only bond or stock factors are sufficient to price \emph{jointly} both asset classes. We answer this question in Figure~\ref{fig:bma_comparison_joint_excess} where we compare the OS pricing performance of the co-pricing BMA-SDF (in red, from Panel A of Table~\ref{tab:tab-is-pricing-excess}) to that of BMA-SDFs constructed separately with only bond (in blue, from Panel B of Table~\ref{tab:tab-is-pricing-excess})  and stock (in yellow, from Panel C of Table \nolinebreak \ref{tab:tab-is-pricing-excess}) factors, respectively. As test assets, we again utilize the 16,383 combinations of our OS bond and stock cross-sections. Throughout, the co-pricing BMA-SDF exhibits significantly lower pricing errors and considerably higher $R^2s$ compared to the bond-only or stock-only BMA-SDFs. That is, in order to price the joint cross-section of bond and stock excess returns, we require information from both factor zoos. 

In Internet Appendix~IA.3.2 we show that the co-pricing BMA-SDF can also effectively price the bond and stock cross-sections \textit{separately}, indicating that the superior performance of the co-pricing BMA-SDF is not simply a result of its ability to price one cross-section better than the other. Furthermore, the \textit{asset-class-specific} BMA-SDFs price their respective cross-sections very well. However, information from the bond factor zoo alone is insufficient to price the cross-section of stock returns, and conversely, information from the stock factor zoo is inadequate to price the cross-section of corporate bond excess returns.

\subsubsection{The saliency of factors over time}\label{subsec:time_varying}
\begin{figure}[tbh!]
\begin{center}
\begin{minipage}{\textwidth}
\begin{subfigure}[b]{\textwidth}
\begin{center}
\includegraphics[scale=.3]{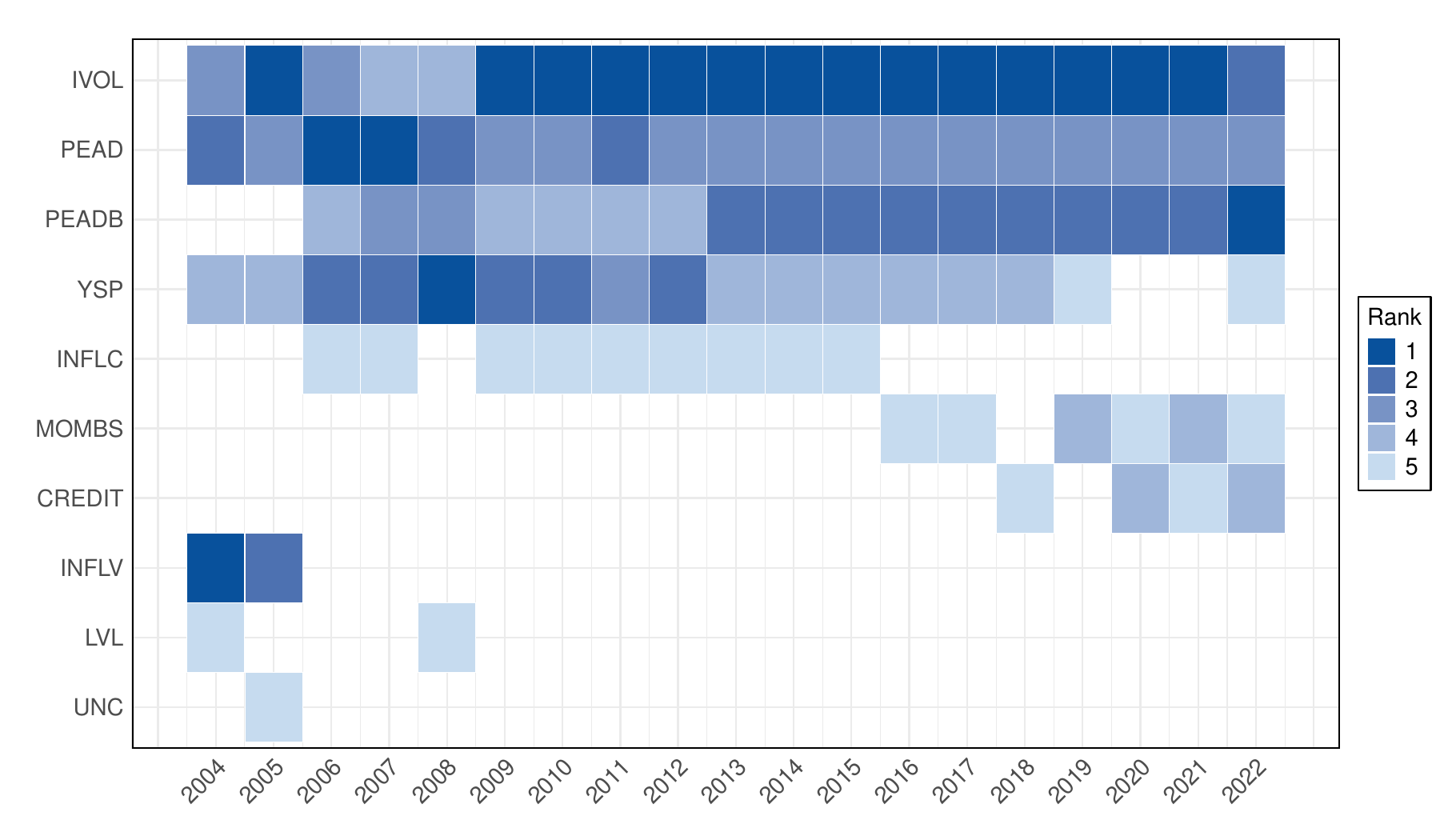}\caption{Expanding forward estimation}

\end{center}
\end{subfigure}

\begin{subfigure}[b]{\textwidth}
\begin{center}
\includegraphics[scale=.3]{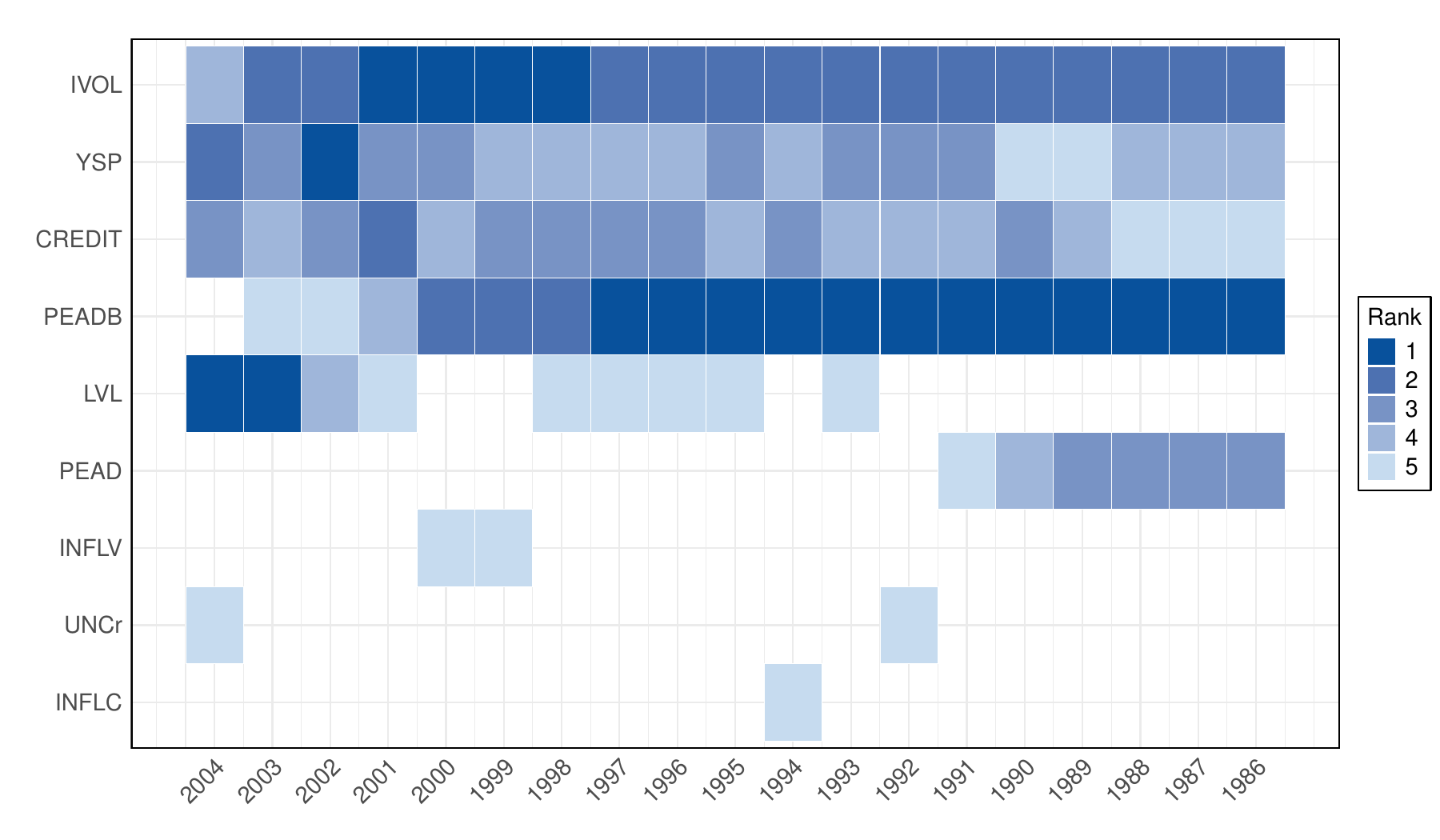}\caption{Expanding backward estimation}
\end{center}
\end{subfigure}
\end{minipage}

\end{center}
\vspace{-4mm}
\caption{Time-varying factor importance.}
\vspace{-2mm}
\begin{justify}
\begin{spacing}{1}
\footnotesize{
The figure highlights the top five factors over time, ordered by their posterior probabilities $\mathbb{E}[\gamma_{j,t}|\text{data}_t]$, and the number of times they are present in the top five, estimated using expanding samples going forward (Panel A) and backward (Panel B) in time.
We use half of the sample as the initial window ($T=222$) and then re-estimate the model every year with an expanding sample. 
The factors are ordered by the total number of times they are present in the `top five.'
The results are shown for prior level of Sharpe ratio shrinkage set to 80\% of the ex post maximum up until year $t$.  
}
\end{spacing}
\end{justify}
\vspace{-4mm}
\label{fig:which_factors_when}
\end{figure}

We now investigate to what extent the relevance of individual factors remains stable over time. To this end, we initially estimate our model for a shorter sample period before subsequently re-estimating the relevant quantities for progressively longer samples. Specifically, we split our sample in half, resulting in two sub-samples with 222 monthly observations each. We first estimate the model for the first subsample spanning July 1986 to June 2004, and then re-estimate it every year, adding twelve new observations at each iteration. Similarly, we estimate backward in time starting with the second subsample from December 2022 to July 2004 and add one year of data at every step. We follow our methodology described in Section~\ref{sec:econ-method} and, throughout, we fix the  shrinkage at 80\% of the corresponding ex post maximum Sharpe ratio for the respective window.  We present the results for the five top factors (based on their posterior probability) in two heatmaps in Figure~\ref{fig:which_factors_when} for the forward (Panel A) and backward estimation (Panel B), respectively, with a higher rank reflected by a darker shade of blue. The top factors ranked by their market prices of risk are also presented in heatmaps in Figure~IA.17 of the Internet Appendix.

Overall, the relevant factors remain remarkably stable. The top five factors from Figure \nolinebreak \ref{Fig:post_probs}, PEADB, IVOL, PEAD, CREDIT, and YSP, all feature prominently in both Panels A and B of Figure \nolinebreak \ref{fig:which_factors_when}. Similarly,  factors that exhibit high market prices of risk in the bottom panel of Figure \nolinebreak \ref{Fig:post_probs_mprs} such as PEADB, CRY, MOMBS, or QMJ, remain highly ranked over a wide range of estimation windows in Figure~IA.17 of the Internet Appendix. When considering rankings based on market prices of risk, the stock market factor MKTS becomes particularly relevant for the backward estimation while it remains just outside the top five for the full sample.  Overall, the results based on time-varying windows largely align with the full sample results presented earlier. 

\subsubsection{Which risks?}\label{sec:Dimensionality}

Next, we further decompose the posterior dimensionality of the SDF and its implied Sharpe ratio to better understand which types of risk are likely to be part of the true latent pricing measure and to what extent different factors capture common information. 

Table~\ref{tab:table-model-dim1} presents the decomposition of the posterior SDF dimensionality and Sharpe ratio split between nontradable and tradable bond and stock factors for different prior values. Panel \nolinebreak A reports results for the pricing of the joint cross-section of stock and corporate bond returns using factors from both zoos to construct the SDF. Instead, Panels B and C focus, respectively, on the separate pricing of corporate bonds and stocks using only factors from their respective zoos. Several salient patterns are evident.

First, Panel A shows that an accurate characterization of the pricing measure requires an SDF that is dense not only in the overall space of observable factors (as per the top Panel of Figure \nolinebreak \ref{Fig:post_dim}), but also over the individual subspaces of nontradable as well as bond and stock tradable factors: the posterior mean number of factors is about 7 for nontradable factors, 6 to 8 for bond, and 9 to 12 for stock tradable factors. Furthermore, this density of the SDF is not driven by the co-pricing task: even pricing only bonds (Panel B) or stocks (Panel C) requires about 7 nontradable, 6 to 8 (for bonds) or 10 to 12 (for stocks) tradable factors, respectively.

\begin{table}[tb!]
\begin{center}
\caption{BMA-SDF dimensionality and Sharpe ratio decomposition by factor type}\label{tab:table-model-dim1}\vspace{-2mm}
\scalebox{.8}{
\begin{tabular}{lcccccccccc} \toprule
& \multicolumn{4}{c}{Total prior SR} &  & \multicolumn{4}{c}{Total prior SR} \\
& 20\% & 40\% & 60\% & 80\% &  & 20\% & 40\% & 60\% & 80\% \\
\midrule
\multicolumn{10}{c}{\textbf{Panel A}: Co-pricing BMA-SDF} \\ \midrule
 & \multicolumn{4}{c}{Nontradable factors} &  & \multicolumn{4}{c}{Tradable factors} \\ \cmidrule(lr){2-5} \cmidrule(lr){7-10}
Mean & 6.97 & 6.94 & 6.97 & 6.80 &  & 19.52 & 18.78 & 17.84 & 15.51 \\
5\% & 4 & 4 & 4 & 4 &  & 14 & 14 & 13 & 10 \\
95\% & 10 & 10 & 10 & 10 &  & 25 & 24 & 23 & 21 \\
$\mathbb{E}[SR_f|\text{data}]$ & 0.21 & 0.43 & 0.70 & 1.12 &  & 0.86 & 1.44 & 1.91 & 2.27 \\
$\mathbb{E}\big[\frac{SR^2_f}{SR^2_m}|\text{data}\big]$ & 0.08 & 0.11 & 0.15 & 0.23 &  & 0.94 & 0.93 & 0.90 & 0.84 \\
 & \multicolumn{4}{c}{Tradable bond factors} &  & \multicolumn{4}{c}{Tradable stock factors} \\ \cmidrule(lr){2-5} \cmidrule(lr){7-10}
Mean & 7.85 & 7.50 & 7.21 & 6.32 &  & 11.67 & 11.28 & 10.63 & 9.19 \\
5\% & 4 & 4 & 4 & 3 &  & 8 & 7 & 7 & 5 \\
95\% & 11 & 11 & 11 & 10 &  & 16 & 15 & 15 & 13 \\
$\mathbb{E}[SR_f|\text{data}]$ & 0.56 & 0.95 & 1.28 & 1.51 &  & 0.66 & 1.13 & 1.50 & 1.77 \\
$\mathbb{E}\big[\frac{SR^2_f}{SR^2_m}|\text{data}\big]$ & 0.43 & 0.43 & 0.43 & 0.39 &  & 0.59 & 0.60 & 0.57 & 0.53 \\
\midrule
\multicolumn{10}{c}{\textbf{Panel B}: Bond BMA-SDF} \\ \midrule
 & \multicolumn{4}{c}{Nontradable factors} &  & \multicolumn{4}{c}{Tradable factors} \\ \cmidrule(lr){2-5} \cmidrule(lr){7-10}
Mean & 6.98 & 6.95 & 7.02 & 7.03 &  & 7.81 & 7.77 & 7.38 & 6.41 \\
5\% & 4 & 4 & 4 & 4 &  & 5 & 5 & 4 & 3 \\
95\% & 10 & 10 & 10 & 10 &  & 11 & 11 & 11 & 10 \\
$\mathbb{E}[SR_f|\text{data}]$ & 0.18 & 0.37 & 0.60 & 0.97 &  & 0.52 & 0.92 & 1.25 & 1.45 \\
$\mathbb{E}\big[\frac{SR^2_f}{SR^2_m}|\text{data}\big]$ & 0.15 & 0.18 & 0.22 & 0.33 &  & 0.86 & 0.83 & 0.78 & 0.66 \\
\midrule
\multicolumn{10}{c}{\textbf{Panel C}: Stock BMA-SDF} \\ \midrule
 & \multicolumn{4}{c}{Nontradable factors} &  & \multicolumn{4}{c}{Tradable factors} \\ \cmidrule(lr){2-5} \cmidrule(lr){7-10}
Mean & 6.98 & 7.02 & 6.92 & 7.02 &  & 11.82 & 11.54 & 11.11 & 9.81 \\
5\% & 4 & 4 & 4 & 4 &  & 8 & 7 & 7 & 6 \\
95\% & 10 & 10 & 10 & 10 &  & 16 & 16 & 15 & 14 \\
$\mathbb{E}[SR_f|\text{data}]$ & 0.14 & 0.29 & 0.47 & 0.79 &  & 0.60 & 1.03 & 1.39 & 1.70 \\
$\mathbb{E}\big[\frac{SR^2_f}{SR^2_m}|\text{data}\big]$ & 0.08 & 0.10 & 0.14 & 0.23 &  & 0.94 & 0.93 & 0.92 & 0.87 \\
\bottomrule
\end{tabular}
}
\end{center}
\begin{spacing}{1}
	{\footnotesize
The table reports posterior means of number of factors (along with the $90\%$ confidence intervals), implied Sharpe ratios $\mathbb{E}[SR_f|\text{data}]$, and the ratio of $SR_f^2$ to the total SDF-implied squared Sharpe ratio $\mathbb{E}\big[SR^2_f/SR^2_m|\text{data}\big]$ for different subsets of factors. Subsets are tradable and nontradable factors, and within tradables we further separate bond and stock factors.
Panels A, B and C report results for the co-pricing, bond-only and stock-only BMA-SDFs, respectively, using the corresponding factor zoos.
}
\end{spacing}
\vspace{-4mm}
\end{table}

Second, each of the three categories of factors is \emph{economically} important.  Focusing on the moderate prior shrinkage case (i.e., 80\% of the ex post achievable Sharpe ratio) in Panel A, the posterior mean of the annualized Sharpe ratio ascribable to the various types of factors ($\mathbb{E}[SR_f|\text{data}]$) is 1.12 for nontradable factors, and 1.51 and 1.77, respectively, for tradable bond and stock factors. Third, there is substantial common priced information across the categories of factors, as the sum of the Sharpe ratios generated by the three sets of factors (for example $1.12 + 1.51 + 1.77 = 4.40$ in Panel A) is much larger than the average posterior SDF-implied Sharpe ratio (which is around $2.5$ in the bottom panel of Figure~\ref{Fig:post_dim}).  This overlap in risks captured by different types of factors is particularly strong among tradable factors, where the sum of the Sharpe ratios of bond and stock factors in the SDF is $1.51+1.77=3.28$, while the posterior mean Sharpe ratio for \textit{all} tradable factors jointly is approximately $2.27$. The degree of common spanning of priced risks can be formally assessed by focusing on the estimated share of the squared Sharpe ratio of the SDF generated by the different types of factors, $\mathbb{E}\big[\frac{SR^2_f}{SR^2_m}|\text{data}\big]$. Summing the shares in Panel A ascribable to, respectively, nontradable (0.23) and tradable bond (0.39) and stock (0.53) factors yields a total of 1.15, i.e., more than 100\%, indicating substantial commonality among the fundamental risks spanned by the different types of factors. Furthermore, the sum of the shares for bond and stock factors ($0.39+0.53= 0.92$) is much larger than the share due to all tradable factors jointly ($0.84$). That is, tradable bond and stock factors capture, at least partially, the same underlying sources of priced risk. Similarly, summing the shares of squared Sharpe ratios ascribable to nontradable and tradable factors in Panels A to C yields 1.07, 0.99, and 1.1, indicating some common spanning between tradable and nontradable factors driven mostly by equity factors.

Since, in the cross-sectional layer of our estimation method (encoded by the likelihood function in equation (\ref{eq:xs-lf})), the ``regressors'' are the loadings in the $N \times K$ matrix of covariances between test assets and factors ($\bm{C}$),  the degree of commonality in pricing implications of the factors in the zoo can be gauged by performing a principal component analysis on the matrix $\bm{C}^T\bm{C}$ (in the OLS case, or $\bm{C}^T \bm{\Sigma}^{-1}\bm{C}$ in the GLS case). In Figure~IA.20 of the Internet Appendix we perform such an analysis and document that the largest five principal components of the factor loadings explain more than 99\% of their \textit{cross-sectional} variation (in the OLS case, and more than 80\% in the GLS case). That is, overall, the findings of this section highlight that the factor zoo is akin to a jungle of noisy proxies for common underlying sources of risk.

Given the salience of tradable factors for the BMA-SDF outlined above, with their share of the squared Sharpe ratio of the SDF in the two-thirds to four-fifths range, a natural question is what types of risks these factors capture. Using the method pioneered by \cite{CampbellShiller_1988_RFS} and extended by \cite{Vuolteenaho_2002}, we classify the tradable factors into those that relate more to discount rate (DR) news and those for which, instead, cash-flow (CF) news is more important.\footnote{See \cite{KoijenVanNieuwerburgh_2011} and more recent work by \cite{Zviadadze_2021}.} Internet Appendix~IA.5 details the empirical (VAR) methodology used for categorizing our 40 tradable bond and stock factors as (mostly) driven by either discount rate or cash-flow news. Therein, we also demonstrate, with extensive robustness tests, that the decomposition remains quite stable across alternative approaches.

\begin{table}[tb!]
\begin{center}
\caption{Discount rate vs. cash-flow news}\label{tab:table-model-dim1-dr-cf}\vspace{-2mm}
\scalebox{.8}{
\begin{tabular}{lcccccccccc} \toprule
  & \multicolumn{4}{c}{Discount rate news} &  & \multicolumn{4}{c}{Cash-flow news} \\\cmidrule(lr){2-5} \cmidrule(lr){7-10}
 & \multicolumn{4}{c}{Total prior SR} &  & \multicolumn{4}{c}{Total prior SR} \\
 & 20\% & 40\% & 60\% & 80\% &  & 20\% & 40\% & 60\% & 80\% \\
\midrule
\multicolumn{10}{c}{\textbf{Panel A}: Co-pricing BMA-SDF, tradable bond and stock factors} \\ \midrule
Mean & 9.81 & 9.60 & 9.28 & 8.20 &  & 9.71 & 9.18 & 8.56 & 7.31 \\
5\% & 6 & 6 & 6 & 5 &  & 6 & 6 & 5 & 4 \\
95\% & 14 & 13 & 13 & 12 &  & 13 & 13 & 12 & 11 \\
$\mathbb{E}[SR_f|\text{data}]$ & 0.65 & 1.19 & 1.70 & 2.10 &  & 0.60 & 1.06 & 1.45 & 1.77 \\
$\mathbb{E}\big[\frac{SR^2_f}{SR^2_m}|\text{data}\big]$ & 0.58 & 0.67 & 0.75 & 0.75 &  & 0.51 & 0.55 & 0.57 & 0.56 \\
\midrule
\multicolumn{10}{c}{\textbf{Panel B}: Bond BMA-SDF, tradable bond factors} \\ \midrule
Mean & 4.97 & 5.05 & 4.94 & 4.43 &  & 2.85 & 2.72 & 2.44 & 1.98 \\
5\% & 2 & 3 & 2 & 2 &  & 1 & 1 & 1 & 0 \\
95\% & 8 & 8 & 7 & 7 &  & 5 & 5 & 4 & 4 \\
$\mathbb{E}[SR_f|\text{data}]$ & 0.44 & 0.85 & 1.21 & 1.43 &  & 0.28 & 0.50 & 0.64 & 0.69 \\
$\mathbb{E}\big[\frac{SR^2_f}{SR^2_m}|\text{data}\big]$ & 0.67 & 0.74 & 0.75 & 0.65 &  & 0.35 & 0.32 & 0.27 & 0.21 \\
 \midrule
\multicolumn{10}{c}{\textbf{Panel C}: Stock BMA-SDF, tradable stock factors} \\ \midrule
Mean & 5.01 & 4.91 & 4.79 & 4.38 &  & 6.81 & 6.63 & 6.31 & 5.43 \\
5\% & 2 & 2 & 2 & 2 &  & 4 & 4 & 3 & 2 \\
95\% & 8 & 7 & 7 & 7 &  & 10 & 10 & 9 & 9 \\
$\mathbb{E}[SR_f|\text{data}]$ & 0.37 & 0.73 & 1.11 & 1.48 &  & 0.47 & 0.83 & 1.16 & 1.44 \\
$\mathbb{E}\big[\frac{SR^2_f}{SR^2_m}|\text{data}\big]$ & 0.44 & 0.54 & 0.65 & 0.72 &  & 0.65 & 0.66 & 0.69 & 0.68 \\
\bottomrule
\end{tabular}
}
\end{center}
\begin{spacing}{1}
	{\footnotesize
The table reports posterior means of number of factors (along with the $90\%$ confidence intervals), implied Sharpe ratios $\mathbb{E}[SR_f|\text{data}]$, and the ratio of $SR_f^2$ to the total SDF-implied squared Sharpe ratio $\mathbb{E}\big[SR^2_f/SR^2_m|\text{data}\big]$ for discount rate and cash-flow news driven tradable factors, respectively.
}
\end{spacing}
\vspace{-4mm}
\end{table}

Table~\ref{tab:table-model-dim1-dr-cf} decomposes, for a range of prior values, the contribution to the SDF dimensionality and Sharpe ratio of the tradable factors, primarily related to DR news on one hand and to CF news on the other. Panel A reports results for the joint pricing of bonds and stocks with all factors, while Panels B and C focus on the two asset classes and factor zoos separately.  The left and right four columns pertain to DR and CF news, respectively. First, DR news factors marginally dominate the composition of the co-pricing BMA-SDF in Panel A. The average factor-implied Sharpe ratios, $\mathbb{E}[SR_f|\text{data}]$, of the DR news-driven factors are consistently higher than those of their CF-driven counterparts. This translates into a significantly higher proportion of the total implied Sharpe ratio being driven by DR-related factors. For a prior level equal to 80\% of the ex post achievable Sharpe ratio, DR-driven factors account for 75\% of the total squared Sharpe ratio of the SDF, compared to 56\% for the CF-driven factors. Second, when considering the corporate bond BMA-SDF (Panel B), the total Sharpe ratio is predominantly driven by bond factors related to DR news. The factor-implied Sharpe ratio $\mathbb{E}[SR_f|\text{data}]$ and $\mathbb{E}\big[\frac{SR^2_f}{SR^2_m}|\text{data}\big]$ for DR-driven factors are nearly double that of the CF-driven factors. Finally, when considering only stock factors (Panel C), both DR and CF news appear to play an equally important role, providing very similar contributions to the Sharpe ratio of the BMA-SDF. 

In Internet Appendix~IA.5.3 we discuss the estimated positioning of the individual factors on the spectrum of DR and CF news. 
Interestingly, the two most likely tradable components of the BMA-SDF, the post-earnings announcement drift factors in bonds and stocks, PEAD and PEADB, are primarily driven by DR news.\footnote{See e.g., \cite{PenmanNir_2019} for a discussion on how earnings reports contain both discount rate and cash-flow news.}

\subsection{Trading the BMA--SDF}\label{sec:trade_BMA}

We now investigate the implementability of the BMA-SDF as a trading strategy and compare its performance to tradable benchmark strategies. 

Portfolio weights for the tradable strategies are constructed by normalizing the posterior means of the MPRs of the SDF representations to sum to one in each specification. Since all benchmark models are exclusively based on tradable factors, we constrain the BMA-SDF to use only such factors. This means that our approach, de facto, focuses on a lower bound for the trading performance of the BMA-SDF since nontradable factors in the BMA-SDF command a non-trivial Sharpe ratio (see Table~\ref{tab:table-model-dim1}). To facilitate comparison, all tradable portfolio strategies are normalized to have the same volatility as the equity market index.

The IS results are presented in Panel A of Table~\ref{tab:tab-fmp}. The IS Sharpe ratio of the tradable BMA-SDF ranges from 1.99 (20\% shrinkage) to 2.85 (80\% shrinkage). The closest competitor is the KNS model, which delivers an IS Sharpe ratio of 2.57. The TOP $\gamma$ and $\lambda$ models, using the top five \textit{tradable} factors by posterior probability and MPR, respectively, also perform well with Sharpe ratios of 2.14 and 2.15 respectively. Note also that the tradable version of the BMA-SDF tends to exhibit much less negative skewness and thinner tails than the other benchmark strategies.

\begin{table}[tbh!]
\begin{center}
\caption{Trading the BMA-SDF and benchmark models}\label{tab:tab-fmp}\vspace{-2mm}
\resizebox{13.5cm}{!}{
\begin{tabular}{lcccc|ccccccccc}\toprule
 & \multicolumn{4}{c}{BMA-SDF prior Sharpe ratio} & TOP $\gamma$ & TOP $\lambda$ & KNS & RPPCA & FF5 & HKM & MKTB & MKTS & EW \\ \cmidrule(lr){2-5}
 & 20\% & 40\% & 60\% & \multicolumn{1}{c}{80\%} &  &  &  &  &  &  &  &  &  \\ \midrule
 \multicolumn{14}{c}{\textbf{Panel A:} In-sample -- 1986:01 to 2022:12 ($T=444$)} \\
 \midrule
Mean & 31.38 & 38.94 & 43.43 & 45.03 & 33.77 & 34.04 & 40.54 & 39.59 & 12.20 & 8.37 & 10.18 & 8.29 & 19.42 \\
SR & 1.99 & 2.46 & 2.75 & 2.85 & 2.14 & 2.15 & 2.57 & 2.51 & 0.77 & 0.53 & 0.64 & 0.52 & 1.23 \\
IR & 1.73 & 2.28 & 2.52 & 2.59 & 1.94 & 1.90 & 2.33 & 2.18 & 0.02 & 0.34 & $-$0.47 & 0.29 & -- \\
Skew & 0.76 & 0.73 & 0.54 & 0.31 & 0.47 & 0.44 & 0.51 & 0.90 & $-$0.70 & $-$0.65 & $-$0.71 & $-$0.78 & $-$0.29 \\
Kurt & 3.55 & 3.08 & 2.47 & 2.00 & 2.53 & 2.54 & 2.98 & 3.07 & 3.41 & 1.91 & 4.68 & 2.22 & 4.63 \\
\midrule
 \multicolumn{14}{c}{\textbf{Panel B:} Out-of-sample -- 2004:07 to 2022:12 ($T=222$)} \\
 \midrule
Mean & 22.72 & 25.73 & 27.17 & 27.90 & 20.59 & 23.41 & 20.36 & 23.01 & 5.90 & 7.12 & 8.22 & 8.71 & 17.15 \\
SR & 1.46 & 1.65 & 1.74 & 1.79 & 1.32 & 1.50 & 1.31 & 1.48 & 0.38 & 0.46 & 0.53 & 0.56 & 1.10 \\
IR & 0.98 & 1.24 & 1.38 & 1.46 & 1.40 & 1.37 & 0.85 & 1.07 & $-$0.27 & $-$0.26 & $-$0.04 & $-$0.21 & -- \\
Skew & 0.30 & 0.04 & $-$0.10 & $-$0.13 & $-$0.62 & 0.17 & $-$1.19 & $-$0.60 & $-$1.59 & $-$0.37 & $-$0.93 & $-$0.54 & $-$1.06 \\
Kurt & 2.39 & 3.59 & 4.06 & 3.77 & 5.77 & 2.38 & 11.97 & 7.74 & 10.60 & 1.51 & 5.42 & 1.28 & 7.22 \\
\midrule
\end{tabular}
}
\end{center}
\begin{spacing}{1}
	{\footnotesize 
    In-sample (Panel A) and out-of-sample (Panel B) performance of the co-pricing BMA-SDF tradable portfolio across prior SR levels, the `TOP' model factors portfolios, the latent co-pricing factor models (KNS and RPPCA), notable benchmark models (FF5, HKM, MKTS, MKTB) and the equally-weighted portfolio (EW) of all (40) tradable factors.
    The in-sample weights for the tradable portfolios are formed scaling the (posterior means of the) MPRs to sum to one in each specification considered.
    The Top $\gamma$ ($\lambda$) model uses the MPRs from the most likely (highest absolute MPRs) factors with 80\% shrinkage.
    These factors are: PEADB, PEAD, CMAs, CRY and MOMBS ($\gamma$) and PEADB, MOMBS, CRY, PEAD and CMAs ($\lambda$).
    For KNS, the weights are obtained directly from the \citet{KozakNagelSantosh_2020} procedure.
    For RPPCA, FF5 and HKM, the weights are estimated via GMM.
    In Panel B, the results are strictly out-of-sample.
    An expanding window is used with an initial window of 222 months to conduct the estimation.
    These weights are then used to invest in the factors over the next 12 months.
    Thereafter, we re-estimate the models in an expanding fashion every year.
    The Top model input factors change dynamically at each estimation.
    For KNS, we re-conduct the two-fold cross-validation at every estimation to pin down the optimal parameters.
    For RPPCA, we re-estimate the PCs at every estimation.
    The Mean is annualized and presented in percent.
    The Sharpe ratio and Information ratio are annualized.
    The benchmark factor to compute the IR is the EW factor.
    Skew and Kurt are skewness and kurtosis, respectively.
    The models are estimated with the 83 bond and stock portfolios and the 40 tradable bond and stock factors as described in Section~\ref{sec:data}. 
    For the BMA-SDFs, we report results for a range of prior Sharpe ratio values that are set as 20\%, 40\%, 60\% and 80\% of the ex post maximum Sharpe ratio of the relevant portfolios and factors. In Panel B, this ratio changes with the expanding window.
    The IS period is 1986:01 to 2022:12 ($T=444$) and the OS period is 2004:07 to 2022:12 ($T=222$).
    }
\end{spacing}
\vspace{-4mm}
\end{table}

\begin{figure}[tb!]
\begin{center}
\vspace{.25cm}
\includegraphics[scale=.4, trim = 0cm 0cm 0cm 1cm,clip]{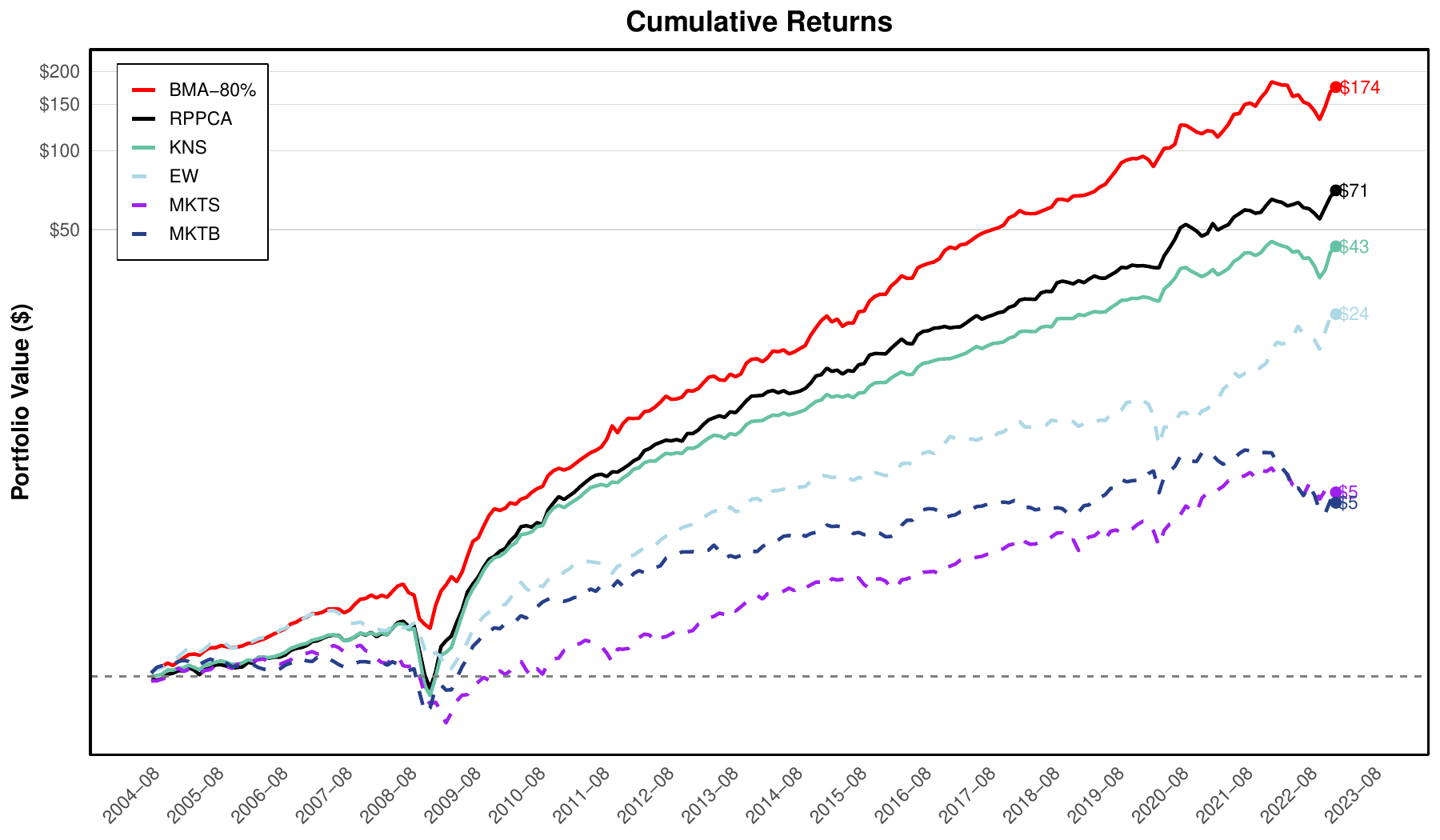}
\end{center}
\vspace{-4mm}
\caption{Out-of-sample investing in the BMA-SDF tradable portfolio and benchmark models.}
\vspace{-2mm}
\begin{justify}
\begin{spacing}{1}
\footnotesize{
     Out-of-sample cumulative return of investing \$1 in the co-pricing BMA-SDF tradable portfolio with 80\% SR prior, the latent factor models KNS and RPPCA, the  stock and bond market factors, MKTS and MKTB, and an equally-weighted factor portfolio EW. 
    An expanding window is used with an initial window of 222 months to conduct the estimation.
    These weights are then used to invest in the factors which are held constant over the next 12 months.
    Thereafter, we re-estimate the models in an expanding fashion every year.
    For KNS, we re-conduct the two-fold cross-validation every estimation to pin down the optimal parameters.
    For RPPCA, we re-estimate the PCs every estimation.
    The models are estimated to price the 83 bond and stock portfolios and the 40 tradable bond and stock factors ($N = 123$) as described in Section~\ref{sec:data}. 
    The out-of-sample evaluation period is 2004:07 to 2022:12 ($T=222$).
}
\end{spacing}
\end{justify}
\vspace{-4mm}
\label{fig:os_tradable_bma}
\end{figure}

In Panel B of Table~\ref{tab:tab-fmp}, we examine the time series OS performance of the same set of tradable portfolios. To conduct this exercise, the out-of-sample period is July 2004 to December 2022---a particularly challenging one as it contains both the Great Recession as well as the contraction during the COVID pandemic.

We use the first half of our baseline data (January 1986 to June 2004) as the training sample for the initial estimation of the tradable portfolio weights. These weights are used to form the portfolios that are held over the first 12 months out-of-sample. Recursively, after one year, the training sample is expanded by twelve months; the portfolio weights are recomputed using the resulting MPRs in the expanded training sample, and the performance of the portfolios is assessed over the following twelve months (yielding, in total, 222 months of OS history).

Strikingly, the OS performance of the BMA-SDF portfolio in Panel B of Table~\ref{tab:tab-fmp} is now significantly greater than any other model considered. The Sharpe ratio of the BMA-SDF portfolio is approximately 1.8 (80\% shrinkage). Moreover, all of the BMA-SDF specifications convincingly outperform the equally weighted (EW) portfolio of tradable factors, which has a SR of 1.1 and is known to be exceedingly difficult to beat \citep{DeMiguelGarlappiUppal_2009}.

But is this robust OS economic performance of the BMA-SDF portfolio due to just a handful of lucky episodes? Figure~\ref{fig:os_tradable_bma} depicts, in log scale, the cumulative returns of investing \$1 in the OS BMA-SDF strategy along with notable benchmarks. For ease of comparison, portfolio returns are scaled to have a constant volatility equal to that of the stock market factor (MKTS). Out-of-sample, the BMA-SDF (80\% shrinkage) tradable portfolio is the clear winner with a cumulative dollar value over the investment period of \$174 versus \$71 for RPPCA. Furthermore, in virtually any multi-year sub-period, the slope (and hence the log return) of the tradable BMA-SDF strategy is higher than that of any of the alternative strategies, stressing that the outperformance is extremely stable out-of-sample, and not just driven by a few lucky events. 

\subsection{The information content of the two factor zoos}\label{sec:Information}

As shown in Section~\ref{sec:xs-pricing} (see Tables \ref{tab:tab-is-pricing-excess} and \ref{tab:tab-os-pricing-excess}), although one can construct well-performing BMA-SDFs to price  bonds and stocks separately using the information in their respective zoos, the joint pricing of these assets requires information from both sets of factors (see Figure~\ref{fig:bma_comparison_joint_excess}). 
In this section, we demonstrate that this result arises from the fact that corporate bond returns reflect not only a component related to compensation for exposure to credit risk, but also a \emph{Treasury term structure} risk premium that is not captured by equity-based factors.

\begin{figure}[tbp!]
\begin{center}
 \includegraphics[scale=.8, trim = 0cm 0cm 0cm 0cm, clip]{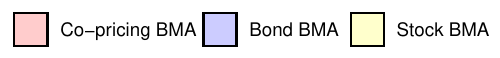}
 
\vspace{.25cm}
\begin{subfigure}[b]{0.46\textwidth}
 \includegraphics[scale=.80]{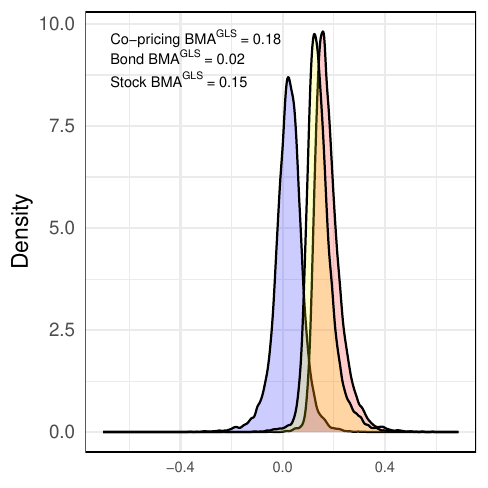}\caption{$R^2_{GLS}$}
\end{subfigure}
\hspace{.2cm}
\begin{subfigure}[b]{0.46\textwidth}
\includegraphics[scale=.80]{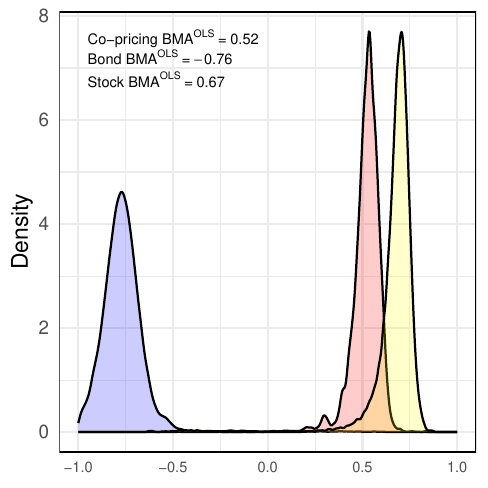}\caption{$R^2_{OLS}$}
\end{subfigure}

\vspace{.5cm} 
\begin{subfigure}[b]{0.46\textwidth}
 \includegraphics[scale=.80]{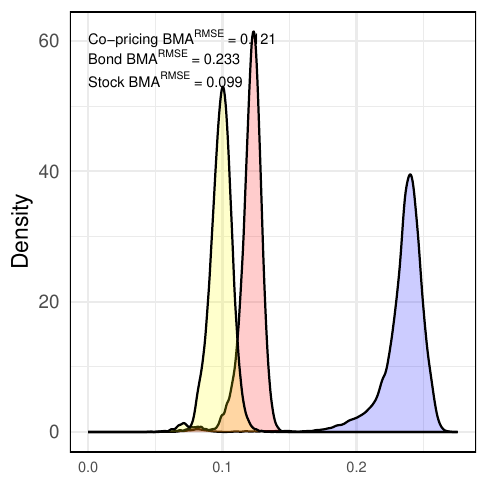}\caption{$R^2_{RMSE}$}
\end{subfigure}
\hspace{.2cm}
\begin{subfigure}[b]{0.46\textwidth}
\includegraphics[scale=.80]{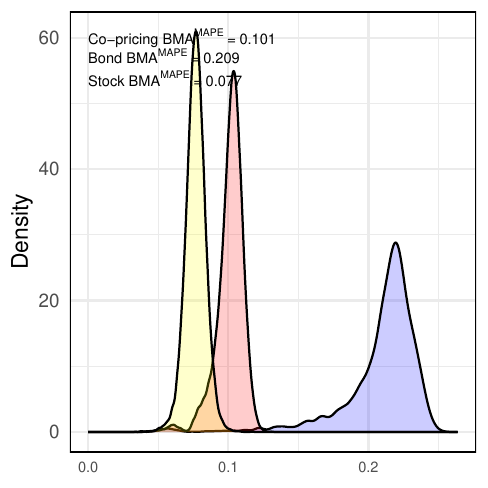}\caption{$R^2_{MAPE}$}
\end{subfigure}
\end{center}
\vspace{-4mm}
\caption{Pricing the joint cross-section of stock and duration-adjusted bond returns.}
\vspace{-2mm}
\begin{justify}
\begin{spacing}{1}
\footnotesize{
This figure plots the distributions of $R^2_{GLS}$, $R^2_{OLS}$, RMSE and MAPE in Panels A, B, C and D respectively across 16,383 possible bond and stock cross-sections using the 14 sets of stock and bond test assets ($2^{14} -1 = 16,383$)  priced using the respective BMA-SDF (the empty set is excluded). All bond test assets (IS and OS) and factors are formed with duration-adjusted returns defined in equation (\ref{eq:dur}).  The models are first estimated using the baseline set of IS test assets and then used to price (with no additional parameter estimation) each set of the 16,383 OS combinations of test assets.  The red distributions correspond to the pricing performance of the co-pricing BMA-SDF. The blue (yellow) distributions correspond to the pricing performance of the bond (stock) only BMA-SDF. The BMA-SDFs are computed with a prior Sharpe ratio value set to 80\% of the ex post maximum Sharpe ratio of the IS test assets. All data are standardized, that is, pricing errors are in Sharpe ratio units. The sample period is  1986:01 to 2022:12 ($T=444$).   
}
\end{spacing}
\end{justify}
\label{fig:bma_comparison_duration}
\end{figure}

To illustrate this point, we now turn our focus to bond returns in excess of duration-matched portfolios of U.S. Treasuries. More precisely, for every bond $i$, we construct the following duration-adjusted return
\begin{equation}\label{eq:dur}
\underset{\text{Duration-adjusted  return}}{\underbrace{R_{bond \, i, t} - R^{Treasury}_{dur \, bond \,i,t } }} \equiv \underset{\text{Excess return}}{\underbrace{R_{bond \, i,t} - R_{f,t}}} - \underset{\text{Treasury component}}{\underbrace{\left( R^{Treasury}_{dur \, bond \,i ,t}  - R_{f,t}\right)}},
\end{equation}
where $R_{bond \, i, t}$ is the return of bond $i$ at time $t$, $R_{f,t}$ denotes the short-term risk-free rate, and $R^{Treasury}_{dur \, bond \,i ,t}$ denotes the return on a portfolio of Treasury securities with the same duration as bond $i$ (constructed as in \citet{vanBinsbergenNozawaSchwert_2025}, see also Internet Appendix~IA.6). As is evident from the right-hand side of equation (\ref{eq:dur}), the duration adjustment removes the implicit Treasury component from the bond excess return, hence isolating the remaining sources of risk compensation that investing in a given bond entails.

Figure~\ref{fig:bma_comparison_duration} reports the distribution of OS measures of fit ($R^2_{GLS}$, $R^2_{OLS}$, RMSE, and MAPE)  across 16,383 possible bond and stock cross-sections using the 14 sets of stock and bond test assets for three different BMA-SDFs based on (i) bond factors only, (ii)  stock factors only, and (iii) both bond and stock factors. The contrast with Figure~\ref{fig:bma_comparison_joint_excess} is stark: once bond returns are adjusted for duration, the BMA-SDF based solely on equity information prices jointly bonds and stocks as effectively as the co-pricing BMA-SDF that additionally includes bond factors. That is, the information content of the bond factor zoo becomes largely irrelevant for co-pricing once the Treasury component of bond returns is removed. 

This last finding raises a natural question: why do we need the bond factors for co-pricing assets in the absence of the duration adjustment? As we are about to demonstrate, bond factors price the Treasury component of corporate bond returns. 

\begin{figure}[tbp]
\begin{center}
\begin{subfigure}[b]{0.45\textwidth}
 \includegraphics[scale=.80]{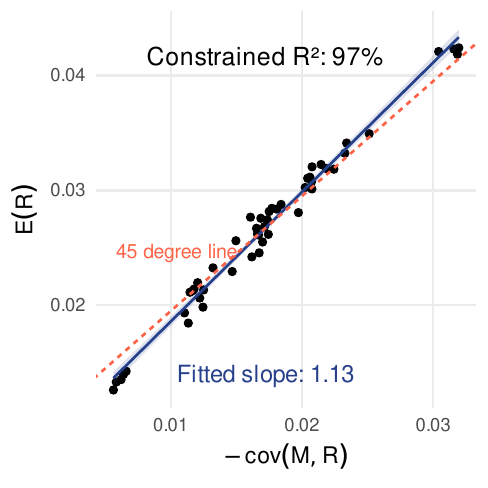}\caption{In-sample with bond factors}
\end{subfigure}
\hspace{.25cm}
\begin{subfigure}[b]{0.45\textwidth}
\includegraphics[scale=.80]{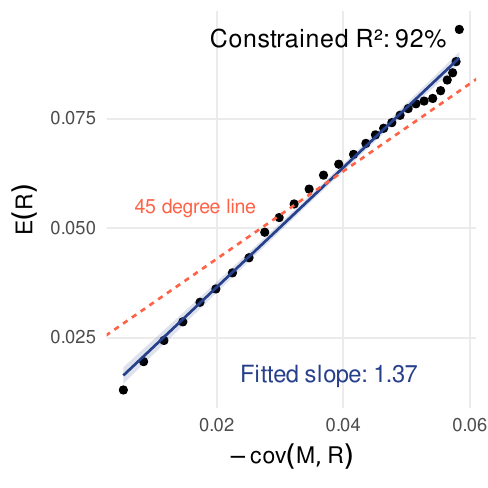}\caption{Out-of-sample with bond factors}
\end{subfigure}

\vspace{.5cm} 
\begin{subfigure}[b]{0.45\textwidth}
 \includegraphics[scale=.80]{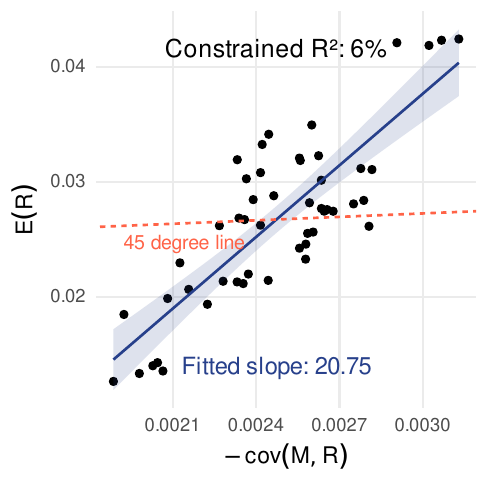}\caption{In-sample with stock factors}
\end{subfigure}
\hspace{.25cm}
\begin{subfigure}[b]{0.45\textwidth}
 \includegraphics[scale=.80]{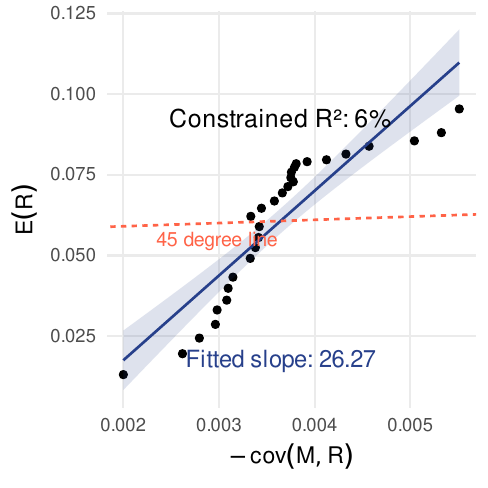}\caption{Out-of-sample with stock factors}
\end{subfigure}
\end{center}
\vspace{-4mm}
    \caption{Pricing the Treasury component of corporate bond returns.}\label{fig:XS_fit_treasuryexcess}
\vspace{-2mm}
\begin{justify}
\begin{spacing}{1}
 \footnotesize{
Plots of sample averages of excess returns for Treasury portfolios, on the $y$-axis, against BMA-SDF-implied risk premia, computed as minus the covariance between portfolio returns and the (posterior mean of the) BMA-SDF, constructed using the nontradable factors plus only bond (Panels A and B) or stock (Panels C and D) factors, on the $x$-axis. Panels A and C: excess returns are the Treasury component from  equation (\ref{eq:dur}), $R^{Treasury}_{dur \, bond \,i ,t}  - R_{f,t}$, using the 50 IS bond portfolio test assets.  Panels B and D: 29 Treasury portfolios of excess returns on Treasury securities with maturities spanning 2 to 30 years. All are  described in Section~\ref{sec:data}. The Treasury component bond and stock BMA-SDFs are estimated using the 50 IS portfolios and the respective bond and stock factors in addition to the 14 nontradable factors described in  \ref{sec:factor_zoo}. For either estimation we do not impose self-pricing for the stock and bond factors. OLS $R^2$s are from a constrained regression that sets the slope coefficient to one. The sample period is 1986:01 to 2022:12 ($T=444$).
}
\end{spacing}
\end{justify}
\end{figure}

Panel A of Figure~\ref{fig:XS_fit_treasuryexcess} summarizes the in-sample pricing of the Treasury component of corporate bond returns using the Treasury component bond BMA-SDF based only on the bond factor zoo. That is, as in-sample test assets we use the Treasury component of the 50 bond portfolios and estimate the model using the 14 nontradable and the 16 tradable bond factors.\footnote{We do not include the factors among the test assets so that the evaluation of fit is based only on the ability to explain the Treasury component. Expected risk premia of portfolios in the figure are proxied by their time series averages (on the vertical axis), while the SDF-implied ones are computed as minus the covariance of (the posterior mean of) the BMA-SDF, hence imposing the theoretical restriction coming from the fundamental asset pricing equation. The constrained $R^2$ is computed imposing a unit slope and zero intercept.} The pricing (evaluated at the posterior mean of the SDF) is nearly perfect, with a cross-sectional (constrained) $R^2_{OLS}$ of about 97\%. Similarly, Panel B shows that the out-of-sample pricing of a cross-section of Treasury excess returns using the same BMA-SDF is also nearly perfect, with a constrained $R^2_{OLS}$ of 92\% and average excess returns and SDF-implied risk premia aligning closely around the 45-degree line.

In contrast, Panels C and D of Figure~\ref{fig:XS_fit_treasuryexcess} report the same cross-sectional pricing exercises using the BMA-SDF based only on stock factors. Clearly, the equity-based BMA-SDF is not able to price the Treasury component of corporate bond returns, neither in- nor out-of-sample, yielding extremely low measures of fit (the constrained $R^2_{OLS}$ is only 6\%) and slope coefficients very far from the theoretical value.

The above highlights that the bond factor zoo is necessary for co-pricing bonds and stocks because the factors proposed in the corporate bond literature price well the Treasury component implicit in corporate bond returns---a component that stock factors fail to price. However, once this component is accounted for---as in the case of duration-adjusted bond returns---co-pricing can effectively be achieved using only equity information.

But does one really need our Bayesian machinery comprising quadrillions of models to uncover this phenomenon? The answer, resoundingly, is yes. As highlighted in Tables IA.XVI and IA.XVII (Panels B and D) of the Internet Appendix, unlike our BMA-SDF, canonical equity-based factor models do quite a poor job in pricing corporate bond returns \textit{even after removing their Treasury component} (with small, and mostly negative, measures of fit, and significantly larger pricing errors than the BMA-SDF). This is due to the fact that both the co-pricing and stock-based BMA-SDFs that price duration-adjusted corporate bonds (and stocks) are \textit{dense} in the space of both tradable and nontradable factors (as per Panel C of Table~\ref{tab:table-model-dim1}). That is, the link between duration-adjusted bond returns and equity market factors extends far beyond the one between these assets and just the equity market index (\citet{vanBinsbergenNozawaSchwert_2025}) or just a handful of factors. Consequently, and importantly, the reward for holding this risk is a multiple of that for the market index alone (with posterior annualized Sharpe ratios of about 1.4 to 1.7 just for the tradable component, as per Panel C of Table~\ref{tab:table-model-dim1}). Furthermore, the high degree of (posterior) factor density of the equity-based SDF that prices duration-adjusted bond returns implies that canonical inference based on low-dimensional models is unreliable (due to misspecification) and affected by a severe omitted variable problem (\cite{GiglioXiu_2018}). For example, in Figure~IA.26 of the Internet Appendix, we test the equity CAPM as a pricing model for the duration-adjusted bond returns. We do so in both SDF and ``beta'' representations using (unlike previous literature) robust estimation methods. As the figure highlights, using robust inference, one would not find a statistically significant link between duration-adjusted bond returns and the equity market index in such a heavily misspecified setting.

Moreover, the Treasury component of corporate bonds is also economically important. The ex post (annualized) maximum Sharpe ratio of the duration-matched Treasury portfolios in equation (\ref{eq:dur}) is approximately 1.48. As illustrated in the bottom panel of Figure~IA.22 of the Internet Appendix, this is roughly the posterior mode of the Sharpe ratio achievable with the  BMA-SDF that prices the Treasury component only with factors in the corporate bond factor zoo. Note also that, as depicted in the top panel of the figure, even for pricing just this Treasury component, the required SDF is quite \textit{dense}, with a median number of factors equal to 14 and a posterior 95\% C.I. ranging from 8 to 19 factors. Furthermore, as shown in Table~IA.XVIII of the Internet Appendix, the required SDF is dense in the space of both nontradable and tradable factors. 

Mirroring the  analysis in Section~\ref{sec:Copricing}, we can assess which factors are more likely to price the Treasury component individually, and how factors should be optimally combined to achieve a portfolio that captures the priced risks in these assets. Figure~IA.23 of the Internet Appendix reports the posterior factor probabilities and market prices of risk implied by the pricing of the Treasury component of corporate bond returns using the corporate bond factor zoo (the prior Sharpe ratio is set to 80\% of the ex post maximum Sharpe ratio).  Overwhelmingly, the most likely factors are nontradable: five out of the six factors with posterior probability higher than the prior value are nontradable.
Furthermore, largely, these factors are the same as those that appear most likely when co-pricing bonds and stocks (the top three being YSP, CREDIT and LVL, followed by the IVOL factor). Moreover, these nontradable factors command large market prices of risk and the probability of zero nontradable factors being in the bond BMA-SDF that prices the Treasury component of corporate bond returns is virtually zero (or 0.014\%).

The bottom panel of Figure~IA.23 of the Internet Appendix tells us which portfolios to form to capture the common risk priced in these cross-sections. Interestingly, in addition to the most likely factors, the bond market index (MKTB) and the traded term structure risk factor (TERM, i.e., the difference between the monthly long-term government bond return and the one-month T-Bill rate of return, \cite{FamaFrench_1992}) feature prominently in the BMA-SDF with, respectively, the second and third largest portfolio weights---and the largest among tradable factors. That is, these factors are not likely fundamental sources of risk, but they appear correlated with the true sources. 

This finding also explains the success of the MKTB factor in \cite{DickersonMuellerRobotti_2023}. As Internet Appendix~IA.7 also confirms, the bond market index commands a large risk premium in its own market, but it is not likely to be part of the true latent SDF. Nevertheless, it commands a substantial risk premium as compensation for being highly correlated with the latent fundamental risks in the bond market, particularly the Treasury component, making it advantageous in a portfolio designed to capture these risks (as per BMA-SDF weights).

Note that, at least in nominal terms, the cash flows of Treasury bonds are perfectly known in advance. Hence, arguably, we would expect discount rate news to be the primary drivers of their priced risk (\cite{ChenZhao_2009}). Given the flexibility of our general prior introduced in Section~\ref{subsec:kappa}, we can use our discount-rate and cash-flow news decomposition of the factors to encode this prior belief about the relative importance of DR versus CF news. In particular, we can use the $\frac{\mathbb{V}(\text{Ndr})}{\mathbb{V}(u)}$ estimates for each factor to compute the (normalized) $\kappa$ weighting vector to inform the prior: DR factors are assigned a positive weight, while CF factors receive a negative weight. This encodes prior beliefs that traded bond factors classified as being driven (relatively) more by DR news, ceteris paribus, explain a larger portion of the squared Sharpe ratio compared to factors driven by CF news.  We report the posterior factor probabilities and market prices of risk implied by the pricing of the Treasury component of corporate bond returns with this DR factor tilt in Figure~IA.24, and the corresponding pricing statistics in Table~IA.XIX of the Internet Appendix. Obviously, this tilt makes the DR factors individually more likely, pushing the likelihood of the MKTB factor above the prior value, but the pricing results are overall very similar to those with the more diffuse prior encoding, with only a very minor improvement in OS pricing performance  and a small perturbation of the portfolio composition as outlined in the bottom panel of Figure~IA.24.  

Our analysis also sheds light on the degree of integration between equity and corporate bond markets. First, as illustrated by the generalized canonical correlation (GCC) analysis in Figure~IA.8 of the Internet Appendix, there is substantial commonality---in the time series dimension---between bond and stock returns, with the first GCC being just under 75\% (Panel C). Furthermore, upon removing the Treasury component from bond returns, the GCC analysis remains virtually unchanged (Panel E), once again suggesting that the Treasury component has distinct drivers compared to the risks spanned in equity markets.

Second, Table~\ref{tab:table-model-dim1} shows both evidence of an overlap between the latent risks captured by equity and bond factors (in Panel A, the sum of the Sharpe ratios achievable with either of the two sets of tradable factors is \textit{larger} than the Sharpe ratios achievable with these tradable factors jointly), but also of separation between the risks priced in the two markets (as the maximum Sharpe ratios achievable with the BMA-SDFs that use only equity or bond factors to separately price the two markets, Panels B and C, are smaller than the Sharpe ratios of the co-pricing BMA-SDF that uses the same factors jointly, in Panel A).

Third, the evidence that equity factors can price corporate bond returns once their Treasury component is accounted for (Figure~\ref{fig:bma_comparison_duration}), and that stock factors cannot price this Treasury component (Figure~\ref{fig:XS_fit_treasuryexcess}), suggests both a segmentation between equity risk and Treasury markets and a commonality between stock markets and credit risk in bonds.

Fourth, in Table~IA.XX of the Internet Appendix, we report the time series correlations between (the posterior means of) BMA-SDFs constructed with equity and bond factors, jointly and separately, to price (jointly and separately) stock returns, bond excess returns, duration-adjusted bond returns, and the Treasury component of corporate bond returns. Therein, the correlation between the bond-factors BMA-SDF that prices the Treasury component of bond returns and the stock-factors BMA-SDF that prices equity returns stands out as particularly low: 0.172 (80\% shrinkage). For comparison, the correlations between the co-pricing BMA-SDF and the bond- and stock-only BMA-SDFs that price these asset classes separately are all well in excess of 70\%. Once again, this suggests that the (partial) evidence of segmentation between equity and bond markets is driven by the Treasury component in the latter.

Hence, overall, we find both evidence of commonality of risks priced in the two markets---net of Treasury effects---and hence of integration, and of a degree of segmentation generated by the implicit loading of corporate bonds on Treasury-related risks.

\subsection{The economic properties of the co-pricing SDF}\label{sec:Properties}

\begin{figure}[tb!]
\begin{center}
  \includegraphics[scale=.4, trim = 0cm .5cm 0cm 0cm,clip]{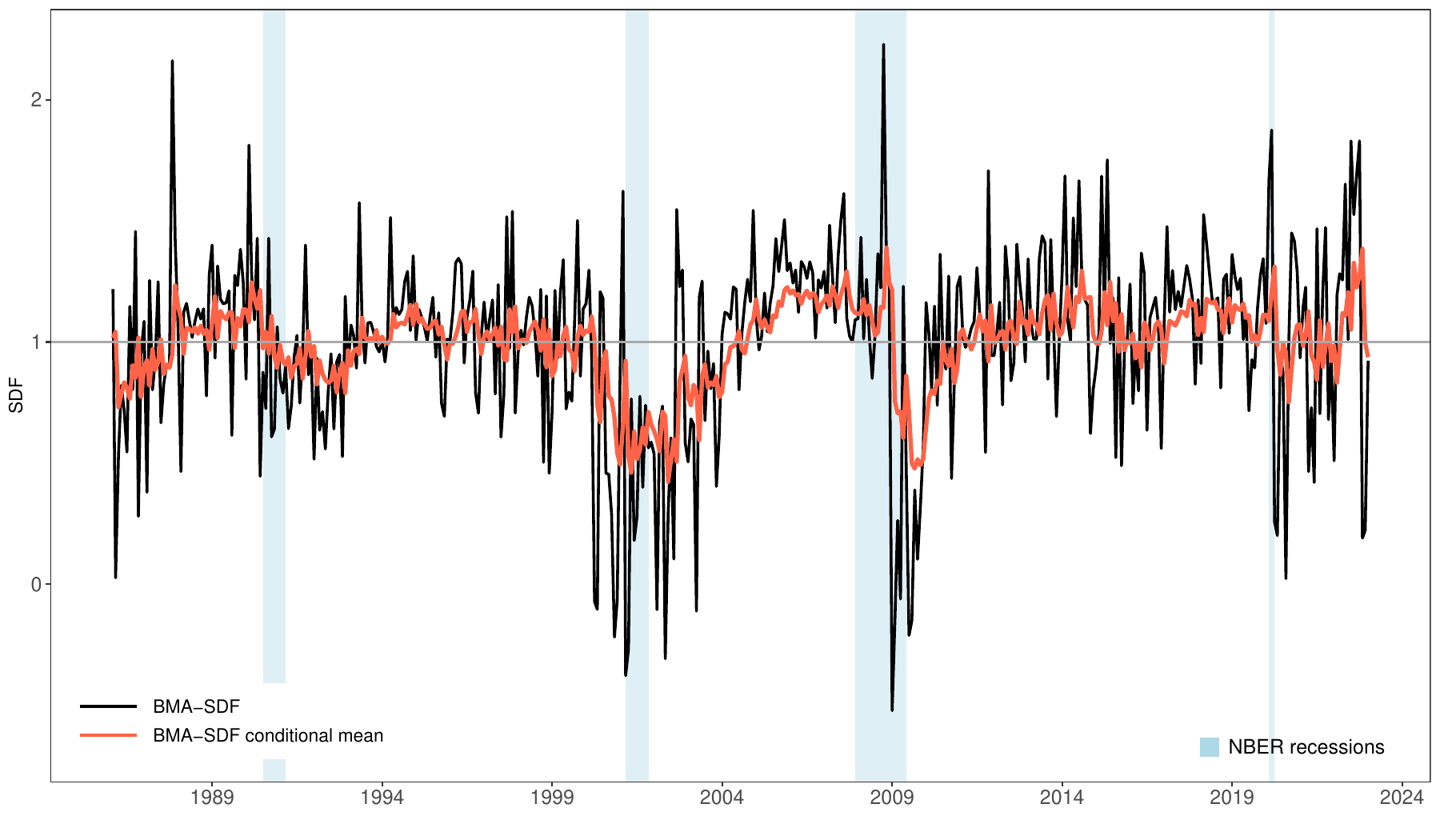} 
\end{center} 
\vspace{-4mm}
\caption{The co-pricing BMA-SDF and its conditional mean.}
\vspace{-2mm}
\begin{justify}
\begin{spacing}{1}
\footnotesize{The figure plots the time series of the (posterior mean of the) co-pricing BMA-SDF and its conditional mean.  The conditional mean is obtained by fitting an ARMA(3,1) to the BMA-SDF whereby the order of the ARMA is selected using the AIC and the BIC. Shaded areas denote NBER recession periods. The sample period is 1986:01 to 2022:12 ($T=444$).
}
\end{spacing}
\end{justify}
\vspace{-10mm}
\label{fig:bma_SDF_TS}
\end{figure}

We now turn to assessing the economic properties of the co-pricing BMA-SDF. Figure~\ref{fig:bma_SDF_TS} depicts the time series of the BMA-SDF (that is, its posterior mean), along with its conditional time series mean (estimated using an ARMA(3,1) model selected via both the Akaike and the Bayesian Information Criteria, AIC and BIC). Both the SDF and its conditional mean exhibit clear business cycle behavior as they increase during expansions and tend to peak right before recessions, being substantially reduced during economic contractions. Moreover, as highlighted in Panel A of Figure~IA.27 of the Internet Appendix, the BMA-SDF is highly predictable: virtually all of its autocorrelation coefficients are statistically significant at the 1\% level up to 20 months ahead, and the $p$-value of the \cite{LjungBox_1978}  test of joint significance is zero at this horizon.  Additionally, about one-fifth of its time series variance is explained by its own lags (23\% for the best AR specification and 19\% for the best ARMA specification according to the BIC). 

Note also that, as shown in Figure~IA.28 of the Internet Appendix, none of the other celebrated SDF models come close to displaying such a level of business cycle variation and persistency: the KNS SDF has about 11\% of its time series variation being predictable by its own history, while this number drops to 4\% for RPPCA, and is only 2\% to 3\% for FF5 and CAPMB, and zero for HKM and CAPM. Remarkably, as shown in Panel A of Table~IA.XXI of the Internet Appendix, the SDFs with a higher degree of persistency, KNS and RPPCA, are exactly the ones with the highest degree of correlation with the BMA-SDF (0.78 and 0.55, respectively), and are the closest competitors for the BMA-SDF in the pricing exercises in Section~\ref{sec:Copricing}. Instead, SDFs that perform significantly worse in cross-sectional pricing have both little time series persistency and correlations with the BMA-SDF in the 0.16 to 0.29 range.

Obviously, the nontradable factors in the BMA-SDF play an important role in generating a pronounced business cycle pattern and a high degree of predictability. Nevertheless, even when removing  the nontradable factors from the BMA-SDF, the resulting SDF remains predictable, with 5\% to 10\% of its time series variation explained by its own lags, and a highly significant  \cite{LjungBox_1978} test statistic even up to 20 months ahead. Furthermore, note that the five most likely factors in the SDF (PEAD, IVOL, PEADB, CREDIT, YSP) explain only about 47\% of the time series variation of the BMA-SDF, further confirming the dense nature of the pricing kernel. Individually, only PEADB and IVOL explain marginally more than 20\% of the time series variation of the SDF, with the other factors accounting individually for 3\% to 7\%.

Recall that the variance of the SDF is equal to the squared Sharpe ratio achievable in the economy. Hence, whether our filtered SDF implies time-varying compensation for risk can be elicited by analyzing the predictability of its volatility. As pointed out in \cite{Engle_1982},  the presence of volatility clustering can be assessed, without taking a parametric stance on the variance process, by simply analyzing the serial correlation of the squared one-step-ahead forecast errors, since these are consistent (yet noisy) estimates of the latent conditional variance. Note that, for instance, such a variance proxy has been used extensively in the macrofinance literature (see, e.g., \citet{BansalKhatchatrianYaron_2005}, \citet{BansalKikuYaron_2012}, \cite{BeelerCampbell_2012}, and \cite{Chen_2017}), and squared forecast errors of returns are commonly used as a proxy for latent conditional variances. 

Panel B of Figure~IA.27 of the Internet Appendix reports the empirical autocorrelation function of the squared forecast errors of the co-pricing BMA-SDF. Most of the autocorrelation coefficients are statistically significant at the 1\% level up to seven months ahead. Moreover, the \cite{LjungBox_1978} test strongly rejects the joint null of zero autocorrelations up to 20 months into the future (the $p$-value of the test is zero). That is, not only does the first moment of our filtered SDF exhibit substantial predictability, but so does its second moment, suggesting time-varying risk compensation in the economy.

\begin{figure}[htb!]
\begin{center}
\includegraphics[scale=.4, trim = 0cm .5cm 0cm 0cm,clip]{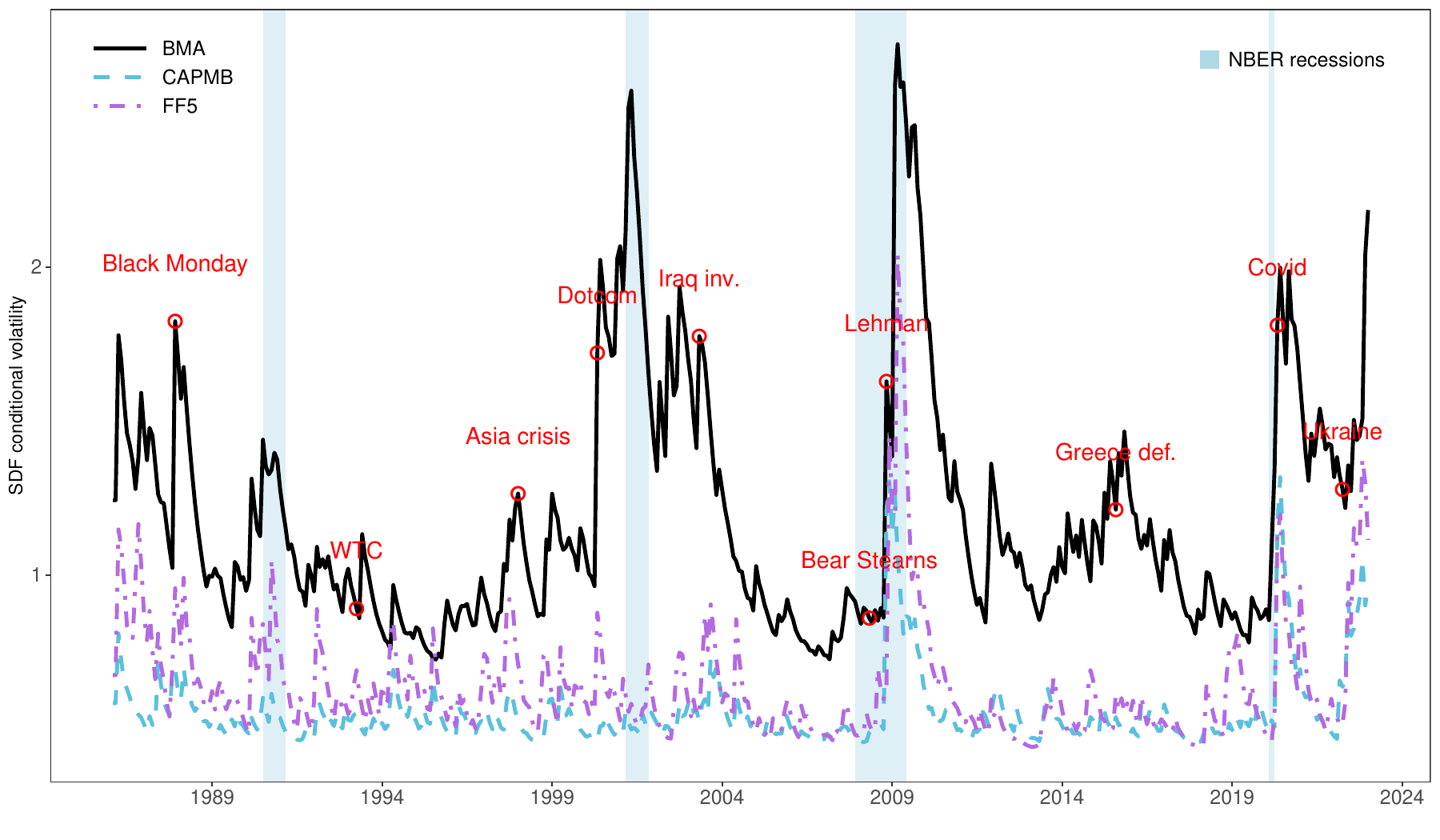}
\end{center} 
\vspace{-4mm}
\caption{Conditional SDF Volatilities.}
\vspace{-2mm}
\begin{justify}
\begin{spacing}{1}
\footnotesize{
The figure plots the annualized volatility of the co-pricing BMA-SDF along with the volatilities of the CAMPB and FF5 SDFs. The volatility of the BMA-SDF is obtained by fitting an ARMA(3,1)-GARCH(1,1) to the posterior mean of the co-pricing BMA-SDF whereby the specification is selected via the AIC and the BIC. 
The GARCH quasi-maximum likelihood coefficient estimates are: 
\vspace{-.2cm}
\begin{center}
\scalebox{.9}{
\begin{tabular}{lccc}
  \multicolumn{4}{c}{$\sigma_{t+1}^2 = \omega + \alpha\epsilon_{t}^2 + \beta \sigma_t^2$} \\
  \hline
  & $\omega$ & $\alpha$ & $\beta $ \\ \hline
  Estimate & 0.01 & 0.15 & 0.81 \\
  Robust SE & 0.00 & 0.04 & 0.06 \\
   \hline
\end{tabular}
}
\end{center}
CAPMB is the bond single-factor model using MKTB, and FF5 is the five-factor model of \cite{FamaFrench_1993}. Estimation details for the benchmark models are given in \ref{sec:benchmark_models}. The volatilities of the SDFs are also computed using a GARCH(1,1) model after selecting an ARMA mean process using the AIC and the BIC. Shaded areas denote NBER recession periods. The sample period is 1986:01 to 2022:12 ($T=444$).
}
\end{spacing}
\end{justify}
\vspace{-10mm}
\label{fig:bma_SDF_vol}
\end{figure}

To tackle the question of whether the SDF-implied time variation in risk compensation (i.e., the economy-wide conditional Sharpe ratio) that we uncover makes economic sense, we fit a simple GARCH(1,1) (see \cite{Bollerslev_1986}) process to our BMA-SDF.\footnote{We estimate the process based on the posterior mean of the BMA-SDF. Ideally, one would estimate the volatility process for each draw of the SDF and for each possible model, and then compute the posterior average of these `draws' for the volatility process. Nevertheless, since GARCH estimation requires numerical optimization, the ideal approach is unfeasible in our model space with quadrillions of models.} Figure~\ref{fig:bma_SDF_vol} presents the estimated conditional volatility of the SDF, revealing striking results. The implied conditional Sharpe ratio is not only highly countercyclical but also exhibits pronounced spikes during periods of market turbulence and heightened economic uncertainty. These include Black Monday, the Asian financial crisis, the burst of the dot-com bubble, the 9/11 terrorist attacks, the Iraq invasion, the great financial crisis, the Greek default and subsequent eurozone debt crisis, the COVID pandemic, and the aftermath of Russia's invasion of Ukraine. Note that the estimated GARCH coefficients imply a highly persistent conditional volatility, with deviations from the mean exhibiting a half-life of approximately 16.6 months.\footnote{Recall that the half-life of a GARCH(1,1) process is defined as $1+\frac{\ln(1/2)}{\ln (\alpha + \beta)} $ where $\alpha$ and $\beta$ denote, respectively, the coefficients on lagged squared error and variance.} 

As per Panel A in Table~\ref{tab:table-model-dim1}, nontradable factors account for about a quarter of the SDF variance. Hence, a natural question is whether the SDF volatility pattern depicted in Figure \nolinebreak \ref{fig:bma_SDF_vol} is simply due to tradable factors. We evaluate this conjecture by removing all tradable factors from the BMA-SDF and re-estimating the volatility process of this nontradable-only SDF. We find that the resulting volatility process remains very persistent (with a half-life of 12.3 months), with pronounced business cycle variation and reaction to periods of heightened economic uncertainty (see Figure~IA.29 of the Internet Appendix).  Moreover, it has a correlation with the volatility of the BMA-SDF in Figure~\ref{fig:bma_SDF_vol} of about 62\%. That is, both tradable and nontradable components of the BMA-SDF are characterized by a very persistent volatility with a clear business cycle pattern.

But is the strong countercyclical behavior of the BMA-SDF volatility, and its sharp increase during periods of economic uncertainty, just a mechanical byproduct of it loading on several tradable factors? Figure~\ref{fig:bma_SDF_vol} suggests that this is not the case. Focusing on the celebrated five-factor model of \cite{FamaFrench_1993} and the bond CAPM (which is not incrementally outperformed by alternative models considered in \citet{DickersonMuellerRobotti_2023}), we apply the same procedure of estimating their SDF coefficients and computing the implied conditional volatilities using a GARCH specification (after fitting a mean model based on the AIC and the BIC). The estimated volatilities for these two SDF models in Figure~\ref{fig:bma_SDF_vol} make clear that the use of tradable factors in the SDF does not mechanically deliver our findings for the BMA-SDF: both the cyclical pattern and the reaction to periods of heightened economic uncertainty are much less pronounced for the FF5 model, and even more so for the CAPMB. This is formally measured in Figure~IA.30  of the Internet Appendix that shows that the half-life of volatility shocks to the FF5 SDF model is only 4.2 months, and for the CAPMB it is just 3 months. Finally, Figure~IA.31 of the Internet Appendix depicts the residual of the linear projection of the BMA-SDF estimated volatility on the estimated volatilities of the KNS, RPPCA, CAPM, CAPMB, HKM and FF5, with the residual showing a strong business cycle pattern and being particularly large and positive during periods of high economic uncertainty, suggesting that these alternative SDF models do not sufficiently capture these states despite being based on tradable factors. 

The observed business cycle variations and predictability in both the first and second moments of the SDF would imply, within a structural model, time-varying and predictable risk premia for tradable assets. Therefore, we now turn to testing this \emph{time series} prediction of our BMA-SDF identified from \emph{cross-sectional} pricing.

The precise functional form of the predictive relation between current SDF moments and future asset returns does depend on the postulated model. Nevertheless, as shown in \citet{BryzgalovaHuangJulliard_2024}, the \cite{HansenJagannathan_1991} conditional SDF projections on the space of returns imply a (log) linear SDF driven by two factors: the innovations to the SDF and the product of the conditional mean of the SDF and the same innovations. Therefore, assuming a contemporaneous linear relationship between asset returns and the SDF yields a simple linear dependence of conditional risk premia on two variables: (i) the conditional variance of the SDF and (ii) the product of this conditional variance with the conditional mean of the SDF.  

\begin{figure}[tb!]
    \centering
    \begin{subfigure}[b]{\textwidth}
        \centering
        \includegraphics[scale=.4, trim = 0cm 0cm 0cm 0cm,clip]{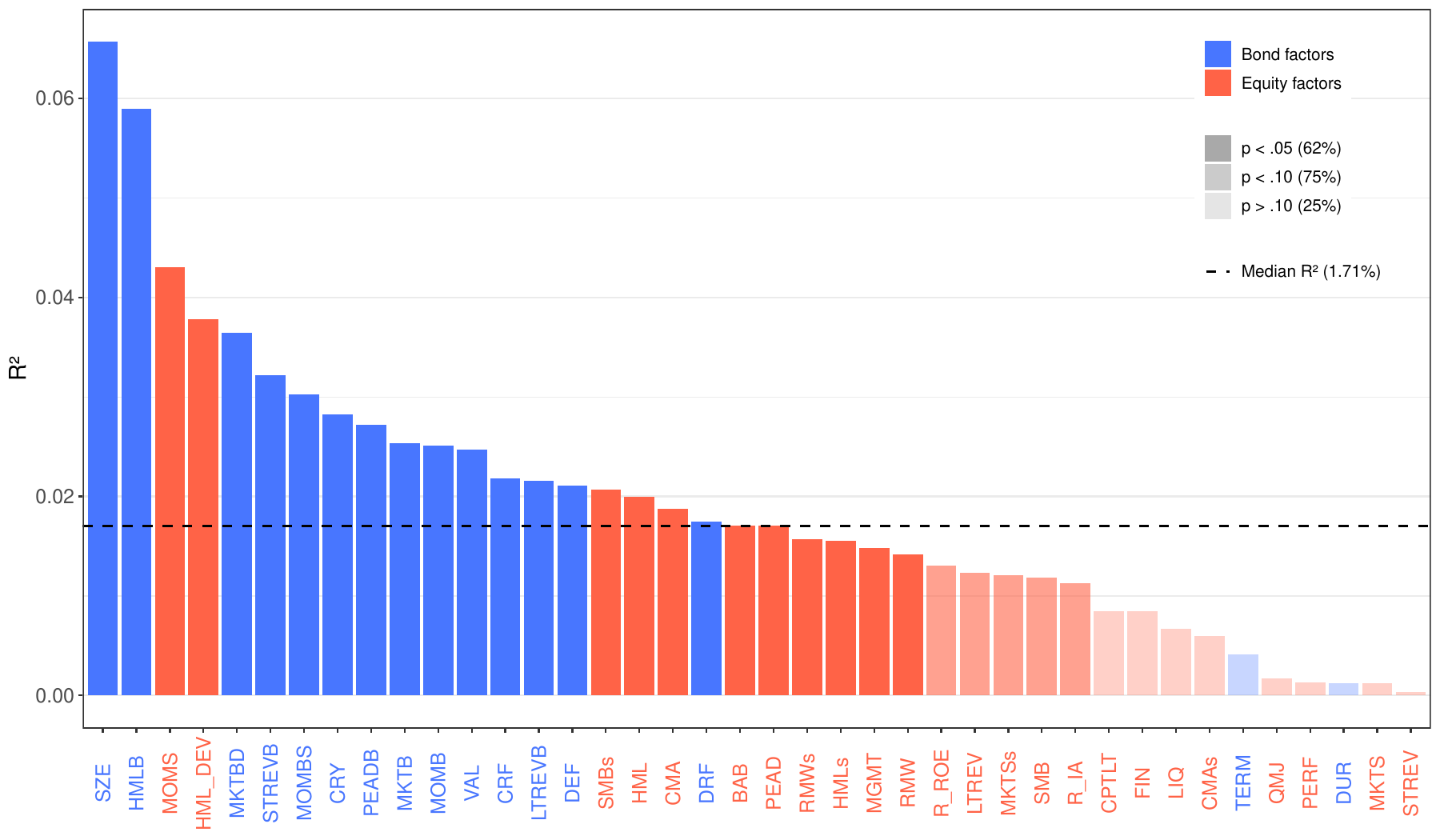}
        \caption{Predictability of monthly log returns}
        \label{fig:first_image}
    \end{subfigure}
    \vfill
    \begin{subfigure}[b]{\textwidth}
        \centering
        \includegraphics[scale=.4, trim = 0cm 0cm 0cm 0cm,clip]{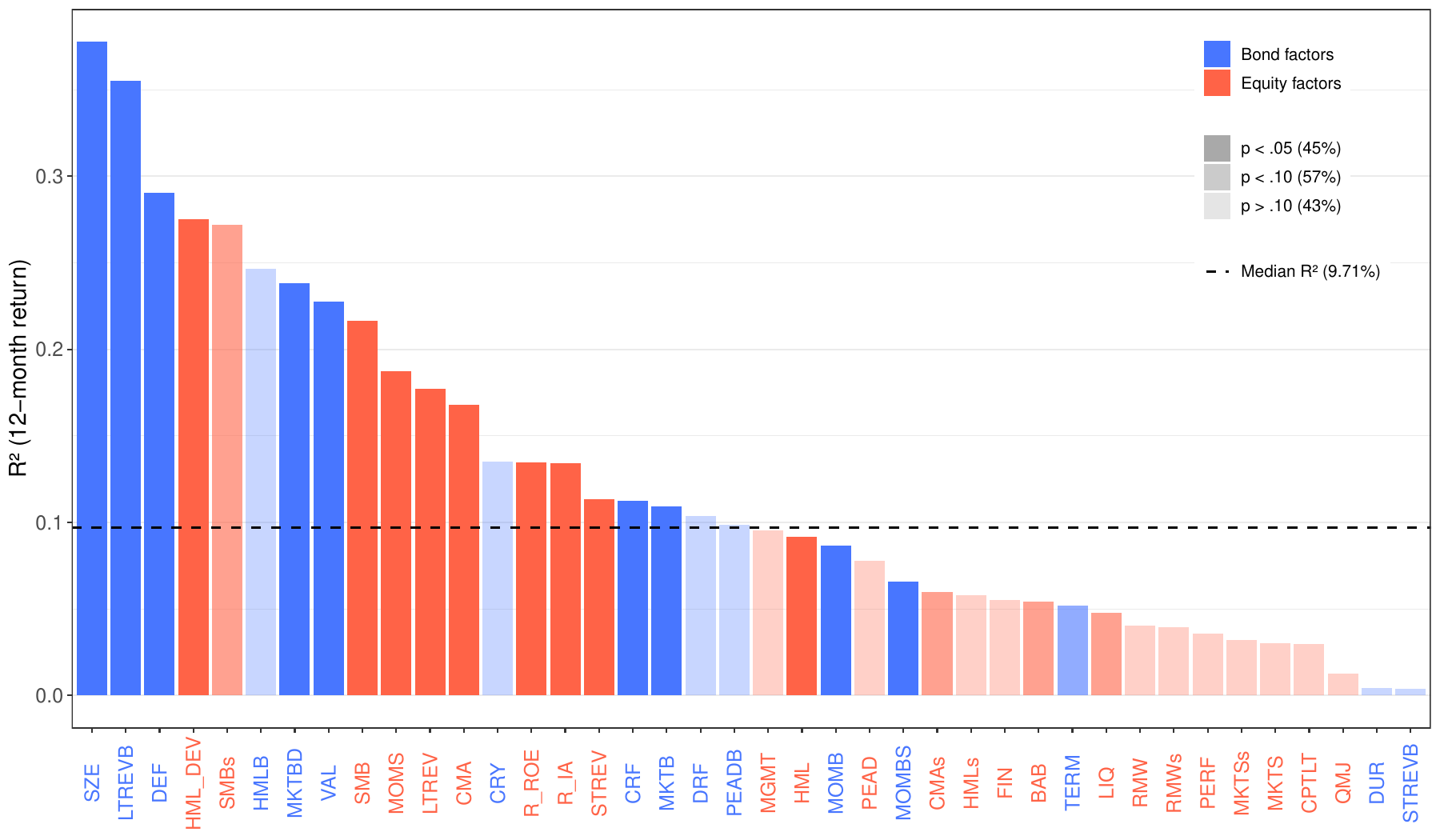}
        \caption{Predictability of twelve months cumulative log returns}
        \label{fig:second_image}
    \end{subfigure}
\vspace{-4mm}
    \caption{Predictability of tradable factors with lagged SDF information.}
\vspace{-2mm}
\begin{justify}
\begin{spacing}{1}
\footnotesize{
The figure shows the $R^2$s of predictive regressions of factor returns on the previous month estimates of the co-pricing BMA-SDF conditional variance and conditional variance interacted with the conditional mean. Panel A shows $R^2$s for one-month ahead predictions while Panel B shows $R^2$s for one-year ahead predictions.  The volatility of the BMA-SDF is obtained by fitting an ARMA(3,1)-GARCH(1,1) to the posterior mean of the co-pricing BMA-SDF whereby the specification is selected via the AIC and the BIC. To account for the overlapping nature of the observations in Panel B, we construct robust standard errors by (i) using a Bartlett kernel (\cite{NeweyWest_1987}) with 15 lags, (ii) constructing a sandwich estimate of the covariance matrix, and (iii) applying a degrees of freedom correction. The 40 predicted tradable factors are described in  \ref{sec:factor_zoo}.
}
\end{spacing}
\end{justify}
\vspace{-10mm}
\label{fig:predict_with_SDF}
\end{figure}

Leveraging this insight, we run predictive regressions of asset (log) returns between time $t-1$ and $t$, as well as cumulated (log) returns between $t-1$ and $t+12$,  on SDF information observed at time $t-1$: $\mathbb{E}_{t-1}[M_t] \times \mathbb{E}_{t-1}[\sigma^2_{t}]$ and  $\mathbb{E}_{t-1}[\sigma^2_{t}]$, where the conditional mean is constructed, as in Figure~\ref{fig:bma_SDF_TS}, by fitting an ARMA(3,1) process (the preferred specification according to the BIC), and the conditional variance is obtained from the GARCH(1,1) estimates depicted in Figure \nolinebreak \ref{fig:bma_SDF_vol} (and also selected via the BIC).

As test assets to be predicted, we employ the bond and stock  factors used in our cross-sectional analysis since these are generally hard to predict and should, according to the previous literature, demand sizable risk premia.

Figure~\ref{fig:predict_with_SDF} summarizes the predictability results. In Panel A, we report the $R^2$ values for the one-month-ahead predictions, and in Panel B the same for the cumulative twelve-month-ahead predictive regressions. We encode, via shading, the joint statistical significance of the regressors as implied by an $F$-test of the regression coefficients. 

The results are striking. For the majority of test assets, we find that information embedded in the lagged SDF significantly predicts future asset returns. At the monthly horizon shown in Panel A, this predictability is statistically significant in 75\% of cases at the 10\% confidence level and in 62\% of cases at the 5\% significance level. Second, the amount of predictability is economically large, albeit not unrealistically so: for the statistically significant cases it ranges from $1.1\%$ to $6\%$ at the monthly horizon (Panel A). At the twelve-month horizon (Panel B) the median $R^2$ is about 10\%, with many factors having more than one-fifth of their time series variation being predictable. Moreover, even with an extremely conservative approach to constructing the covariance matrix, the $F$-test yields statistically significant results in about 60\% of cases at the 10\% level and 45\% of cases at the 5\% level.\footnote{We construct conservative standard errors by (i) using a Bartlett kernel (\cite{NeweyWest_1987}) with 15 lags, (ii) constructing a sandwich estimate of the covariance matrix, and (iii) applying a degrees of freedom correction to account for the relatively small sample of independent observations.}

\FloatBarrier

\section{Robustness}\label{sec:robustness}

In this section, we discuss an extensive array of robustness exercises that all confirm our main findings.

\subsection{Factor tilting}\label{sec:factor_tilt}

Our novel spike-and-slab prior in Section \ref{subsec:kappa} allows us to assign a heterogeneous degree of prior shrinkage to the different types of factors by setting the hyper-parameter $\kappa$ to values different from zero. This parameter encodes our prior belief about the share of the SDF Sharpe ratio generated by the respective types of factors.

Consider $\kappa \in \{-0.5,0.5\}$. Setting $\kappa = 0.5$ for bond factors implies the belief that, ceteris paribus, they explain a share of the squared Sharpe ratio of the SDF that is $\frac{1+\kappa}{1-\kappa} =$ 3 times as large as the share of stock factors. This represents a substantial departure from the homogeneous prior setting (i.e., $\kappa =0$). Nevertheless, since our prior is not dogmatic and does not impose a hard threshold, it can be falsified if the data do not conform with it.

Figure IA.32 of the Internet Appendix reports the posterior factor probabilities estimated with the tilted prior in favor of either bond or stock factors. Remarkably, the factors that we identify as more likely given the data in Section \ref{sec:Copricing_SDF} still have posterior probabilities above the prior value in 9 out of 10 cases.
That is, the likelihood of the data is quite informative for these more likely factors, and the prior perturbation has only a limited effect on the posterior probabilities.  Similarly, the posterior market prices of risk depicted in Figure IA.33 of the Internet Appendix highlight that the set of factors that features more prominently in the co-pricing BMA-SDF is largely unchanged, albeit their individual posterior $\lambda$s do vary in the expected directions.

For a \emph{sparse} SDF, we would expect these perturbations of the posterior $\lambda$s to have a substantial impact on its pricing ability. For a \emph{dense} SDF that combines multiple noisy proxies for common underlying sources of risk, we should expect instead a much more muted effect (as also implied by our simulation results in Section \ref{sec:Simulation}). Table IA.XXII of the Internet Appendix summarizes the resulting in- and out-of-sample pricing performance of the tilted BMA-SDF for our baseline cross-section of test assets. Overall, the effect of the prior tilting is quite small but unambiguous in direction: as we tilt toward \emph{either} type of factor, the out-of-sample pricing ability deteriorates. This strengthens the results in Section \ref{sec:Information}: for the co-pricing of bond and stock excess returns, we need information from both factor zoos. Consequently, over-reliance on either type of factor worsens the BMA-SDF performance.

Interestingly, as shown in Table IA.XXIII of the Internet Appendix, where we consider the separate pricing of bond and stock excess returns, the deterioration in pricing performance is stronger for equities when tilting the prior in favor of bond factors---again reinforcing the result in Section \ref{sec:Information} of a much more limited information content in the bond factor zoo relative to the equity one.

Finally, we revisit our findings on co-pricing when bond returns are duration-adjusted. The results in Section \ref{sec:Information} strongly suggest that, once the Treasury component of bond returns is accounted for, the bond factor zoo becomes largely redundant. If this were truly the case, we would expect that tilting the prior in favor of stock (bond) factors should actually improve (worsen) the pricing ability of the BMA-SDF. This is exactly what Figure IA.34 of the Internet Appendix highlights. Unambiguously, as we tilt the prior away from bond factors, the out-of-sample measures of cross-sectional fit improve. Furthermore, an extreme tilt in favor of stock factors (see Figure IA.35 of the Internet Appendix) maximizes the pricing ability of the BMA-SDF. This reinforces our previous finding: bond factors are largely redundant for co-pricing bond and stock portfolios once the Treasury component of the latter is accounted for.

\subsection{Imposing sparsity}\label{sec:sparsity}

Recall that in our method, beliefs regarding SDF density are encoded through a Beta-distributed prior probability of factor inclusion: $\pi(\gamma_j=1|\omega_j) = \omega_j\sim Beta(a_{\omega},b_{\omega})$. In our baseline estimations, we assign a $\mathrm{Beta}(1,1)$ prior distribution to this quantity---equivalent to a uniform prior on $[0,\,1]$ and analogous to the flat prior implicit in canonical frequentist inference. This specification reflects our decision not to take an ex ante stance on whether the SDF should be sparse or dense.

However, the literature commonly assumes a high degree of sparsity, either ex ante or through specification selection, typically favoring factor models with approximately five factors. Our framework accommodates such beliefs in a flexible, non-dogmatic manner by choosing the prior mean and variance of $\omega_j$, $\mathbb{E}[\omega_j]=\frac{a_{\omega}}{a_{\omega}+b_{\omega}}$ and $\operatorname{Var}(\omega_j)=\frac{a_{\omega}b_{\omega}}{(a_{\omega}+b_{\omega})^{2}(a_{\omega}+b_{\omega}+1)}$. 

Using appropriate $a_{\omega}$ and $b_{\omega}$, we can concentrate the prior on model dimensions typical in the literature. Specifically, we set $a_{\omega}\approx 3.54$ and $b_{\omega}\approx 34.66$ to achieve two objectives: (i) the prior expectation of included factors, $\mathbb{E}[\omega_j]\times K$, yields the canonical five-factor model, and (ii) the prior two standard deviation credible interval encompasses models with zero to ten factors (since $\operatorname{Var}(\omega_j) = (2.5/K)^2$).

Results using this sparsity-favoring prior appear in Internet Appendix IA.9.2. Three key findings emerge. First, Table IA.XXIV of the Internet Appendix shows that the factors with posterior probabilities exceeding the prior value (that is, 9.26\%) are essentially identical to those in our baseline estimates. The only exception occurs under the lowest prior shrinkage, where PEAD's posterior probability drops below this threshold---an expected outcome given this prior's reduced ability to control confounding effects from weak factors. Second, Table IA.XXV of the Internet Appendix demonstrates that the BMA-SDF's pricing performance under the sparsity-favoring prior remains superior to alternative specifications in the literature, particularly out-of-sample. Third, despite this relative advantage, imposing sparsity degrades the performance of the BMA-SDF compared to our baseline findings in Tables \ref{tab:tab-is-pricing-excess} and \ref{tab:tab-os-pricing-excess}. This deterioration is expected: as Figure \ref{Fig:post_dim} and Table \ref{tab:table-model-dim1} demonstrate, the data strongly support a dense SDF. Consequently, artificially imposing sparsity necessarily worsens the performance of BMA-SDF, as our results confirm. These findings highlight once more that the quest for a pricing kernel in the previous literature, focusing on low-dimensional observable factor models, relies on a stringent assumption not supported by the data.

\subsection{Estimation \emph{excluding} the most likely factors}\label{sec:remove_top}

The empirical findings in Section \ref{sec:Copricing} strongly suggest that the joint zoo of bond and stock factors resembles a jungle of noisy proxies for common underlying sources of risk. As we show theoretically in Section \ref{sec:ModelSel} and in simulation in Section \ref{sec:Simulation}, if this characterization is accurate, our BMA-SDF method should provide a good approximation of the true latent SDF even when factors capturing fundamental risk sources are removed from the candidate set. 

To assess whether this robustness property holds in the data, we remove the factors identified as most salient for characterizing the true latent SDF, construct a BMA-SDF using the remaining factors, and evaluate its pricing ability both in- and out-of-sample. We perform this exercise by removing three different factor sets: (i) the top five factors ranked by posterior probabilities; (ii) the top five factors ranked by posterior weights in the BMA-SDF (i.e., factors with the largest posterior market prices of risk); and (iii) the union of factors from sets (i) and (ii). This constitutes a stringent test of our method, as we remove the factors individually identified as the most informative about priced risk in the economy.

Internet Appendix IA.9.3 reports the empirical results. Remarkably, the BMA-SDF constructed with this limited information set still strongly outperforms canonical models from the literature both in- and out-of-sample (see Table IA.XXVI of the Internet Appendix). As shown in Figures IA.36 to IA.38 of the Internet Appendix, this performance is achieved by increasing the posterior weights, $\mathbb{E}[\lambda_j|\text{data}]$, of several noisy proxies in the BMA-SDF---precisely what our theoretical and simulation results in Section \ref{sec:ModelSel} predict. 

However, we do observe some minor degree of deterioration in the performance of the BMA-SDF, particularly when minimal prior shrinkage is applied. This is again an expected outcome given this prior's reduced ability to control confounding effects from weak factors. However, even in the most extreme case, this reduction remains moderate, with out-of-sample $R^2$ measures dropping by only 8\% in the worst-case scenario.

Overall, these results confirm both the soundness and robustness of our method in recovering pricing information from the factor zoo and our finding that most factors are noisy proxies of common underlying risk sources.

\subsection{Estimation uncertainty}\label{sec:robustness_data}

Finally, we show that our asset pricing results are robust across (i)  different corporate bond data, (ii) varying bond and stock cross-sections and (iii) different factor zoos and sample periods. The detailed results are presented in Internet Appendix IA.10.

\subsubsection{Varying corporate bond data}\label{sec:vary_corporate_data}

In Internet Appendix IA.1 we describe different sources for corporate bond data used for academic research and, in particular, we show the robustness of the bond factors with respect to the data source and calculation method. In this section we confirm that the asset pricing implications are also robust to the choice of corporate bond data. Detailed results are presented in Internet Appendix IA.10.1.

In particular, we compare the pricing performance of the co-pricing BMA-SDF across five different sets of corporate bond data: (i) our baseline LBFI/BAML ICE bond-level data, (ii) the LBFI/BAML ICE firm-level data, (iii) the LBFI/BAML ICE bond-level data but using only quotes (i.e., removing matrix prices), (iv) the transaction-based WRDS TRACE data, and (v) the transaction-based DFPS TRACE data. That is, with each of these datasets, we re-estimate the co-pricing BMA-SDF using the 83 test assets and 54 tradable and nontradable factors. Across estimations, only the 50 IS bond test assets and tradable bond factors change. 

First, the ex post Sharpe ratios across all shrinkage levels are very closely aligned, as shown in Table IA.XXVII of the Internet Appendix. Second, the posterior probabilities across the data sets are very consistent. On average, eight out of the ten most likely factors (including the top five) match the baseline results from Section \ref{sec:Copricing_SDF} (see Figure IA.39 of the Internet Appendix). Finally, the in- and out-of-sample asset pricing performance of the BMA-SDF is fairly consistent across corporate bond data sets and, most importantly, the BMA-SDF still emerges as the dominant model across all estimations, with a tight spread between min and max values (see Figures IA.40 (IS)  and IA.41 (OS) of the Internet Appendix). 

\subsubsection{Varying cross-sections}\label{sec:vary_cross_section} 

Our baseline estimate of the BMA-SDF is specific to the test assets that we describe in Section \nolinebreak \ref{sec:data}. We now vary the cross-section of test assets and re-estimate the co-pricing BMA-SDF for \textit{hundreds} of alternative sets of test assets across bonds and stocks. 

Specifically, we include the 153 long-short equity anomalies provided by \citet{JensenKellyPedersen_2023} and the corporate bond counterparts from \citet{DickNielsenFeldhuetterPedersenStolborg_2024} for a joint corporate bond and stock cross-section of 306 anomalies. From this very large cross-section, we then randomly sample anomaly pairs to generate 100 in-sample co-pricing cross-sections. Each sampled cross-section consists of 25 bond and 25 stock portfolios from the same underlying anomaly characteristic. Together with the 40 tradable bond and stock factors, we use 90 IS test assets for the estimation. In Figure IA.42 of the Internet Appendix we present the average posterior probabilities (Panel A) and the market prices of risk (Panel B), along with their respective minimum and maximum values across the 100 estimations, with the Sharpe ratio shrinkage set to 80\% of the ex post maximum. IVOL, PEADB, and PEAD still emerge as the most probable factors for inclusion in the SDF---consistent with the results documented in Figure \ref{Fig:post_probs} and Table \ref{tab:table-app-probs} of \ref{sec:all_Sharpe_priors}.

In Figures IA.43 (IS) and IA.44 (OS) of the Internet Appendix we present the averages, minima and maxima of the $R^2_{GLS}$ (Panel A) and $R^2_{OLS}$ (Panel B) values across the 100 sets of test assets for the BMA-SDF across our four Sharpe ratio shrinkage levels and the additional models we consider in Tables  \ref{tab:tab-is-pricing-excess} and \ref{tab:tab-os-pricing-excess}. The BMA-SDF with 60\% and 80\% Sharpe ratio shrinkage, as well as the `TOP' model, outperform all other models, confirming the results presented in Section \ref{sec:xs-pricing}. Finally, we also obtain similar results when we switch the IS and OS test assets (i.e., instead of evaluating the pricing performance on the OS test assets, we use them to estimate the BMA-SDF); these results are presented in Figure IA.45 and Table IA.XXVIII of the Internet Appendix. For all sets of IS test assets, our results remain materially the same. That is, we identify a similar set of factors that should be included in the co-pricing SDF, estimate consistent market prices of risk, and obtain similar in- and out-of-sample asset pricing performance for the BMA-SDF.

To further assess the OS performance of our approach, we evaluate its pricing ability across \textit{millions} of potential OS cross-sections of bond and stock portfolios. Again, we use the \citet{JensenKellyPedersen_2023} and the \citet{DickNielsenFeldhuetterPedersenStolborg_2024} anomaly data. We form OS cross-sections with 50 and 100 portfolios (i.e., 25 and 50 anomaly pairs, for stocks and bonds), respectively. From the 306 anomalies, we sample the respective cross-section one million times and evaluate the OS pricing performance using the BMA-SDFs estimated with the baseline set of test assets in Panel A of Table \ref{tab:tab-is-pricing-excess}. The results are presented in Table IA.XXIX of the Internet Appendix with the BMA-SDF and the TOP factors model once again outperforming their competitors. 

\subsubsection{Varying factor zoos and sample periods}\label{sec:vary_zoo} 

Finally, we check the robustness of our results regarding the expansion of the factor zoo and the alteration of the sample periods. First, in order to expand the set of stock and nontradable factors included in our analysis, we consider a shorter sample (ending in December 2016) to include all 51 stock factors considered in \citet{BryzgalovaHuangJulliard_2023} as well as their stock portfolio of IS test assets to re-estimate the co-pricing BMA-SDF. Second, we extend the corporate bond factor zoo from 16 to 29 factors by adding the 13 \citet{DickNielsenFeldhuetterPedersenStolborg_2024} composite bond return factors formed with equity characteristics. Third, we restrict the sample period to the TRACE era (from 2002 onward only) and include the tradable liquidity factor (LRF) from \citet{BaiBaliWen_2019} and the two nontradable illiquidity factors from \cite{LinWangWu_2011}. Fourth, we estimate the models for the pre-TRACE period (1986 to 2002) and repeat the analysis using the split used by \citet{vanBinsbergenNozawaSchwert_2025} who consider pre- and post-2000 data. Finally, we estimate the models on an extended time series starting in 1977, resulting in a total of 549 observations in the time series. 

The posterior probabilities and market prices of risk for these estimations are reported in Figures IA.46 to IA.48 of the Internet Appendix, with associated asset pricing results documented in Tables IA.XXX (IS) and IA.XXXI (OS) of the Internet Appendix. The IS and OS asset pricing results for the pre- and post-TRACE and pre- and post-2000 sample splits are reported in Tables IA.XXXII and IA.XXXIII of the Internet Appendix, respectively. 

Overall, the results are remarkably robust, and our BMA-SDF generally outperforms competing models, independent of how we cut the data.

As we show, both theoretically and in simulation in Section \ref{sec:ModelSel}, the stability of the findings is to be expected given the robust inference method we use: if individual factors are combinations of signal (about the fundamental sources of risk) plus ``noise'' (their unpriced component, see, e.g., \citet{DanielMotaRottkeSantos_2020}), the BMA-SDF provides an optimal aggregation scheme that maximizes the signal-to-noise ratio of the resulting SDF. Hence, albeit perturbations of the data might alter the signal-to-noise ratio of individual factors, this effect is largely mitigated in the BMA-SDF that our method delivers, rendering such issues, as the data confirms, of second order concern for our analysis.

\section{Conclusion}\label{sec:conclusion}

We generalize the Bayesian estimation method of \citet{BryzgalovaHuangJulliard_2023} to handle multiple asset classes, developing a novel understanding of factor posterior probabilities and model averaging in asset pricing, and we apply it to the study of over 18 quadrillion linear factor models for the joint pricing of corporate bond and stock returns.

Strikingly, decomposing bond excess returns into their credit and Treasury components reveals that nontradable and tradable \emph{stock} factors are largely \emph{sufficient} for pricing the credit component, making the bond factor literature effectively redundant for this purpose. Conversely, tradable \emph{bond} factors (along with nontradable ones) remain necessary for pricing the Treasury component---a risk that stock factors do not seem to capture.

Overall, we find that the true latent SDF is \emph{dense} in the space of observable nontradable and tradable bond and stock factors. Importantly, this implies that \emph{all} low dimensional observable factor models proposed to date are affected by severe misspecification and rejected by the data.

Individually, only very few factors should be included in the SDF with high probability. Most notably, two tradable behavioral factors capturing the post-earnings announcement drift in bonds and stocks exhibit posterior probabilities above their prior value, along with nontradable factors such as the slope of the Treasury yield curve, the AAA/BAA yield spread, and the idiosyncratic equity volatility. However, these factors capture only a fraction of the risks priced in the joint cross-section of bonds and stocks, and literally dozens of other factors, both tradable and nontradable, are necessary---jointly---to span the risks driving asset prices. Nevertheless, the SDF-implied maximum Sharpe ratio is not extreme because the many factors necessary for an accurate characterization of the latent SDF are multiple noisy proxies for common underlying sources of risk. 

A Bayesian Model Averaging over the space of all possible Stochastic Discount Factor models aggregates this diffuse pricing information optimally and outperforms all existing models in explaining---jointly and individually---the cross-section of corporate bond and stock returns, both in- and out-of-sample. Furthermore, leveraging the fact that the Bayesian averaging over the space of models is equivalent to an averaging over the space of factors, we show that the BMA-SDF yields a \emph{tradable} strategy with a time-series \emph{out-of-sample} Sharpe ratio of 1.5 to 1.8, with only yearly rebalancing, in the challenging evaluation period spanning July 2004 to December 2022.

The BMA-SDF exhibits a distinctive business cycle behavior, and persistent and cyclical first and second moments. Furthermore, its volatility increases sharply during recessions and at times of heightened economic uncertainty, suggesting time variation in conditional risk premia. And indeed, we find that lagged BMA-SDF information is a strong and significant predictor of future asset returns.

\appendix

\renewcommand{\theequation}{A.\arabic{equation}}%
\renewcommand{\theHequation}{A.\arabic{equation}}%
\renewcommand{\thefigure}{A.\arabic{figure}} \setcounter{figure}{0}
\renewcommand{\theHfigure}{A.\arabic{figure}}%
\renewcommand{\thetable}{A.\arabic{table}} \setcounter{table}{0}
\renewcommand{\theHtable}{A.\arabic{table}}%

\section{The factor zoo list} \label{sec:factor_zoo}

We list all 54 bond, stock and nontradable factors we consider in Table~\ref{tab:table-001-appendix} along with a detailed description of their construction, associated reference, and data source.

\begin{scriptsize}
\begin{longtable}{p{1.1cm}p{6.3cm}p{2.6cm}p{1.9cm}}
\caption{List of factors for cross-sectional asset pricing.}\label{tab:table-001-appendix}\\
\toprule
{\textbf{Factor ID}} & {\textbf{Factor name and description}} & {\textbf{Reference}} & {\textbf{Source}} \\ 
\midrule
\endfirsthead
\toprule
{\textbf{Factor ID}} & {\textbf{Factor name and description}} & {\textbf{Reference}} & {\textbf{Source}} \\ 
\midrule
\endhead
\bottomrule
\multicolumn{4}{p{1\textwidth}}{The table lists all tradable bond, stock as well as the nontradable factors used in the main paper. For each of the factors, we present their identification index (Factor ID), a description of the factor construction, and the source of the data for downloading and/or constructing the factor time series.} \\
\endlastfoot

\multicolumn{4}{c}{ \textbf{Panel A: Tradable corporate bond factors} } \\
\midrule
CRF & Credit risk factor. Equally-weighted average return on two `credit portfolios': CRF$_{VaR}$, and CRF$_{REV}$. CRF$_{VaR}$ is the average return difference between the lowest-rating (i.e., highest credit risk) portfolio and the highest-rating (i.e., lowest credit risk) portfolio across the VaR95 portfolios. CRF$_{REV}$ is the average return difference between the lowest-rating portfolio and the highest-rating portfolio across quintiles sorted on bond short-term reversal. & \cite{BaiBaliWen_2019} & \href{https://openbondassetpricing.com/corporate-bond-factor-zoo/}{Open Source Bond Asset Pricing} \\
CRY & Bond carry factor. Independent sort ($5 \times 5$) to form 25 portfolios according to ratings and bond credit spreads (CS). For each rating quintile, calculate the weighted average return difference between the highest CS quintile and the lowest CS quintile. CRY is computed as the average long-short portfolio return across all rating quintiles. & \cite{HottingaVanLeeuwenVanIjserloo_2001}, \cite{HouwelingVanZundert_2017} & \href{https://openbondassetpricing.com/corporate-bond-factor-zoo/}{Open Source Bond Asset Pricing} \\
DEF & Bond default risk factor. The difference between the return on the market portfolio of long-term corporate bond returns (the Composite portfolio on the corporate bond module of Ibbotson Associates) and the long-term government bond return. & \cite{FamaFrench_1992} and \cite{GebhardtHvidkjaerSwaminathan_2005a}. & \href{https://docs.google.com/spreadsheets/d/1g4LOaRj4TvwJr9RIaA_nwrXXWTOy46bP/edit#gid=2070662242}{Amit Goyal website} \\
DRF & Downside risk factor. Independent sort ($5 \times 5$) to form 25 portfolios according to ratings and 95\% value-at-risk (VaR95). For each rating quintile, calculate the weighted average return difference between the highest VaR5 quintile and the lowest VaR5 quintile. DRF is computed as the average long-short portfolio return across all rating quintiles. & \cite{BaiBaliWen_2019} & \href{https://openbondassetpricing.com/corporate-bond-factor-zoo/}{Open Source Bond Asset Pricing} \\
DUR & Bond duration factor. Independent sort ($5 \times 5$) to form 25 portfolios according to ratings and bond duration (DUR$^B$). For each rating quintile, calculate the weighted average return difference between the highest DUR$^B$ quintile and the lowest DUR$^B$ quintile. DUR is computed as the average long-short portfolio return across all rating quintiles. & \cite{GebhardtHvidkjaerSwaminathan_2005a} and \cite{DangHollsteinProkopczuk_2023}. & \href{https://openbondassetpricing.com/corporate-bond-factor-zoo/}{Open Source Bond Asset Pricing} \\
HMLB & Bond book-to-market factor. Independent sort ($2 \times 3$) to form 6 portfolios according to bond size and bond book-to-market (BBM), defined as bond principal value scaled by market value. For each size portfolio, calculate the weighted average return difference between the lowest BBM tercile and the highest BBM tercile. HMLB is computed as the average long-short portfolio return across the two size portfolios. & \cite{BartramGrinblattNozawa_2025} & \href{https://openbondassetpricing.com/corporate-bond-factor-zoo/}{Open Source Bond Asset Pricing} \\
LTREVB & Bond long-term reversal factor. Dependent sort ($3 \times 3 \times 3$) to form 27 portfolios according to ratings, maturity, and the 48-13 cumulative previous bond return (LTREV$^B$). For each rating quintile, the factor is computed as the average return differential between the portfolio with the lowest LTREV$^B$ and the one with the highest LTREV$^B$ within the rating and maturity portfolios. LTREVB is computed as the average long-short portfolio return across the nine rating-maturity terciles. & \cite{BaliSubrahmanyamWen_2021} & \href{https://openbondassetpricing.com/corporate-bond-factor-zoo/}{Open Source Bond Asset Pricing} \\
MKTB & Corporate Bond Market excess return. Constructed using bond returns in excess of the one-month risk-free rate of return. & \cite{DickersonMuellerRobotti_2023} & \href{https://openbondassetpricing.com/corporate-bond-factor-zoo/}{Open Source Bond Asset Pricing} \\
MKTBD & Corporate Bond Market duration-adjusted return. Constructed using bond returns in excess of their duration-matched U.S. Treasury bond rate of return. & \cite{vanBinsbergenNozawaSchwert_2025} & \href{https://openbondassetpricing.com/corporate-bond-factor-zoo/}{Open Source Bond Asset Pricing} \\
MOMB & Bond momentum factor formed with bond momentum. Independent sort ($5 \times 5$) to form 25 portfolios according to ratings and the 12-2 cumulative previous bond return (MOM). For each rating quintile, calculate the weighted average return difference between the highest MOM quintile and the lowest MOM quintile. MOMB is computed as the average long-short portfolio return across all rating quintiles. & \cite{GebhardtHvidkjaerSwaminathan_2005} & \href{https://openbondassetpricing.com/corporate-bond-factor-zoo/}{Open Source Bond Asset Pricing} \\
MOMBS & Bond momentum factor formed with equity momentum. Independent sort ($5 \times 5$) to form 25 portfolios according to ratings and the 6-1 cumulative previous equity return (MOMs). For each rating quintile, calculate the weighted average return difference between the highest MOMs quintile and the lowest MOMs quintile. MOMBS is computed as the average long-short portfolio return across all rating quintiles. & \cite{HottingaVanLeeuwenVanIjserloo_2001}, \cite{GebhardtHvidkjaerSwaminathan_2005} and \cite{DangHollsteinProkopczuk_2023} & \href{https://openbondassetpricing.com/corporate-bond-factor-zoo/}{Open Source Bond Asset Pricing} \\
PEADB & Bond earnings announcement drift factor. Independent sort ($2 \times 3$) to form 6 portfolios according to market equity and earnings surprises (CAR), computed according to \cite{ChanJegadeeshLakonishok_1996}. For each firm size portfolio, calculate the weighted average return difference between the highest CAR terciles and the lowest CAR tercile. PEADB is computed as the average long-short portfolio return across the two firm size portfolios. & \cite{NozawaQiuXiong_2025} & \href{https://openbondassetpricing.com/corporate-bond-factor-zoo/}{Open Source Bond Asset Pricing} \\
STREVB & Bond short-term reversal factor. Independent sort ($5 \times 5$) to form 25 portfolios according to ratings and the prior month's bond return (REV). For each rating quintile, calculate the weighted average return difference between the lowest REV quintile and the highest REV quintile. STREVB is computed as the average long-short portfolio return across all rating quintiles. & \cite{KhangKing_2004} and \cite{BaliSubrahmanyamWen_2021} & \href{https://openbondassetpricing.com/corporate-bond-factor-zoo/}{Open Source Bond Asset Pricing} \\
SZE & Bond size factor. Dependent sort ($3 \times 3$) to form 3 portfolios according to ratings and then with each rating tercile another 3 portfolios on bond size (SIZE). Bond size is defined as bond price multiplied by issue size (amount outstanding). For each rating tercile, calculate the weighted average return difference between the lowest SIZE tercile and the highest SIZE tercile. SZE is computed as the average long-short portfolio return across all rating terciles. & \cite{HottingaVanLeeuwenVanIjserloo_2001} and \cite{HouwelingVanZundert_2017} & \href{https://openbondassetpricing.com/corporate-bond-factor-zoo/}{Open Source Bond Asset Pricing} \\
TERM & Bond term structure risk factor. The difference between the monthly long-term government bond return and the one-month T-Bill rate of return. & \cite{FamaFrench_1992} and \cite{GebhardtHvidkjaerSwaminathan_2005a}. & \href{https://docs.google.com/spreadsheets/d/1g4LOaRj4TvwJr9RIaA_nwrXXWTOy46bP/edit#gid=2070662242}{Amit Goyal website} \\
VAL & Bond value factor. Independent sort ($2 \times 3$) to form 6 portfolios according to bond size and bond value (VAL$^B$). VAL$^B$ is computed via cross-sectional regressions of credit spreads on ratings, maturity, and the 3-month change in credit spread. The percentage difference between the actual credit spread and the fitted ('fair') credit spread for each bond is the VAL$^B$ characteristic. For each size portfolio, calculate the weighted average return difference between the highest VAL$^B$ tercile and the lowest VAL$^B$ tercile. VAL is computed as the average long-short portfolio return across the two size portfolios. & \cite{CorreiaRichardsonTuna_2012} and \cite{HouwelingVanZundert_2017} & \href{https://openbondassetpricing.com/corporate-bond-factor-zoo/}{Open Source Bond Asset Pricing} \\
\midrule
\multicolumn{4}{c}{ \textbf{Panel B: Tradable stock factors} } \\
\midrule
BAB		& Betting-against-beta factor, constructed as a portfolio that holds low-beta assets, leveraged to a beta of 1, and that shorts high-beta assets, de-leveraged to a beta of 1. & \cite{FrazziniPedersen_2014} & \href{https://www.aqr.com/Insights/Datasets}{AQR data library}\\
CMA		& Investment factor, constructed as a long-short portfolio of stocks sorted by their investment activity.  	& \cite{FamaFrench_2015}& \href{https://mba.tuck.dartmouth.edu/pages/faculty/ken.french/data_library.html}{Kenneth French website}\\
CMAs & CMA with a hedged unpriced component.& \cite{DanielMotaRottkeSantos_2020}& \href{http://www.kentdaniel.net/data.php}{Kent Daniel website}\\
CPTLT	& The value-weighted equity return for the New York Fed's primary dealer sector not including new equity issuance.  &  \cite{HeKellyManela_2017} & \href{https://voices.uchicago.edu/zhiguohe/data-and-empirical-patterns/intermediary-capital-ratio-and-risk-factor/}{Zhiguo He website}\\
FIN	& Long-term behavioral factor, predominantly capturing the impact of share issuance and correction. & \cite{DanielHirshleiferSun_2020} & \href{http://www.kentdaniel.net/data.php}{Kent Daniel website}\\
HML		& Value factor, constructed as a long-short portfolio of stocks sorted by their book-to-market ratio.  & \cite{FamaFrench_1992}& \href{https://mba.tuck.dartmouth.edu/pages/faculty/ken.french/data_library.html}{Kenneth French website}\\
HML\_DEV	& A version of the HML factor that relies on the current price level to sort the stocks into long and short legs. & \cite{AsnessFrazzini_2013}& \href{https://www.aqr.com/Insights/Datasets}{AQR data library}\\
HMLs & HML with a hedged unpriced component.& \cite{DanielMotaRottkeSantos_2020}& \href{http://www.kentdaniel.net/data.php}{Kent Daniel website}\\
LIQ	& Liquidity factor, constructed as a long-short portfolio of stocks sorted by their exposure to LIQ\_NT. & \cite{PastorStambaugh_2003}& \href{http://finance.wharton.upenn.edu/~stambaug/?_ga=2.74993170.1245141189.1568745149-1845734336.1509072906}{Robert Stambaugh website}\\
LTREV	& Long-term reversal factor, constructed as a long-short portfolio of stocks sorted by their cumulative return accrued in the previous 60-13 months.& \cite{JegadeeshTitman_2001}& \href{https://mba.tuck.dartmouth.edu/pages/faculty/ken.french/data_library.html}{Kenneth French website}\\
MGMT	& Management performance mispricing factor. &  \cite{StambaughYuan_2017}  & \href{https://jkpfactors.com/}{Global factor data website}\\
MKTS	& Market excess return. &  \cite{Sharpe_1964} and \cite{Lintner_1965} & \href{https://mba.tuck.dartmouth.edu/pages/faculty/ken.french/data_library.html}{Kenneth French website}\\
MKTSs & Market factor with a hedged unpriced component.& \cite{DanielMotaRottkeSantos_2020}&  \href{http://www.kentdaniel.net/data.php}{Kent Daniel website}\\
MOMS		& Momentum factor, constructed as a long-short portfolio of stocks sorted by their 12-2 months cumulative previous return.  & \cite{Carhart_1997}, \cite{JegadeeshTitman_1993}& \href{https://mba.tuck.dartmouth.edu/pages/faculty/ken.french/data_library.html}{Kenneth French website}\\
PEAD	& Short-term behavioral factor, reflecting post-earnings announcement drift. & \cite{DanielHirshleiferSun_2020} & \href{http://www.kentdaniel.net/data.php}{Kent Daniel website}\\
PERF	& Firm performance mispricing factor. & \cite{StambaughYuan_2017} & \href{https://jkpfactors.com/}{Global factor data website}\\
QMJ		& Quality-minus-junk factor, constructed as a long-short portfolio of stocks sorted by the combination of their safety, profitability, growth, and the quality of management practices.& \cite{AsnessFrazziniPedersen_2019}& \href{https://www.aqr.com/Insights/Datasets}{AQR data library}\\
RMW		& Profitability factor, constructed as a long-short portfolio of stocks sorted by their profitability.  & \cite{FamaFrench_2015}& \href{https://mba.tuck.dartmouth.edu/pages/faculty/ken.french/data_library.html}{Kenneth French website}\\
RMWs & RMW with a hedged unpriced component.& \cite{DanielMotaRottkeSantos_2020}& \href{http://www.kentdaniel.net/data.php}{Kent Daniel website}\\
R\_IA	& Investment factor, constructed as a long-short portfolio of stocks sorted by their investment-to-capital. & \cite{HouXueZhang_2015}& \href{http://global-q.org/index.html}{Lu Zhang website}\\
R\_ROE	& Profitability factor, constructed as a long-short portfolio of stocks sorted by their return on equity. & \cite{HouXueZhang_2015}&  \href{http://global-q.org/index.html}{Lu Zhang website}\\
SMB		& Size factor, constructed as a long-short portfolio of stocks sorted by their market cap.  & \cite{FamaFrench_1992} & \href{https://mba.tuck.dartmouth.edu/pages/faculty/ken.french/data_library.html}{Kenneth French website}\\
SMBs & SMB with a hedged unpriced component.& \cite{DanielMotaRottkeSantos_2020}& \href{http://www.kentdaniel.net/data.php}{Kent Daniel website}\\
STREV	& Short-term reversal factor, constructed as a long-short portfolio of stocks sorted by their previous month return. & \cite{JegadeeshTitman_1993}& \href{https://mba.tuck.dartmouth.edu/pages/faculty/ken.french/data_library.html}{Kenneth French website}\\
\midrule
\multicolumn{4}{c}{ \textbf{Panel C: Nontradable corporate bond and stock factors} } \\
\midrule
CPTL	& Intermediary capital nontradable risk factor. Constructed using AR(1) innovations to the market-based capital ratio of primary dealers, scaled by
the lagged capital ratio.  &  \cite{HeKellyManela_2017} & \href{https://voices.uchicago.edu/zhiguohe/data-and-empirical-patterns/intermediary-capital-ratio-and-risk-factor/}{Zhiguo He website}\\
CREDIT	& Bond credit risk factor. Difference between the yields of BAA and AAA indices from Moody's. Also computed with our own data as the difference between the average yield of BAA and (AAA$+$AA) rated bonds. See Internet Appendix~IA.11 for further computational details.  &  \cite{FamaFrench_1993} & \href{https://docs.google.com/spreadsheets/d/1g4LOaRj4TvwJr9RIaA_nwrXXWTOy46bP/edit#gid=2070662242}{Amit Goyal website} or FRED for \href{https://fred.stlouisfed.org/series/AAA}{AAA} and \href{https://fred.stlouisfed.org/series/BAA}{BAA} indices.\\
EPU	& Economic Policy Uncertainty. First difference in the economic policy uncertainty index.  &  \cite{BakerBloomDavis_2016} and \cite{DangHollsteinProkopczuk_2023} & \href{https://fred.stlouisfed.org/series/USEPUINDXM}{FRED}\\
EPUT	& Economic Tax Policy Uncertainty. First difference in the economic tax policy uncertainty index.  &  \cite{BakerBloomDavis_2016} and \cite{DangHollsteinProkopczuk_2023} & \href{https://fred.stlouisfed.org/series/EPUTAXES/?utm_source=rss&utm_medium=rss&utm_campaign=fred-updates}{FRED}\\
INFLC	& Shocks to core inflation. Unexpected core inflation component captured by an ARMA(1,1) model. Monthly core inflation is calculated as the percentage change in the seasonally adjusted Consumer Price Index for All Urban Consumers: All Items Less Food and Energy which is lagged by one-month to account for the inflation data release lag.  &  \cite{FangLiuRoussanov_2025} & \href{https://fred.stlouisfed.org/series/CPIAUCSL}{FRED}\\
INFLV	& Inflation volatility. Computed as the 6-month volatility of the unexpected inflation component captured by an ARMA(1,1) model. Monthly inflation is calculated as the percentage change in the seasonally adjusted Consumer Price Index for All Urban Consumers (CPI) which is lagged by one-month to account for the inflation data release lag.  &  \cite{KangPflueger_2015} and \cite{Ceballos_2023} & \href{https://fred.stlouisfed.org/series/CPIAUCSL}{FRED}\\
IVOL	& Idiosyncratic equity volatility factor. Cross-sectional volatility of all firms in the CRSP database in each month \textit{t}. &  \cite{CampbellTaksler_2003} & CRSP \\
LVL	& Level term structure factor. Constructed as the
first principal component of the one- through 30-year CRSP Fixed Term Indices U.S. Treasury Bond yields. &  \cite{KoijenLustigVanNieuwerburgh_2017} & CRSP Indices\\
LIQNT	& Liquidity factor, computed as the average of individual-stock measures estimated with daily data (residual predictability, controlling for the market factor)& \cite{PastorStambaugh_2003}& \href{http://finance.wharton.upenn.edu/~stambaug/?_ga=2.74993170.1245141189.1568745149-1845734336.1509072906}{Robert Stambaugh website}\\
UNC	& First difference in the Macroeconomic uncertainty index.  &  \cite{LudvigsonJuradoNg_2015} and \cite{BaliSubrahmanyamWen_2021_JFQA} & \href{https://www.sydneyludvigson.com/macro-and-financial-uncertainty-indexes}{Sydney Ludvigson website}\\
UNCf	& First difference in the Financial economic uncertainty index.  &  \cite{LudvigsonJuradoNg_2015} & \href{https://www.sydneyludvigson.com/macro-and-financial-uncertainty-indexes}{Sydney Ludvigson website}\\
UNCr	& First difference in the Real economic uncertainty index.  &  \cite{LudvigsonJuradoNg_2015} & \href{https://www.sydneyludvigson.com/macro-and-financial-uncertainty-indexes}{Sydney Ludvigson website}\\
VIX	& First difference in the CBOE VIX.  &  \cite{ChungWangWu_2019} & \href{https://fred.stlouisfed.org/series/VIXCLS}{FRED}\\
YSP	& Slope term structure factor. Constructed as the difference in the five and one-year U.S. Treasury Bond yields. &  \cite{KoijenLustigVanNieuwerburgh_2017} & CRSP Indices\\
\end{longtable}
\end{scriptsize}

\vspace{-.25cm}
\section{Posterior sampling}\label{post_samp}
\vspace{-.25cm}
The posterior of the time series parameters follows the canonical Normal-inverse-Wishart distribution (see, e.g., \citet*{BauwensLubranoRichard_1999}) given by
\begin{equation}\label{eq:muY}
	\bm{\mu_Y} | \bm{\Sigma_Y}, \mathbf{Y} \sim \normal \left(\bm{\hat{\mu}_Y}, \,\,\, \bm{\Sigma_Y}/T\right), 
\end{equation}
\begin{equation}\label{eq:SigmaY}
	\bm{\Sigma_Y} | \mathbf{Y} \sim \mathcal{W}^{-1} \left(T-1, \,\,\,\,\sum_{t=1}^T \left(\bm{Y_t}-\bm{\widehat \mu_Y}\right) \left(\bm{Y_t}-\bm{\widehat \mu_Y}\right)^\top\right), 
\end{equation}
where $\bm{\hat{\mu}_Y} \equiv \frac{1}{T} \sum_{t=1}^T \bm{Y_t}$, $\mathcal{W}^{-1}$ is the inverse-Wishart distribution, $\bm{Y} \equiv \{\bm{Y}_t\}_{t=1}^T$, and note that the covariance matrix of factors and test assets, $\bm{C_f}$, is contained within $\bm{\Sigma_Y}$.

Define $\bm{D} = \bm{\tilde{D}} \times \bm{\kappa}$  where $ \bm{\tilde{D}} $ is a diagonal matrix with elements $c$, $(r(\gamma_1) \psi_1)^{-1}$, ..., $(r(\gamma_{K}) \psi_{K})^{-1} $ and $\bm{\kappa}$ is a conformable column vector with elements $1, \, 1+\kappa_1, \, \dots, 1+\kappa_K$ such that $\sum_{k=1}^{K} \kappa_j =0$ and $0<|\kappa_j|<1$ $\forall j$.
It then follows that, given our prior formulations, the posterior distributions of the parameters in the cross-sectional layer $(\bm{\lambda}, \bm{\gamma}, \bm{\omega}, \sigma^2)$,  conditional on the draws of $\bm{\mu_R}$, $\bm{\Sigma_R}$, and $\bm{C}$ from the time series layer, are:
	\begin{equation}\label{lambda_post_continuous_gls}
		\bm{\lambda}|\text{data},\sigma^2,\bm{\gamma},\bm{\omega} \sim \normal\big(\bm{\hat{\lambda}},\hat{\sigma}^2 (\bm{\hat{\lambda}})\big),  
	\end{equation}
	\begin{equation}\label{BF_continuous}
		\frac{p(\gamma_j=1| \text{data},\bm{\lambda},\bm{\omega},\sigma^2,\bm{\gamma_{-j}})}{p(\gamma_j=0| \text{data},\bm{\lambda},\bm{\omega},\sigma^2,\bm{\gamma_{-j}})} = \frac{\omega_j}{1-\omega_j} \frac{p(\lambda_j|\gamma_j=1,\sigma^2)}{p(\lambda_j|\gamma_j=0,\sigma^2)}, 
	\end{equation}
	\begin{equation}\label{omega_post}
		\omega_j |\text{data},\bm{\lambda},\bm{\gamma},\sigma^2 \sim Beta\left(\gamma_j+a_\omega,1-\gamma_j+b_\omega\right),
	\end{equation} 
	\begin{equation}\label{sigma_post_continuous_gls}
		\sigma^2 | \text{data}, \bm{\omega},\bm{\lambda},\bm{\gamma} \sim \mathcal{IG} \left(\frac{N+K+1}{2},\frac{(\bm{\mu_R}-\bm{C} \bm{\lambda})^\top \bm{\Sigma_R}^{-1} (\bm{\mu_R}-\bm{C} \bm{\lambda}) + \bm{\lambda}^\top \bm{D} \bm{\lambda}}{2}\right),
	\end{equation}
	where $\bm{\hat{\lambda}}=(\bm{C}^\top \bm{\Sigma_R}^{-1} \bm{C} + \bm{D})^{-1} \bm{C}^\top \bm{\Sigma_R}^{-1} \bm{\mu_R}$, $\hat{\sigma}^2 (\bm{\hat{\lambda}})= \sigma^2 (\bm{C}^\top \bm{\Sigma_R}^{-1} \bm{C} + \bm{D})^{-1}$ and $\mathcal{IG}$ denotes the inverse-Gamma distribution.
	
Hence, posterior sampling is achieved with a Gibbs sampler that draws sequentially the time series layer parameters ($\bm{\mu_R}$, $\bm{\Sigma_R}$, and $\bm{C}$) from equations (\ref{eq:muY}) and (\ref{eq:SigmaY}), and then, conditional on these realizations, draws sequentially from equations (\ref{lambda_post_continuous_gls}) to (\ref{sigma_post_continuous_gls}).

\section{Probabilities and risk prices across prior Sharpe ratios}\label{sec:all_Sharpe_priors}

We report the full list of posterior probabilities and the associated annualized risk premia (in Sharpe ratio units) which complements the results from Figure~\ref{Fig:post_probs} in Table~\ref{tab:table-app-probs}.

\begin{table}[tbp!]  
\caption{Posterior factor probabilities and risk prices for the co-pricing factor zoo}\label{tab:table-app-probs}
\vspace{-.5cm}
	\begin{center}
		\scalebox{0.65}{
\begin{tabular}{lrrrrcrrrr} \toprule
& \multicolumn{4}{c}{Factor prob., $\mathbb{E}[\gamma_j|\text{data}]$} &  & \multicolumn{4}{c}{Price of risk,   $\mathbb{E}[\lambda_j|\text{data}]$} \\
\cmidrule(lr){2-5} \cmidrule(lr){7-10}
& \multicolumn{4}{c}{Total prior Sharpe ratio} &  & \multicolumn{4}{c}{Total prior Sharpe ratio} \\
Factors & 20\% & 40\% & 60\% & 80\% &  & 20\% & 40\% & 60\% & 80\% \\ \midrule
\rowcolor{gray!20}PEADB  & 0.555 & 0.629 & 0.713 & 0.711 &  & 0.054 & 0.213 & 0.446 & 0.645 \\
\rowcolor{gray!20}PEAD  & 0.523 & 0.559 & 0.618 & 0.614 &  & 0.035 & 0.138 & 0.297 & 0.449 \\
\rowcolor{gray!20}IVOL  & 0.502 & 0.529 & 0.567 & 0.623 &  & 0.010 & 0.043 & 0.108 & 0.265 \\
\rowcolor{gray!20}CREDIT  & 0.498 & 0.497 & 0.530 & 0.557 &  & 0.008 & 0.033 & 0.084 & 0.191 \\
\rowcolor{gray!20}YSP  & 0.507 & 0.502 & 0.504 & 0.519 &  & 0.003 & 0.014 & 0.034 & 0.088 \\
MOMBS  & 0.492 & 0.518 & 0.543 & 0.476 &  & 0.059 & 0.200 & 0.366 & 0.432 \\
INFLV  & 0.509 & 0.514 & 0.511 & 0.484 &  & 0.002 & 0.007 & 0.014 & 0.022 \\
INFLC  & 0.500 & 0.501 & 0.494 & 0.492 &  & $-$0.001 & $-$0.004 & $-$0.011 & $-$0.028 \\
CMAs  & 0.489 & 0.500 & 0.502 & 0.480 &  & 0.015 & 0.061 & 0.131 & 0.215 \\
LVL  & 0.495 & 0.493 & 0.491 & 0.493 &  & 0.000 & 0.002 & 0.006 & 0.019 \\
EPU  & 0.509 & 0.503 & 0.498 & 0.457 &  & 0.001 & 0.004 & 0.008 & 0.009 \\
UNCr  & 0.494 & 0.490 & 0.499 & 0.480 &  & 0.001 & 0.004 & 0.012 & 0.032 \\
MKTS  & 0.496 & 0.510 & 0.494 & 0.458 &  & 0.055 & 0.173 & 0.289 & 0.391 \\
EPUT  & 0.500 & 0.492 & 0.497 & 0.462 &  & 0.003 & 0.009 & 0.016 & 0.019 \\
LIQNT  & 0.501 & 0.482 & 0.492 & 0.475 &  & $-$0.003 & $-$0.013 & $-$0.039 & $-$0.095 \\
CRY  & 0.483 & 0.463 & 0.501 & 0.479 &  & 0.049 & 0.151 & 0.334 & 0.500 \\
QMJ  & 0.499 & 0.501 & 0.487 & 0.438 &  & 0.072 & 0.193 & 0.321 & 0.412 \\
RMWs  & 0.500 & 0.501 & 0.481 & 0.438 &  & 0.025 & 0.077 & 0.141 & 0.205 \\
UNCf  & 0.499 & 0.492 & 0.479 & 0.446 &  & $-$0.002 & $-$0.001 & 0.018 & 0.065 \\
UNC  & 0.487 & 0.484 & 0.480 & 0.445 &  & $-$0.001 & -0.000 & 0.005 & 0.014 \\
VIX  & 0.482 & 0.485 & 0.468 & 0.452 &  & 0.000 & 0.002 & 0.005 & 0.010 \\
SZE  & 0.502 & 0.465 & 0.464 & 0.421 &  & 0.006 & 0.026 & 0.061 & 0.104 \\
CPTL  & 0.487 & 0.480 & 0.457 & 0.411 &  & 0.016 & 0.046 & 0.067 & 0.074 \\
MKTB  & 0.521 & 0.482 & 0.439 & 0.376 &  & 0.091 & 0.188 & 0.248 & 0.278 \\
MKTSs  & 0.494 & 0.478 & 0.447 & 0.397 &  & 0.015 & 0.038 & 0.064 & 0.103 \\
LTREVB  & 0.500 & 0.482 & 0.437 & 0.387 &  & 0.016 & 0.051 & 0.079 & 0.094 \\
SMBs  & 0.491 & 0.476 & 0.450 & 0.384 &  & 0.004 & 0.016 & 0.029 & 0.034 \\
CPTLT  & 0.478 & 0.459 & 0.456 & 0.406 &  & 0.023 & 0.068 & 0.130 & 0.186 \\
LIQ  & 0.475 & 0.476 & 0.443 & 0.390 &  & 0.005 & 0.025 & 0.053 & 0.082 \\
BAB  & 0.485 & 0.492 & 0.435 & 0.372 &  & 0.021 & 0.054 & 0.076 & 0.097 \\
VAL  & 0.501 & 0.469 & 0.426 & 0.378 &  & 0.016 & 0.056 & 0.099 & 0.126 \\
STREV  & 0.487 & 0.476 & 0.445 & 0.365 &  & 0.009 & 0.034 & 0.071 & 0.101 \\
LTREV  & 0.498 & 0.473 & 0.432 & 0.357 &  & 0.009 & 0.031 & 0.052 & 0.057 \\
PERF  & 0.503 & 0.469 & 0.433 & 0.343 &  & 0.048 & 0.104 & 0.120 & 0.093 \\
R\_ROE  & 0.490 & 0.465 & 0.416 & 0.357 &  & 0.049 & 0.103 & 0.135 & 0.159 \\
MGMT  & 0.490 & 0.475 & 0.420 & 0.338 &  & 0.058 & 0.125 & 0.162 & 0.173 \\
CRF  & 0.494 & 0.454 & 0.421 & 0.349 &  & 0.015 & 0.052 & 0.093 & 0.123 \\
HMLs  & 0.478 & 0.461 & 0.411 & 0.357 &  & 0.004 & 0.011 & 0.021 & 0.026 \\
CMA  & 0.469 & 0.464 & 0.421 & 0.351 &  & 0.028 & 0.063 & 0.077 & 0.063 \\
HML\_DEV  & 0.492 & 0.446 & 0.414 & 0.353 &  & 0.001 & 0.002 & 0.014 & 0.041 \\
HMLB  & 0.475 & 0.464 & 0.438 & 0.326 &  & 0.038 & 0.104 & 0.148 & 0.120 \\
MOMB  & 0.472 & 0.459 & 0.424 & 0.346 &  & $-$0.002 & $-$0.007 & $-$0.005 & $-$0.003 \\
MOMS  & 0.464 & 0.445 & 0.422 & 0.365 &  & 0.020 & 0.057 & 0.095 & 0.139 \\
STREVB  & 0.478 & 0.449 & 0.414 & 0.349 &  & 0.003 & 0.007 & 0.011 & 0.007 \\
MKTBD  & 0.487 & 0.442 & 0.403 & 0.351 &  & 0.014 & 0.029 & 0.029 & 0.015 \\
R\_IA  & 0.473 & 0.437 & 0.418 & 0.349 &  & 0.034 & 0.079 & 0.120 & 0.140 \\
TERM  & 0.474 & 0.443 & 0.397 & 0.354 &  & 0.027 & 0.058 & 0.085 & 0.116 \\
SMB  & 0.476 & 0.434 & 0.410 & 0.331 &  & 0.010 & 0.044 & 0.079 & 0.086 \\
HML  & 0.477 & 0.435 & 0.405 & 0.327 &  & 0.003 & $-$0.016 & $-$0.037 & $-$0.040 \\
DUR  & 0.475 & 0.422 & 0.393 & 0.352 &  & 0.010 & $-$0.021 & $-$0.081 & $-$0.146 \\
DRF  & 0.471 & 0.435 & 0.401 & 0.330 &  & 0.039 & 0.068 & 0.069 & 0.034 \\
DEF  & 0.467 & 0.421 & 0.395 & 0.333 &  & 0.000 & $-$0.007 & $-$0.021 & $-$0.030 \\
FIN  & 0.476 & 0.424 & 0.392 & 0.311 &  & 0.034 & 0.035 & 0.015 & $-$0.004 \\
RMW  & 0.473 & 0.428 & 0.381 & 0.315 &  & 0.027 & 0.019 & $-$0.018 & $-$0.055 \\
\bottomrule
\end{tabular}
		}
	\end{center}
	\vspace{-0.2cm}
 	\begin{spacing}{1}
	{\footnotesize
The table reports posterior probabilities, $\mathbb{E}[\gamma_j|\text{data}]$, and posterior means of annualized market prices of risk, $\mathbb{E}[\lambda_j|\text{data}]$, of the 54 bond and stock factors described in \ref{sec:factor_zoo}. The prior for each factor inclusion is a Beta(1, 1), yielding a prior expectation for $\gamma_j$ of 50\%. Results are tabulated for different values of the prior Sharpe ratio, $\sqrt{\mathbb{E}_\pi [SR^2_{\bm{f}} \mid \sigma^2]}$, with values set to 20\%, 40\%, 60\% and 80\% of the ex post maximum Sharpe ratio of the test assets. The factors are ordered by the average posterior probability across the four levels of shrinkage. Test assets are the 83 bond and stock portfolios and 40 tradable bond and stock factors described in Section~\ref{sec:data}. The sample period is 1986:01 to 2022:12 ($T = 444$).    
}  
\end{spacing}
\end{table}

\section{Benchmark asset pricing models}\label{sec:benchmark_models}

We benchmark the performance of the BMA-SDF against several frequentist asset pricing models as well as other latent factor models. In the following, we provide the estimation details for the models that are compared to the BMA-SDF in Section~\ref{sec:Copricing}. A larger set of comparison benchmark models is considered in Internet Appendix~IA.3.2.

\paragraph{CAPM and CAPMB} The single-factor equity CAPM and the bond equivalent CAPMB. The CAPM is the value-weighted equity market factor from  \href{https://mba.tuck.dartmouth.edu/pages/faculty/ken.french/data_library.html}{Kenneth French's webpage.} The bond CAPM (CAPMB) is the value-weighted corporate bond market factor. We estimate factor risk prices using a GLS version of GMM (see, e.g., \citet[pp. 256--258]{Cochrane_2005}).

\paragraph{FF5} The original five-factor model of \cite{FamaFrench_1993} that includes the MKTS, SMB and HML factors from \cite{FamaFrench_1992} and the default (DEF) and term structure (TERM) factors introduced in \cite{FamaFrench_1993}. We estimate factor risk prices using a GLS version of GMM (see, e.g., \citet[pp. 256--258]{Cochrane_2005}).

\paragraph{HKM} The intermediary capital two-factor asset pricing model of \citet*{HeKellyManela_2017}. Includes the MKTS factor from \cite{FamaFrench_1992} and the value-weighted (tradable version) of the intermediary capital factor, CPTLT in excess of the one-month risk-free rate. We estimate factor risk prices using a GLS version of GMM (see, e.g., \citet[pp. 256--258]{Cochrane_2005}).

\paragraph{KNS} The latent factor model approach of \citet{KozakNagelSantosh_2020}.  For each in-sample bond, stock or co-pricing cross-section, we select the optimal shrinkage level and number of factors chosen by twofold cross-validation. Given our data has a time series length of $T = 444$, the first sample is simply January 1986 to June 2004 and the second sample is July 2004 to December 2022.

\paragraph{RPPCA} The risk premia PCA methodology of \cite{LettauPelger_2020}. We use five principal components. In our main estimation used for the baseline results, we set $\gamma$ from their equation (4) equal to 20. Changing this parameter to 10, or a lower value, does not quantitatively affect pricing performance.

\putbib
\end{bibunit}

\typeout{get arXiv to do 4 passes: Label(s) may have changed. Rerun}

\begin{thebibliography}{129}
\expandafter\ifx\csname natexlab\endcsname\relax\def\natexlab#1{#1}\fi
\providecommand{\url}[1]{\texttt{#1}}
\providecommand{\href}[2]{#2}
\providecommand{\path}[1]{#1}
\providecommand{\DOIprefix}{doi:}
\providecommand{\ArXivprefix}{arXiv:}
\providecommand{\URLprefix}{URL: }
\providecommand{\Pubmedprefix}{pmid:}
\providecommand{\doi}[1]{\href{http://dx.doi.org/#1}{\path{#1}}}
\providecommand{\Pubmed}[1]{\href{pmid:#1}{\path{#1}}}
\providecommand{\bibinfo}[2]{#2}
\ifx\xfnm\relax \def\xfnm[#1]{\unskip,\space#1}\fi
\bibitem[{Asness and Frazzini(2013)}]{AsnessFrazzini_2013}
\bibinfo{author}{Asness, C.}, \bibinfo{author}{Frazzini, A.},
  \bibinfo{year}{2013}.
\newblock \bibinfo{title}{The devil in {HML}'s details}.
\newblock \bibinfo{journal}{Journal of Portfolio Management}
  \bibinfo{volume}{39}, \bibinfo{pages}{49--68}.
\bibitem[{Asness et~al.(2019)Asness, Frazzini and
  Pedersen}]{AsnessFrazziniPedersen_2019}
\bibinfo{author}{Asness, C.S.}, \bibinfo{author}{Frazzini, A.},
  \bibinfo{author}{Pedersen, L.H.}, \bibinfo{year}{2019}.
\newblock \bibinfo{title}{Quality minus junk}.
\newblock \bibinfo{journal}{Review of Accounting Studies} \bibinfo{volume}{24},
  \bibinfo{pages}{34--112}.
\bibitem[{Avramov et~al.(2023)Avramov, Cheng, Metzker and
  Voigt}]{AvramovChengMetzkerVoigt_2023}
\bibinfo{author}{Avramov, D.}, \bibinfo{author}{Cheng, S.},
  \bibinfo{author}{Metzker, L.}, \bibinfo{author}{Voigt, S.},
  \bibinfo{year}{2023}.
\newblock \bibinfo{title}{Integrating factor models}.
\newblock \bibinfo{journal}{The Journal of Finance} \bibinfo{volume}{78},
  \bibinfo{pages}{1593--1646}.
\bibitem[{Bai et~al.(2019)Bai, Bali and Wen}]{BaiBaliWen_2019}
\bibinfo{author}{Bai, J.}, \bibinfo{author}{Bali, T.G.}, \bibinfo{author}{Wen,
  Q.}, \bibinfo{year}{2019}.
\newblock \bibinfo{title}{{RETRACTED}: Common risk factors in the cross-section
  of corporate bond returns}.
\newblock \bibinfo{journal}{Journal of Financial Economics}
  \bibinfo{volume}{131}, \bibinfo{pages}{619--642}.
\bibitem[{Baker et~al.(2016)Baker, Bloom and Davis}]{BakerBloomDavis_2016}
\bibinfo{author}{Baker, S.R.}, \bibinfo{author}{Bloom, N.},
  \bibinfo{author}{Davis, S.J.}, \bibinfo{year}{2016}.
\newblock \bibinfo{title}{Measuring economic policy uncertainty}.
\newblock \bibinfo{journal}{Quarterly Journal of Economics}
  \bibinfo{volume}{131}, \bibinfo{pages}{1593--1636}.
\bibitem[{Bali et~al.(2021a)Bali, Subrahmanyam and
  Wen}]{BaliSubrahmanyamWen_2021}
\bibinfo{author}{Bali, T.G.}, \bibinfo{author}{Subrahmanyam, A.},
  \bibinfo{author}{Wen, Q.}, \bibinfo{year}{2021}a.
\newblock \bibinfo{title}{Long-term reversals in the corporate bond market}.
\newblock \bibinfo{journal}{Journal of Financial Economics}
  \bibinfo{volume}{139}, \bibinfo{pages}{656--677}.
\bibitem[{Bali et~al.(2021b)Bali, Subrahmanyam and
  Wen}]{BaliSubrahmanyamWen_2021_JFQA}
\bibinfo{author}{Bali, T.G.}, \bibinfo{author}{Subrahmanyam, A.},
  \bibinfo{author}{Wen, Q.}, \bibinfo{year}{2021}b.
\newblock \bibinfo{title}{The macroeconomic uncertainty premium in the
  corporate bond market}.
\newblock \bibinfo{journal}{Journal of Financial and Quantitative Analysis}
  \bibinfo{volume}{56}, \bibinfo{pages}{1653--1678}.
\bibitem[{Bansal et~al.(2005)Bansal, Khatchatrian and
  Yaron}]{BansalKhatchatrianYaron_2005}
\bibinfo{author}{Bansal, R.}, \bibinfo{author}{Khatchatrian, V.},
  \bibinfo{author}{Yaron, A.}, \bibinfo{year}{2005}.
\newblock \bibinfo{title}{Interpretable asset markets?}
\newblock \bibinfo{journal}{European Economic Review} \bibinfo{volume}{49},
  \bibinfo{pages}{531--560}.
\bibitem[{Bansal et~al.(2012)Bansal, Kiku and Yaron}]{BansalKikuYaron_2012}
\bibinfo{author}{Bansal, R.}, \bibinfo{author}{Kiku, D.},
  \bibinfo{author}{Yaron, A.}, \bibinfo{year}{2012}.
\newblock \bibinfo{title}{An empirical evaluation of the long-run risks model
  for asset prices}.
\newblock \bibinfo{journal}{Critical Finance Review} \bibinfo{volume}{1},
  \bibinfo{pages}{183--221}.
\bibitem[{Barillas and Shanken(2017)}]{BarillasShanken_2017}
\bibinfo{author}{Barillas, F.}, \bibinfo{author}{Shanken, J.},
  \bibinfo{year}{2017}.
\newblock \bibinfo{title}{Which alpha?}
\newblock \bibinfo{journal}{The Review of Financial Studies}
  \bibinfo{volume}{30}, \bibinfo{pages}{1316--1338}.
\bibitem[{Barillas and Shanken(2018)}]{BarillasShanken_2018}
\bibinfo{author}{Barillas, F.}, \bibinfo{author}{Shanken, J.},
  \bibinfo{year}{2018}.
\newblock \bibinfo{title}{Comparing asset pricing models}.
\newblock \bibinfo{journal}{The Journal of Finance} \bibinfo{volume}{73},
  \bibinfo{pages}{715--754}.
\bibitem[{Bartram et~al.(2025)Bartram, Grinblatt and
  Nozawa}]{BartramGrinblattNozawa_2025}
\bibinfo{author}{Bartram, S.M.}, \bibinfo{author}{Grinblatt, M.},
  \bibinfo{author}{Nozawa, Y.}, \bibinfo{year}{2025}.
\newblock \bibinfo{title}{Book-to-market, mispricing, and the cross section of
  corporate bond returns}.
\newblock \bibinfo{journal}{Journal of Financial and Quantitative Analysis}
  \bibinfo{volume}{60}, \bibinfo{pages}{1185--1233}.
\bibitem[{Bauwens et~al.(1999)Bauwens, Lubrano and
  Richard}]{BauwensLubranoRichard_1999}
\bibinfo{author}{Bauwens, L.}, \bibinfo{author}{Lubrano, M.},
  \bibinfo{author}{Richard, J.F.}, \bibinfo{year}{1999}.
\newblock \bibinfo{title}{Bayesian Inference in Dynamic Econometric Models}.
\newblock \bibinfo{publisher}{Oxford University Press},
  \bibinfo{address}{Oxford}.
\bibitem[{Beeler and Campbell(2012)}]{BeelerCampbell_2012}
\bibinfo{author}{Beeler, J.}, \bibinfo{author}{Campbell, J.Y.},
  \bibinfo{year}{2012}.
\newblock \bibinfo{title}{The long-run risks model and aggregate asset prices:
  An empirical assessment}.
\newblock \bibinfo{journal}{Critical Finance Review} \bibinfo{volume}{1},
  \bibinfo{pages}{141--182}.
\bibitem[{Belloni et~al.(2014)Belloni, Chernozhukov and
  Hansen}]{BelloniChernozhukovHansen_2014}
\bibinfo{author}{Belloni, A.}, \bibinfo{author}{Chernozhukov, V.},
  \bibinfo{author}{Hansen, C.}, \bibinfo{year}{2014}.
\newblock \bibinfo{title}{Inference on treatment effects after selection among
  high-dimensional controls}.
\newblock \bibinfo{journal}{Review of Economic Studies} \bibinfo{volume}{81},
  \bibinfo{pages}{608--650}.
\bibitem[{Bhamra et~al.(2010)Bhamra, Kuehn and
  Strebulaev}]{BhamraKuehnStrebualev_2010}
\bibinfo{author}{Bhamra, H.S.}, \bibinfo{author}{Kuehn, L.A.},
  \bibinfo{author}{Strebulaev, I.A.}, \bibinfo{year}{2010}.
\newblock \bibinfo{title}{The levered equity risk premium and credit spreads: A
  unified framework}.
\newblock \bibinfo{journal}{The Review of Financial Studies}
  \bibinfo{volume}{23}, \bibinfo{pages}{645--703}.
\bibitem[{van Binsbergen et~al.(2025)van Binsbergen, Nozawa and
  Schwert}]{vanBinsbergenNozawaSchwert_2025}
\bibinfo{author}{van Binsbergen, J.H.}, \bibinfo{author}{Nozawa, Y.},
  \bibinfo{author}{Schwert, M.}, \bibinfo{year}{2025}.
\newblock \bibinfo{title}{Duration-based valuation of corporate bonds}.
\newblock \bibinfo{journal}{The Review of Financial Studies}
  \bibinfo{volume}{38}, \bibinfo{pages}{158--191}.
\bibitem[{Blume and Keim(1987)}]{BlumeKeim_1987}
\bibinfo{author}{Blume, M.E.}, \bibinfo{author}{Keim, D.B.},
  \bibinfo{year}{1987}.
\newblock \bibinfo{title}{Lower-grade bonds: Their risks and returns}.
\newblock \bibinfo{journal}{Financial Analysts Journal} \bibinfo{volume}{43},
  \bibinfo{pages}{26--66}.
\bibitem[{Bollerslev(1986)}]{Bollerslev_1986}
\bibinfo{author}{Bollerslev, T.}, \bibinfo{year}{1986}.
\newblock \bibinfo{title}{Generalized autoregressive conditional
  heteroskedasticity}.
\newblock \bibinfo{journal}{Journal of Econometrics} \bibinfo{volume}{31},
  \bibinfo{pages}{307--327}.
\bibitem[{Bollerslev and Wooldridge(1992)}]{BollerslevWooldridge_1992}
\bibinfo{author}{Bollerslev, T.}, \bibinfo{author}{Wooldridge, J.M.},
  \bibinfo{year}{1992}.
\newblock \bibinfo{title}{Quasi-maximum likelihood estimation and inference in
  dynamic models with time-varying covariances}.
\newblock \bibinfo{journal}{Econometric Reviews} \bibinfo{volume}{11},
  \bibinfo{pages}{143--172}.
\newblock \DOIprefix\doi{10.1080/07474939208800229}.
\bibitem[{Bryzgalova et~al.(2023)Bryzgalova, Huang and
  Julliard}]{BryzgalovaHuangJulliard_2023}
\bibinfo{author}{Bryzgalova, S.}, \bibinfo{author}{Huang, J.},
  \bibinfo{author}{Julliard, C.}, \bibinfo{year}{2023}.
\newblock \bibinfo{title}{Bayesian solutions for the factor zoo: {W}e just ran
  two quadrillion models}.
\newblock \bibinfo{journal}{The Journal of Finance} \bibinfo{volume}{78},
  \bibinfo{pages}{487--557}.
\bibitem[{Bryzgalova et~al.(2024)Bryzgalova, Huang and
  Julliard}]{BryzgalovaHuangJulliard_2024}
\bibinfo{author}{Bryzgalova, S.}, \bibinfo{author}{Huang, J.},
  \bibinfo{author}{Julliard, C.}, \bibinfo{year}{2024}.
\newblock \bibinfo{title}{Macro strikes back: Term structure of risk premia and
  market segmentation}.
\newblock \bibinfo{note}{Working Paper, London School of Economics}.
\bibitem[{Campbell and Shiller(1988)}]{CampbellShiller_1988_RFS}
\bibinfo{author}{Campbell, J.Y.}, \bibinfo{author}{Shiller, R.J.},
  \bibinfo{year}{1988}.
\newblock \bibinfo{title}{The dividend-price ratio and expectations of future
  dividends and discount factors}.
\newblock \bibinfo{journal}{The Review of Financial Studies}
  \bibinfo{volume}{1}, \bibinfo{pages}{195--228}.
\bibitem[{Campbell and Taksler(2003)}]{CampbellTaksler_2003}
\bibinfo{author}{Campbell, J.Y.}, \bibinfo{author}{Taksler, G.B.},
  \bibinfo{year}{2003}.
\newblock \bibinfo{title}{Equity volatility and corporate bond yields}.
\newblock \bibinfo{journal}{The Journal of Finance} \bibinfo{volume}{58},
  \bibinfo{pages}{2321--2349}.
\bibitem[{Carhart(1997)}]{Carhart_1997}
\bibinfo{author}{Carhart, M.M.}, \bibinfo{year}{1997}.
\newblock \bibinfo{title}{On persistence in mutual fund performance}.
\newblock \bibinfo{journal}{The Journal of Finance} \bibinfo{volume}{52},
  \bibinfo{pages}{57--82}.
\bibitem[{Ceballos(2023)}]{Ceballos_2023}
\bibinfo{author}{Ceballos, L.}, \bibinfo{year}{2023}.
\newblock \bibinfo{title}{Inflation volatility risk and the cross-section of
  corporate bond returns}.
\newblock \bibinfo{note}{Working Paper, University of San Diego}.
\bibitem[{Chan et~al.(1996)Chan, Jegadeesh and
  Lakonishok}]{ChanJegadeeshLakonishok_1996}
\bibinfo{author}{Chan, L.K.}, \bibinfo{author}{Jegadeesh, N.},
  \bibinfo{author}{Lakonishok, J.}, \bibinfo{year}{1996}.
\newblock \bibinfo{title}{Momentum strategies}.
\newblock \bibinfo{journal}{The Journal of Finance} \bibinfo{volume}{51},
  \bibinfo{pages}{1681--1713}.
\bibitem[{Chen(2017)}]{Chen_2017}
\bibinfo{author}{Chen, A.Y.}, \bibinfo{year}{2017}.
\newblock \bibinfo{title}{{External Habit in a Production Economy: A Model of
  Asset Prices and Consumption Volatility Risk}}.
\newblock \bibinfo{journal}{The Review of Financial Studies}
  \bibinfo{volume}{30}, \bibinfo{pages}{2890--2932}.
\bibitem[{Chen and Zimmermann(2022)}]{ChenZimmermann_2021}
\bibinfo{author}{Chen, A.Y.}, \bibinfo{author}{Zimmermann, T.},
  \bibinfo{year}{2022}.
\newblock \bibinfo{title}{Open source cross-sectional asset pricing}.
\newblock \bibinfo{journal}{Critical Finance Review} \bibinfo{volume}{27},
  \bibinfo{pages}{207--264}.
\bibitem[{Chen et~al.(2018)Chen, Cui, He and Milbradt}]{ChenCuiHeMilbradt_2018}
\bibinfo{author}{Chen, H.}, \bibinfo{author}{Cui, R.}, \bibinfo{author}{He,
  Z.}, \bibinfo{author}{Milbradt, K.}, \bibinfo{year}{2018}.
\newblock \bibinfo{title}{{Quantifying Liquidity and Default Risks of Corporate
  Bonds over the Business Cycle}}.
\newblock \bibinfo{journal}{The Review of Financial Studies}
  \bibinfo{volume}{31}, \bibinfo{pages}{852--897}.
\bibitem[{Chen and Zhao(2009)}]{ChenZhao_2009}
\bibinfo{author}{Chen, L.}, \bibinfo{author}{Zhao, X.}, \bibinfo{year}{2009}.
\newblock \bibinfo{title}{Return decomposition}.
\newblock \bibinfo{journal}{The Review of Financial Studies}
  \bibinfo{volume}{22}, \bibinfo{pages}{5213--5249}.
\bibitem[{Chen et~al.(2024)Chen, Roussanov, Wang and
  Zou}]{ChenRoussanovWangZou_2024}
\bibinfo{author}{Chen, Z.}, \bibinfo{author}{Roussanov, N.L.},
  \bibinfo{author}{Wang, X.}, \bibinfo{author}{Zou, D.}, \bibinfo{year}{2024}.
\newblock \bibinfo{title}{Common risk factors in the returns on stocks, bonds
  (and options), redux}.
\newblock \bibinfo{note}{Working Paper, The Wharton School}.
\bibitem[{Chib et~al.(2020)Chib, Zeng and Zhao}]{ChibZengZhao_2020}
\bibinfo{author}{Chib, S.}, \bibinfo{author}{Zeng, X.}, \bibinfo{author}{Zhao,
  L.}, \bibinfo{year}{2020}.
\newblock \bibinfo{title}{On comparing asset pricing models}.
\newblock \bibinfo{journal}{The Journal of Finance} \bibinfo{volume}{75},
  \bibinfo{pages}{551--577}.
\bibitem[{Choi and Kim(2018)}]{ChoiKim_2018}
\bibinfo{author}{Choi, J.}, \bibinfo{author}{Kim, Y.}, \bibinfo{year}{2018}.
\newblock \bibinfo{title}{Anomalies and market (dis)integration}.
\newblock \bibinfo{journal}{Journal of Monetary Economics}
  \bibinfo{volume}{100}, \bibinfo{pages}{16--34}.
\bibitem[{Chordia et~al.(2017)Chordia, Goyal, Nozawa, Subrahmanyam and
  Tong}]{ChordiaGoyalNozawaSubrahmanyamTong_2017}
\bibinfo{author}{Chordia, T.}, \bibinfo{author}{Goyal, A.},
  \bibinfo{author}{Nozawa, Y.}, \bibinfo{author}{Subrahmanyam, A.},
  \bibinfo{author}{Tong, Q.}, \bibinfo{year}{2017}.
\newblock \bibinfo{title}{Are capital market anomalies common to equity and
  corporate bond markets? {A}n empirical investigation}.
\newblock \bibinfo{journal}{Journal of Financial and Quantitative Analysis}
  \bibinfo{volume}{52}, \bibinfo{pages}{1301--1342}.
\bibitem[{Chung et~al.(2019)Chung, Wang and Wu}]{ChungWangWu_2019}
\bibinfo{author}{Chung, K.H.}, \bibinfo{author}{Wang, J.}, \bibinfo{author}{Wu,
  C.}, \bibinfo{year}{2019}.
\newblock \bibinfo{title}{Volatility and the cross-section of corporate bond
  returns}.
\newblock \bibinfo{journal}{Journal of Financial Economics}
  \bibinfo{volume}{133}, \bibinfo{pages}{397--417}.
\bibitem[{Cochrane(2005)}]{Cochrane_2005}
\bibinfo{author}{Cochrane, J.H.}, \bibinfo{year}{2005}.
\newblock \bibinfo{title}{Asset Pricing}. volume~\bibinfo{volume}{1}.
\newblock \bibinfo{publisher}{Princeton University Press Princeton, NJ}.
\bibitem[{Cochrane(2011)}]{Cochrane_2011_Pres}
\bibinfo{author}{Cochrane, J.H.}, \bibinfo{year}{2011}.
\newblock \bibinfo{title}{Presidential address: {D}iscount rate}.
\newblock \bibinfo{journal}{The Journal of Finance} \bibinfo{volume}{66},
  \bibinfo{pages}{1047--1108}.
\bibitem[{Cohen et~al.(2002)Cohen, Gompers and
  Vuolteenaho}]{CohenGompersVuolteenaho_2002}
\bibinfo{author}{Cohen, R.B.}, \bibinfo{author}{Gompers, P.A.},
  \bibinfo{author}{Vuolteenaho, T.}, \bibinfo{year}{2002}.
\newblock \bibinfo{title}{Who underreacts to cash-flow news? {E}vidence from
  trading between individuals and institutions}.
\newblock \bibinfo{journal}{Journal of Financial Economics}
  \bibinfo{volume}{66}, \bibinfo{pages}{409--462}.
\bibitem[{Correia et~al.(2012)Correia, Richardson and
  Tuna}]{CorreiaRichardsonTuna_2012}
\bibinfo{author}{Correia, M.}, \bibinfo{author}{Richardson, S.},
  \bibinfo{author}{Tuna, {\.I}.}, \bibinfo{year}{2012}.
\newblock \bibinfo{title}{Value investing in credit markets}.
\newblock \bibinfo{journal}{Review of Accounting Studies} \bibinfo{volume}{17},
  \bibinfo{pages}{572--609}.
\bibitem[{Dang et~al.(2023)Dang, Hollstein and
  Prokopczuk}]{DangHollsteinProkopczuk_2023}
\bibinfo{author}{Dang, T.D.}, \bibinfo{author}{Hollstein, F.},
  \bibinfo{author}{Prokopczuk, M.}, \bibinfo{year}{2023}.
\newblock \bibinfo{title}{Which factors for corporate bond returns?}
\newblock \bibinfo{journal}{The Review of Asset Pricing Studies}
  \bibinfo{volume}{13}, \bibinfo{pages}{615--652}.
\bibitem[{Daniel et~al.(2020a)Daniel, Hirshleifer and
  Sun}]{DanielHirshleiferSun_2020}
\bibinfo{author}{Daniel, K.}, \bibinfo{author}{Hirshleifer, D.},
  \bibinfo{author}{Sun, L.}, \bibinfo{year}{2020}a.
\newblock \bibinfo{title}{Short- and long-horizon behavioral factors}.
\newblock \bibinfo{journal}{The Review of Financial Studies}
  \bibinfo{volume}{33}, \bibinfo{pages}{1673--1736}.
\bibitem[{Daniel et~al.(2020b)Daniel, Mota, Rottke and
  Santos}]{DanielMotaRottkeSantos_2020}
\bibinfo{author}{Daniel, K.}, \bibinfo{author}{Mota, L.},
  \bibinfo{author}{Rottke, S.}, \bibinfo{author}{Santos, T.},
  \bibinfo{year}{2020}b.
\newblock \bibinfo{title}{The cross-section of risk and returns}.
\newblock \bibinfo{journal}{The Review of Financial Studies}
  \bibinfo{volume}{33}, \bibinfo{pages}{1927--1979}.
\bibitem[{De~Long et~al.(1990)De~Long, Shleifer, Summers and
  Waldman}]{DeLongShleiferSummersWaldmann_1990}
\bibinfo{author}{De~Long, B.}, \bibinfo{author}{Shleifer, A.},
  \bibinfo{author}{Summers, L.C.}, \bibinfo{author}{Waldman, R.},
  \bibinfo{year}{1990}.
\newblock \bibinfo{title}{Noise trader risk in financial markets}.
\newblock \bibinfo{journal}{Journal of Political Economy} \bibinfo{volume}{98},
  \bibinfo{pages}{703--738}.
\bibitem[{Delao et~al.(2025)Delao, Han and Myers}]{DelaoHanMyers_2025}
\bibinfo{author}{Delao, R.}, \bibinfo{author}{Han, X.}, \bibinfo{author}{Myers,
  S.}, \bibinfo{year}{2025}.
\newblock \bibinfo{title}{The return of return dominance: Decomposing the
  cross-section of prices}.
\newblock \bibinfo{journal}{Journal of Financial Economics}
  \bibinfo{volume}{169}, \bibinfo{pages}{104059}.
\bibitem[{Della~Vigna and Pollet(2009)}]{DellaVignaPollet_2009}
\bibinfo{author}{Della~Vigna, S.}, \bibinfo{author}{Pollet, J.M.},
  \bibinfo{year}{2009}.
\newblock \bibinfo{title}{Investor inattention and {F}riday earnings
  announcements}.
\newblock \bibinfo{journal}{The Journal of Finance} \bibinfo{volume}{64},
  \bibinfo{pages}{709--749}.
\bibitem[{Dello~Preite et~al.(2025)Dello~Preite, Uppal, Zaffaroni and
  Zviadadze}]{DelloPreiteUppalZaffaroniZviadadze_2024}
\bibinfo{author}{Dello~Preite, M.}, \bibinfo{author}{Uppal, R.},
  \bibinfo{author}{Zaffaroni, P.}, \bibinfo{author}{Zviadadze, I.},
  \bibinfo{year}{2025}.
\newblock \bibinfo{title}{Cross-sectional asset pricing with unsystematic
  risk}.
\newblock \bibinfo{note}{Working Paper, EDHEC Business School}.
\bibitem[{DeMiguel et~al.(2009)DeMiguel, Garlappi and
  Uppal}]{DeMiguelGarlappiUppal_2009}
\bibinfo{author}{DeMiguel, V.}, \bibinfo{author}{Garlappi, L.},
  \bibinfo{author}{Uppal, R.}, \bibinfo{year}{2009}.
\newblock \bibinfo{title}{Optimal versus naive diversification: How inefficient
  is the {1/N} portfolio strategy?}
\newblock \bibinfo{journal}{The Review of Financial studies}
  \bibinfo{volume}{22}, \bibinfo{pages}{1915--1953}.
\bibitem[{Dick-Nielsen et~al.(2025)Dick-Nielsen, Feldh{\"u}tter, Pedersen and
  Stolborg}]{DickNielsenFeldhuetterPedersenStolborg_2024}
\bibinfo{author}{Dick-Nielsen, J.}, \bibinfo{author}{Feldh{\"u}tter, P.},
  \bibinfo{author}{Pedersen, L.H.}, \bibinfo{author}{Stolborg, C.},
  \bibinfo{year}{2025}.
\newblock \bibinfo{title}{Corporate bond factors: Replication failures and a
  new framework}.
\newblock \bibinfo{note}{Working Paper, Copenhagen Business School}.
\bibitem[{Dickerson et~al.(2025)Dickerson, Fournier, Jeanneret and
  Mueller}]{DickersonFournierJeanneretMueller_2025}
\bibinfo{author}{Dickerson, A.}, \bibinfo{author}{Fournier, M.},
  \bibinfo{author}{Jeanneret, A.}, \bibinfo{author}{Mueller, P.},
  \bibinfo{year}{2025}.
\newblock \bibinfo{title}{A credit risk explanation of the correlation between
  corporate bonds and stocks}.
\newblock \bibinfo{note}{Working Paper, UNSW}.
\bibitem[{Dickerson et~al.(2023)Dickerson, Mueller and
  Robotti}]{DickersonMuellerRobotti_2023}
\bibinfo{author}{Dickerson, A.}, \bibinfo{author}{Mueller, P.},
  \bibinfo{author}{Robotti, C.}, \bibinfo{year}{2023}.
\newblock \bibinfo{title}{Priced risk in corporate bonds}.
\newblock \bibinfo{journal}{Journal of Financial Economics}
  \bibinfo{volume}{150}, \bibinfo{pages}{103707}.
\bibitem[{Dickerson et~al.(2024)Dickerson, Robotti and
  Rossetti}]{DickersonRobottiRossetti_2024}
\bibinfo{author}{Dickerson, A.}, \bibinfo{author}{Robotti, C.},
  \bibinfo{author}{Rossetti, G.}, \bibinfo{year}{2024}.
\newblock \bibinfo{title}{Common pitfalls in the evaluation of corporate bond
  strategies}.
\newblock \bibinfo{note}{Working Paper, Warwick Business School}.
\bibitem[{Duarte et~al.(2025)Duarte, Jones, Mo and
  Khorram}]{DuarteJonesMoKhorram_2024}
\bibinfo{author}{Duarte, J.}, \bibinfo{author}{Jones, C.S.},
  \bibinfo{author}{Mo, H.}, \bibinfo{author}{Khorram, M.},
  \bibinfo{year}{2025}.
\newblock \bibinfo{title}{Too good to be true: Look-ahead bias in empirical
  option research}.
\newblock \bibinfo{note}{Working Paper, USC Marshall}.
\bibitem[{Elkamhi et~al.(2023)Elkamhi, Jo and Nozawa}]{ElkamhiJoNozawa_2023}
\bibinfo{author}{Elkamhi, R.}, \bibinfo{author}{Jo, C.},
  \bibinfo{author}{Nozawa, Y.}, \bibinfo{year}{2023}.
\newblock \bibinfo{title}{A one-factor model of corporate bond premia}.
\newblock \bibinfo{journal}{Management Science} \bibinfo{volume}{70},
  \bibinfo{pages}{1875--1900}.
\bibitem[{Elton et~al.(2001)Elton, Gruber, Agrawal and
  Mann}]{EltonGruberAgrawalMann_2001}
\bibinfo{author}{Elton, E.J.}, \bibinfo{author}{Gruber, M.J.},
  \bibinfo{author}{Agrawal, D.}, \bibinfo{author}{Mann, C.},
  \bibinfo{year}{2001}.
\newblock \bibinfo{title}{Explaining the rate spread on corporate bonds}.
\newblock \bibinfo{journal}{The Journal of Finance} \bibinfo{volume}{56},
  \bibinfo{pages}{247--277}.
\bibitem[{Elton et~al.(1995)Elton, Gruber and Blake}]{EltonGruberBlake_1995}
\bibinfo{author}{Elton, E.J.}, \bibinfo{author}{Gruber, M.J.},
  \bibinfo{author}{Blake, C.R.}, \bibinfo{year}{1995}.
\newblock \bibinfo{title}{Fundamental economic variables, expected returns, and
  bond fund performance}.
\newblock \bibinfo{journal}{The Journal of Finance} \bibinfo{volume}{50},
  \bibinfo{pages}{1229--1256}.
\bibitem[{Engle(1982)}]{Engle_1982}
\bibinfo{author}{Engle, R.F.}, \bibinfo{year}{1982}.
\newblock \bibinfo{title}{Autoregressive conditional heteroskedasticity with
  estimates of the variance of united kingdom inflation}.
\newblock \bibinfo{journal}{Econometrica} \bibinfo{volume}{50},
  \bibinfo{pages}{987--1007}.
\bibitem[{Fama and French(1992)}]{FamaFrench_1992}
\bibinfo{author}{Fama, E.F.}, \bibinfo{author}{French, K.R.},
  \bibinfo{year}{1992}.
\newblock \bibinfo{title}{The cross-section of expected stock returns}.
\newblock \bibinfo{journal}{The Journal of Finance} \bibinfo{volume}{47},
  \bibinfo{pages}{427--465}.
\bibitem[{Fama and French(1993)}]{FamaFrench_1993}
\bibinfo{author}{Fama, E.F.}, \bibinfo{author}{French, K.R.},
  \bibinfo{year}{1993}.
\newblock \bibinfo{title}{Common risk factors in the returns on stocks and
  bonds}.
\newblock \bibinfo{journal}{Journal of Financial Economics}
  \bibinfo{volume}{33}, \bibinfo{pages}{3--56}.
\bibitem[{Fama and French(2015)}]{FamaFrench_2015}
\bibinfo{author}{Fama, E.F.}, \bibinfo{author}{French, K.R.},
  \bibinfo{year}{2015}.
\newblock \bibinfo{title}{A five-factor asset pricing model}.
\newblock \bibinfo{journal}{Journal of Financial Economics}
  \bibinfo{volume}{116}, \bibinfo{pages}{1--22}.
\bibitem[{Fang et~al.(2024)Fang, Liu and Roussanov}]{FangLiuRoussanov_2025}
\bibinfo{author}{Fang, X.}, \bibinfo{author}{Liu, Y.},
  \bibinfo{author}{Roussanov, N.}, \bibinfo{year}{2024}.
\newblock \bibinfo{title}{Getting to the core: Inflation risks within and
  across asset classes}.
\newblock \bibinfo{journal}{forthcoming, Review of Financial Studies} .
\bibitem[{Favilukis et~al.(2020)Favilukis, Lin and
  Zhao}]{FavilukisLinZhao_2020}
\bibinfo{author}{Favilukis, J.}, \bibinfo{author}{Lin, X.},
  \bibinfo{author}{Zhao, X.}, \bibinfo{year}{2020}.
\newblock \bibinfo{title}{The elephant in the room: The impact of labor
  obligations on credit markets}.
\newblock \bibinfo{journal}{American Economic Review} \bibinfo{volume}{110},
  \bibinfo{pages}{1673--1712}.
\bibitem[{Feng et~al.(2020)Feng, Giglio and Xiu}]{FengGiglioXiu_2020}
\bibinfo{author}{Feng, G.}, \bibinfo{author}{Giglio, S.}, \bibinfo{author}{Xiu,
  D.}, \bibinfo{year}{2020}.
\newblock \bibinfo{title}{Taming the factor zoo: A test of new factors}.
\newblock \bibinfo{journal}{The Journal of Finance} \bibinfo{volume}{75},
  \bibinfo{pages}{1327--1370}.
\bibitem[{Fisher(1959)}]{Fisher_1959}
\bibinfo{author}{Fisher, L.}, \bibinfo{year}{1959}.
\newblock \bibinfo{title}{Determinants of risk premiums on corporate bonds}.
\newblock \bibinfo{journal}{Journal of Political Economy} \bibinfo{volume}{67},
  \bibinfo{pages}{217--237}.
\bibitem[{Frazzini and Pedersen(2014)}]{FrazziniPedersen_2014}
\bibinfo{author}{Frazzini, A.}, \bibinfo{author}{Pedersen, L.H.},
  \bibinfo{year}{2014}.
\newblock \bibinfo{title}{Betting against beta}.
\newblock \bibinfo{journal}{Journal of Financial Economics}
  \bibinfo{volume}{111}, \bibinfo{pages}{1--25}.
\bibitem[{Gebhardt et~al.(2005a)Gebhardt, Hvidkjaer and
  Swaminathan}]{GebhardtHvidkjaerSwaminathan_2005a}
\bibinfo{author}{Gebhardt, W.R.}, \bibinfo{author}{Hvidkjaer, S.},
  \bibinfo{author}{Swaminathan, B.}, \bibinfo{year}{2005}a.
\newblock \bibinfo{title}{The cross-section of expected corporate bond returns:
  Betas or characteristics?}
\newblock \bibinfo{journal}{Journal of Financial Economics}
  \bibinfo{volume}{75}, \bibinfo{pages}{85--114}.
\bibitem[{Gebhardt et~al.(2005b)Gebhardt, Hvidkjaer and
  Swaminathan}]{GebhardtHvidkjaerSwaminathan_2005}
\bibinfo{author}{Gebhardt, W.R.}, \bibinfo{author}{Hvidkjaer, S.},
  \bibinfo{author}{Swaminathan, B.}, \bibinfo{year}{2005}b.
\newblock \bibinfo{title}{Stock and bond market interaction: Does momentum
  spill over?}
\newblock \bibinfo{journal}{Journal of Financial Economics}
  \bibinfo{volume}{75}, \bibinfo{pages}{651--690}.
\bibitem[{Gebhardt et~al.(2001)Gebhardt, Lee and
  Swaminathan}]{GebhardtLeeSwaminathan_2001}
\bibinfo{author}{Gebhardt, W.R.}, \bibinfo{author}{Lee, C.M.C.},
  \bibinfo{author}{Swaminathan, B.}, \bibinfo{year}{2001}.
\newblock \bibinfo{title}{Toward an implied cost of capital}.
\newblock \bibinfo{journal}{Journal of Accounting Research}
  \bibinfo{volume}{39}, \bibinfo{pages}{135--176}.
\bibitem[{Giesecke et~al.(2011)Giesecke, Longstaff, Schaefer and
  Strebulaev}]{GieseckeLongstaffSchaeferStrebulaev_2011}
\bibinfo{author}{Giesecke, K.}, \bibinfo{author}{Longstaff, F.A.},
  \bibinfo{author}{Schaefer, S.}, \bibinfo{author}{Strebulaev, I.},
  \bibinfo{year}{2011}.
\newblock \bibinfo{title}{Corporate bond default risk: A 150-year perspective}.
\newblock \bibinfo{journal}{Journal of Financial Economics}
  \bibinfo{volume}{102}, \bibinfo{pages}{233--250}.
\bibitem[{Giglio and Xiu(2021)}]{GiglioXiu_2018}
\bibinfo{author}{Giglio, S.}, \bibinfo{author}{Xiu, D.}, \bibinfo{year}{2021}.
\newblock \bibinfo{title}{Asset pricing with omitted factors}.
\newblock \bibinfo{journal}{Journal of Political Economy}
  \bibinfo{volume}{129}, \bibinfo{pages}{1947--1990}.
\bibitem[{Gomes and Schmid(2021)}]{GomesSchmid_2021}
\bibinfo{author}{Gomes, J.F.}, \bibinfo{author}{Schmid, L.},
  \bibinfo{year}{2021}.
\newblock \bibinfo{title}{Equilibrium asset pricing with leverage and default}.
\newblock \bibinfo{journal}{The Journal of Finance} \bibinfo{volume}{76},
  \bibinfo{pages}{977--1018}.
\bibitem[{Gospodinov et~al.(2014)Gospodinov, Kan and
  Robotti}]{GospodinovKanRobotti_2014}
\bibinfo{author}{Gospodinov, N.}, \bibinfo{author}{Kan, R.},
  \bibinfo{author}{Robotti, C.}, \bibinfo{year}{2014}.
\newblock \bibinfo{title}{Misspecification-robust inference in linear
  asset-pricing models with irrelevant risk factors}.
\newblock \bibinfo{journal}{The Review of Financial Studies}
  \bibinfo{volume}{27}, \bibinfo{pages}{2139--2170}.
\bibitem[{Gospodinov et~al.(2019)Gospodinov, Kan and
  Robotti}]{GospodinovKanRobotti_2019}
\bibinfo{author}{Gospodinov, N.}, \bibinfo{author}{Kan, R.},
  \bibinfo{author}{Robotti, C.}, \bibinfo{year}{2019}.
\newblock \bibinfo{title}{Too good to be true? fallacies in evaluating risk
  factor models}.
\newblock \bibinfo{journal}{Journal of Financial Economics}
  \bibinfo{volume}{132}, \bibinfo{pages}{451--471}.
\bibitem[{Gospodinov and Robotti(2021)}]{GospodinovRobotti_2021}
\bibinfo{author}{Gospodinov, N.}, \bibinfo{author}{Robotti, C.},
  \bibinfo{year}{2021}.
\newblock \bibinfo{title}{Common pricing across asset classes: Empirical
  evidence revisited}.
\newblock \bibinfo{journal}{Journal of Financial Economics}
  \bibinfo{volume}{140}, \bibinfo{pages}{292--324}.
\bibitem[{Hansen and Jagannathan(1991)}]{HansenJagannathan_1991}
\bibinfo{author}{Hansen, L.}, \bibinfo{author}{Jagannathan, R.},
  \bibinfo{year}{1991}.
\newblock \bibinfo{title}{Implications of security market data for models of
  dynamic economies}.
\newblock \bibinfo{journal}{Journal of Political Economy} \bibinfo{volume}{99},
  \bibinfo{pages}{225--262}.
\bibitem[{Hansen(1982)}]{Hansen_1982}
\bibinfo{author}{Hansen, L.P.}, \bibinfo{year}{1982}.
\newblock \bibinfo{title}{Large sample properties of method of moments
  estimators}.
\newblock \bibinfo{journal}{Econometrica} \bibinfo{volume}{50},
  \bibinfo{pages}{1029--1054}.
\bibitem[{Harvey(2017)}]{Harvey_2017}
\bibinfo{author}{Harvey, C.R.}, \bibinfo{year}{2017}.
\newblock \bibinfo{title}{Presidential address: The scientific outlook in
  financial economics}.
\newblock \bibinfo{journal}{The Journal of Finance} \bibinfo{volume}{72},
  \bibinfo{pages}{1399--1440}.
\bibitem[{Harvey et~al.(2016)Harvey, Liu and Zhu}]{HarveyLiuZhu_2016}
\bibinfo{author}{Harvey, C.R.}, \bibinfo{author}{Liu, Y.},
  \bibinfo{author}{Zhu, H.}, \bibinfo{year}{2016}.
\newblock \bibinfo{title}{...and the cross-section of expected returns}.
\newblock \bibinfo{journal}{The Review of Financial Studies}
  \bibinfo{volume}{29}, \bibinfo{pages}{5--68}.
\bibitem[{Hayashi(2000)}]{Hayashi_2000}
\bibinfo{author}{Hayashi, F.}, \bibinfo{year}{2000}.
\newblock \bibinfo{title}{Econometrics}.
\newblock \bibinfo{publisher}{Princeton University Press},
  \bibinfo{address}{Princeton, NJ}.
\bibitem[{He et~al.(2017)He, Kelly and Manela}]{HeKellyManela_2017}
\bibinfo{author}{He, Z.}, \bibinfo{author}{Kelly, B.}, \bibinfo{author}{Manela,
  A.}, \bibinfo{year}{2017}.
\newblock \bibinfo{title}{Intermediary asset pricing: New evidence from many
  asset classes}.
\newblock \bibinfo{journal}{Journal of Financial Economics}
  \bibinfo{volume}{126}, \bibinfo{pages}{1--35}.
\bibitem[{Heyerdahl-Larsen et~al.(2023)Heyerdahl-Larsen, Illeditsch and
  Walden}]{HeyerdahlLarsenIlleditschWalden_2023}
\bibinfo{author}{Heyerdahl-Larsen, C.}, \bibinfo{author}{Illeditsch, P.K.},
  \bibinfo{author}{Walden, J.}, \bibinfo{year}{2023}.
\newblock \bibinfo{title}{Model selection by market selection}.
\newblock \bibinfo{note}{SSRN Working Paper No 4401170}.
\bibitem[{Hirshleifer et~al.(2011)Hirshleifer, Lim and
  Teoh}]{HirshleiferLimTeoh_2011}
\bibinfo{author}{Hirshleifer, D.}, \bibinfo{author}{Lim, S.S.},
  \bibinfo{author}{Teoh, S.H.}, \bibinfo{year}{2011}.
\newblock \bibinfo{title}{Limited investor attention and stock market
  misreactions to accounting information}.
\newblock \bibinfo{journal}{Review of Asset Pricing Studies}
  \bibinfo{volume}{1}, \bibinfo{pages}{35--73}.
\bibitem[{Hirshleifer and Teoh(2003)}]{HirshleiferTeoh_2003}
\bibinfo{author}{Hirshleifer, D.}, \bibinfo{author}{Teoh, S.H.},
  \bibinfo{year}{2003}.
\newblock \bibinfo{title}{Limited attention, information disclosure, and
  financial reporting}.
\newblock \bibinfo{journal}{Journal of Accounting and Economics}
  \bibinfo{volume}{36}, \bibinfo{pages}{337--386}.
\bibitem[{Hoeting et~al.(1999)Hoeting, Madigan, Raftery and
  Volinsky}]{HoetingMadiganRafteryVolinsky_1999}
\bibinfo{author}{Hoeting, J.A.}, \bibinfo{author}{Madigan, D.},
  \bibinfo{author}{Raftery, A.E.}, \bibinfo{author}{Volinsky, C.T.},
  \bibinfo{year}{1999}.
\newblock \bibinfo{title}{Bayesian model averaging: A tutorial}.
\newblock \bibinfo{journal}{Statistical Science} \bibinfo{volume}{14},
  \bibinfo{pages}{382--401}.
\bibitem[{Hottinga et~al.(2001)Hottinga, van Leeuwen and van
  Ijserloo}]{HottingaVanLeeuwenVanIjserloo_2001}
\bibinfo{author}{Hottinga, J.}, \bibinfo{author}{van Leeuwen, E.},
  \bibinfo{author}{van Ijserloo, J.}, \bibinfo{year}{2001}.
\newblock \bibinfo{title}{Successful factors to select outperforming corporate
  bonds}.
\newblock \bibinfo{journal}{Journal of Portfolio Management}
  \bibinfo{volume}{28}, \bibinfo{pages}{88--101}.
\bibitem[{Hou et~al.(2015)Hou, Xue and Zhang}]{HouXueZhang_2015}
\bibinfo{author}{Hou, K.}, \bibinfo{author}{Xue, C.}, \bibinfo{author}{Zhang,
  L.}, \bibinfo{year}{2015}.
\newblock \bibinfo{title}{Digesting anomalies: An investment approach}.
\newblock \bibinfo{journal}{The Review of Financial Studies}
  \bibinfo{volume}{28}, \bibinfo{pages}{650--705}.
\bibitem[{Houweling and Van~Zundert(2017)}]{HouwelingVanZundert_2017}
\bibinfo{author}{Houweling, P.}, \bibinfo{author}{Van~Zundert, J.},
  \bibinfo{year}{2017}.
\newblock \bibinfo{title}{Factor investing in the corporate bond market}.
\newblock \bibinfo{journal}{Financial Analysts Journal} \bibinfo{volume}{73},
  \bibinfo{pages}{100--115}.
\bibitem[{Jeffreys(1961)}]{Jeffreys_1961}
\bibinfo{author}{Jeffreys, H.}, \bibinfo{year}{1961}.
\newblock \bibinfo{title}{Theory of Probability}.
\newblock \bibinfo{edition}{3rd} ed., \bibinfo{publisher}{Oxford University
  Press}, \bibinfo{address}{Oxford}.
\bibitem[{Jegadeesh and Titman(1993)}]{JegadeeshTitman_1993}
\bibinfo{author}{Jegadeesh, N.}, \bibinfo{author}{Titman, S.},
  \bibinfo{year}{1993}.
\newblock \bibinfo{title}{Returns to buying winners and selling losers:
  Implications for stock market efficiency}.
\newblock \bibinfo{journal}{The Journal of Finance} \bibinfo{volume}{48},
  \bibinfo{pages}{65--91}.
\bibitem[{Jegadeesh and Titman(2001)}]{JegadeeshTitman_2001}
\bibinfo{author}{Jegadeesh, N.}, \bibinfo{author}{Titman, S.},
  \bibinfo{year}{2001}.
\newblock \bibinfo{title}{Profitability of momentum strategies: An evaluation
  of alternative explanations}.
\newblock \bibinfo{journal}{The Journal of Finance} \bibinfo{volume}{56},
  \bibinfo{pages}{699--720}.
\bibitem[{Jensen et~al.(2023)Jensen, Kelly and
  Pedersen}]{JensenKellyPedersen_2023}
\bibinfo{author}{Jensen, T.I.}, \bibinfo{author}{Kelly, B.},
  \bibinfo{author}{Pedersen, L.H.}, \bibinfo{year}{2023}.
\newblock \bibinfo{title}{Is there a replication crisis in finance?}
\newblock \bibinfo{journal}{The Journal of Finance} \bibinfo{volume}{78},
  \bibinfo{pages}{2465--2518}.
\bibitem[{Kan and Zhang(1999a)}]{KanZhang_1999gmm}
\bibinfo{author}{Kan, R.}, \bibinfo{author}{Zhang, C.}, \bibinfo{year}{1999}a.
\newblock \bibinfo{title}{{GMM} tests of stochastic discount factor models with
  useless factors}.
\newblock \bibinfo{journal}{Journal of Financial Economics}
  \bibinfo{volume}{54}, \bibinfo{pages}{103--127}.
\bibitem[{Kan and Zhang(1999b)}]{KanZhang_1999two}
\bibinfo{author}{Kan, R.}, \bibinfo{author}{Zhang, C.}, \bibinfo{year}{1999}b.
\newblock \bibinfo{title}{Two-pass tests of asset pricing models with useless
  factors}.
\newblock \bibinfo{journal}{The Journal of Finance} \bibinfo{volume}{54},
  \bibinfo{pages}{203--235}.
\bibitem[{Kang and Pflueger(2015)}]{KangPflueger_2015}
\bibinfo{author}{Kang, J.}, \bibinfo{author}{Pflueger, C.E.},
  \bibinfo{year}{2015}.
\newblock \bibinfo{title}{Inflation risk in corporate bonds}.
\newblock \bibinfo{journal}{The Journal of Finance} \bibinfo{volume}{70},
  \bibinfo{pages}{115--162}.
\bibitem[{Khan and Thomas(2013)}]{KhanThomas_2013}
\bibinfo{author}{Khan, A.}, \bibinfo{author}{Thomas, J.K.},
  \bibinfo{year}{2013}.
\newblock \bibinfo{title}{Credit shocks and aggregate fluctuations in an
  economy with production heterogeneity}.
\newblock \bibinfo{journal}{Journal of Political Economy}
  \bibinfo{volume}{121}, \bibinfo{pages}{1055--1107}.
\bibitem[{Khang and King(2004)}]{KhangKing_2004}
\bibinfo{author}{Khang, K.}, \bibinfo{author}{King, T.H.D.},
  \bibinfo{year}{2004}.
\newblock \bibinfo{title}{Return reversals in the bond market: Evidence and
  causes}.
\newblock \bibinfo{journal}{Journal of Banking and Finance}
  \bibinfo{volume}{28}, \bibinfo{pages}{569--593}.
\bibitem[{Kleibergen(2009)}]{Kleibergen_2009}
\bibinfo{author}{Kleibergen, F.}, \bibinfo{year}{2009}.
\newblock \bibinfo{title}{Tests of risk premia in linear factor models}.
\newblock \bibinfo{journal}{Journal of Econometrics} \bibinfo{volume}{149},
  \bibinfo{pages}{149--173}.
\bibitem[{Kleibergen and Zhan(2020)}]{KleibergenZhan_2020}
\bibinfo{author}{Kleibergen, F.}, \bibinfo{author}{Zhan, Z.},
  \bibinfo{year}{2020}.
\newblock \bibinfo{title}{Robust inference for consumption-based asset
  pricing}.
\newblock \bibinfo{journal}{The Journal of Finance} \bibinfo{volume}{75},
  \bibinfo{pages}{507--550}.
\bibitem[{Koijen et~al.(2017)Koijen, Lustig and
  Van~Nieuwerburgh}]{KoijenLustigVanNieuwerburgh_2017}
\bibinfo{author}{Koijen, R.S.}, \bibinfo{author}{Lustig, H.},
  \bibinfo{author}{Van~Nieuwerburgh, S.}, \bibinfo{year}{2017}.
\newblock \bibinfo{title}{The cross-section and time series of stock and bond
  returns}.
\newblock \bibinfo{journal}{Journal of Monetary Economics}
  \bibinfo{volume}{88}, \bibinfo{pages}{50--69}.
\bibitem[{Koijen and Van~Nieuwerburgh(2011)}]{KoijenVanNieuwerburgh_2011}
\bibinfo{author}{Koijen, R.S.}, \bibinfo{author}{Van~Nieuwerburgh, S.},
  \bibinfo{year}{2011}.
\newblock \bibinfo{title}{Predictability of returns and cash flows}.
\newblock \bibinfo{journal}{Annual Review of Financial Economics}
  \bibinfo{volume}{3}, \bibinfo{pages}{467--491}.
\bibitem[{Kozak et~al.(2020)Kozak, Nagel and Santosh}]{KozakNagelSantosh_2020}
\bibinfo{author}{Kozak, S.}, \bibinfo{author}{Nagel, S.},
  \bibinfo{author}{Santosh, S.}, \bibinfo{year}{2020}.
\newblock \bibinfo{title}{Shrinking the cross-section}.
\newblock \bibinfo{journal}{Journal of Financial Economics}
  \bibinfo{volume}{135}, \bibinfo{pages}{271--292}.
\bibitem[{Lettau et~al.(2014)Lettau, Maggiori and
  Weber}]{LettauMaggioriWeber_2014}
\bibinfo{author}{Lettau, M.}, \bibinfo{author}{Maggiori, M.},
  \bibinfo{author}{Weber, M.}, \bibinfo{year}{2014}.
\newblock \bibinfo{title}{Conditional risk premia in currency markets and other
  asset classes}.
\newblock \bibinfo{journal}{Journal of Financial Economics}
  \bibinfo{volume}{114}, \bibinfo{pages}{197--225}.
\bibitem[{Lettau and Pelger(2020)}]{LettauPelger_2020}
\bibinfo{author}{Lettau, M.}, \bibinfo{author}{Pelger, M.},
  \bibinfo{year}{2020}.
\newblock \bibinfo{title}{Estimating latent asset-pricing factors}.
\newblock \bibinfo{journal}{Journal of Econometrics} \bibinfo{volume}{218},
  \bibinfo{pages}{1--31}.
\bibitem[{Lewellen et~al.(2010)Lewellen, Nagel and
  Shanken}]{LewellenNagelShanken_2010}
\bibinfo{author}{Lewellen, J.}, \bibinfo{author}{Nagel, S.},
  \bibinfo{author}{Shanken, J.}, \bibinfo{year}{2010}.
\newblock \bibinfo{title}{A skeptical appraisal of asset pricing tests}.
\newblock \bibinfo{journal}{Journal of Financial Economics}
  \bibinfo{volume}{96}, \bibinfo{pages}{175--194}.
\bibitem[{Lin et~al.(2011)Lin, Wang and Wu}]{LinWangWu_2011}
\bibinfo{author}{Lin, H.}, \bibinfo{author}{Wang, J.}, \bibinfo{author}{Wu,
  C.}, \bibinfo{year}{2011}.
\newblock \bibinfo{title}{Liquidity risk and expected corporate bond returns}.
\newblock \bibinfo{journal}{Journal of Financial Economics}
  \bibinfo{volume}{99}, \bibinfo{pages}{628--650}.
\bibitem[{Lintner(1965)}]{Lintner_1965}
\bibinfo{author}{Lintner, J.}, \bibinfo{year}{1965}.
\newblock \bibinfo{title}{Security prices, risk, and maximal gains from
  diversification}.
\newblock \bibinfo{journal}{The Journal of Finance} \bibinfo{volume}{20},
  \bibinfo{pages}{587--615}.
\bibitem[{Liu and Wu(2021)}]{LiuWu_2021}
\bibinfo{author}{Liu, Y.}, \bibinfo{author}{Wu, J.C.}, \bibinfo{year}{2021}.
\newblock \bibinfo{title}{Reconstructing the yield curve}.
\newblock \bibinfo{journal}{Journal of Financial Economics}
  \bibinfo{volume}{142}, \bibinfo{pages}{1395--1425}.
\bibitem[{Ljung and Box(1978)}]{LjungBox_1978}
\bibinfo{author}{Ljung, G.M.}, \bibinfo{author}{Box, G.E.P.},
  \bibinfo{year}{1978}.
\newblock \bibinfo{title}{On a measure of lack of fit in time series models}.
\newblock \bibinfo{journal}{Biometrika} \bibinfo{volume}{65},
  \bibinfo{pages}{297--303}.
\bibitem[{Ludvigson et~al.(2015)Ludvigson, Jurado and
  Ng}]{LudvigsonJuradoNg_2015}
\bibinfo{author}{Ludvigson, S.C.}, \bibinfo{author}{Jurado, K.},
  \bibinfo{author}{Ng, S.}, \bibinfo{year}{2015}.
\newblock \bibinfo{title}{Measuring uncertainty}.
\newblock \bibinfo{journal}{The American Economic Review}
  \bibinfo{volume}{105}, \bibinfo{pages}{1177--1216}.
\bibitem[{Martineau(2022)}]{Martineau_2022}
\bibinfo{author}{Martineau, C.}, \bibinfo{year}{2022}.
\newblock \bibinfo{title}{Rest in peace post-earnings announcement drift}.
\newblock \bibinfo{journal}{Critical Finance Review} \bibinfo{volume}{11},
  \bibinfo{pages}{613--646}.
\bibitem[{McCullough(1830)}]{McCullough_1830}
\bibinfo{author}{McCullough, J.R.}, \bibinfo{year}{1830}.
\newblock \bibinfo{title}{The Principles of Political Economy: With a Sketch of
  the Rise and Progress of the Science (2nd ed.)}.
\newblock \bibinfo{publisher}{Edinburgh, London, and Dublin}.
\bibitem[{Merton(1974)}]{Merton_1974}
\bibinfo{author}{Merton, R.C.}, \bibinfo{year}{1974}.
\newblock \bibinfo{title}{On the pricing of corporate debt: The risk structure
  of interest rates}.
\newblock \bibinfo{journal}{The Journal of Finance} \bibinfo{volume}{29},
  \bibinfo{pages}{449--470}.
\bibitem[{Newey and McFadden(1994)}]{NeweyMcFadden_1994}
\bibinfo{author}{Newey, W.K.}, \bibinfo{author}{McFadden, D.},
  \bibinfo{year}{1994}.
\newblock \bibinfo{title}{Large sample estimation and hypothesis testing}, in:
  \bibinfo{editor}{Engle, R.F.}, \bibinfo{editor}{McFadden, D.} (Eds.),
  \bibinfo{booktitle}{Handbook of Econometrics}. \bibinfo{publisher}{Elsevier
  Press}. volume~\bibinfo{volume}{4}, pp. \bibinfo{pages}{2111--2245}.
\bibitem[{Newey and West(1987)}]{NeweyWest_1987}
\bibinfo{author}{Newey, W.K.}, \bibinfo{author}{West, K.D.},
  \bibinfo{year}{1987}.
\newblock \bibinfo{title}{A simple, positive semi-definite, heteroskedasticity
  and autocorrelation consistent covariance matrix}.
\newblock \bibinfo{journal}{Econometrica} \bibinfo{volume}{55},
  \bibinfo{pages}{703--708}.
\bibitem[{Nozawa et~al.(2025)Nozawa, Qiu and Xiong}]{NozawaQiuXiong_2025}
\bibinfo{author}{Nozawa, Y.}, \bibinfo{author}{Qiu, Y.},
  \bibinfo{author}{Xiong, Y.}, \bibinfo{year}{2025}.
\newblock \bibinfo{title}{Disagreement and bond pead}.
\newblock \bibinfo{note}{Working Paper, University of Toronto}.
\bibitem[{Parker and Julliard(2003)}]{ParkerJulliard_2003}
\bibinfo{author}{Parker, J.A.}, \bibinfo{author}{Julliard, C.},
  \bibinfo{year}{2003}.
\newblock \bibinfo{title}{Consumption Risk and Cross-Sectional Returns}.
\newblock \bibinfo{type}{Working Paper} \bibinfo{number}{9538}. National Bureau
  of Economic Research.
\bibitem[{Parker and Julliard(2005)}]{ParkerJulliard_2005}
\bibinfo{author}{Parker, J.A.}, \bibinfo{author}{Julliard, C.},
  \bibinfo{year}{2005}.
\newblock \bibinfo{title}{Consumption risk and the cross section of expected
  returns}.
\newblock \bibinfo{journal}{Journal of Political Economy}
  \bibinfo{volume}{113}, \bibinfo{pages}{185--222}.
\bibitem[{P{\'a}stor(2000)}]{Pastor_2000}
\bibinfo{author}{P{\'a}stor, L.}, \bibinfo{year}{2000}.
\newblock \bibinfo{title}{Portfolio selection and asset pricing models}.
\newblock \bibinfo{journal}{The Journal of Finance} \bibinfo{volume}{55},
  \bibinfo{pages}{179--223}.
\bibitem[{P\'{a}stor and Stambaugh(2000)}]{PastorStambaugh_2000}
\bibinfo{author}{P\'{a}stor, L.}, \bibinfo{author}{Stambaugh, R.F.},
  \bibinfo{year}{2000}.
\newblock \bibinfo{title}{Comparing asset pricing models: {An} investment
  perspective}.
\newblock \bibinfo{journal}{Journal of Financial Economics}
  \bibinfo{volume}{56}, \bibinfo{pages}{335--381}.
\bibitem[{P\'{a}stor and Stambaugh(2003)}]{PastorStambaugh_2003}
\bibinfo{author}{P\'{a}stor, L.}, \bibinfo{author}{Stambaugh, R.F.},
  \bibinfo{year}{2003}.
\newblock \bibinfo{title}{Liquidity risk and expected stock returns}.
\newblock \bibinfo{journal}{Journal of Political Economy}
  \bibinfo{volume}{111}, \bibinfo{pages}{642--685}.
\bibitem[{Penman and Yehuda(2019)}]{PenmanNir_2019}
\bibinfo{author}{Penman, S.H.}, \bibinfo{author}{Yehuda, N.},
  \bibinfo{year}{2019}.
\newblock \bibinfo{title}{A matter of principle: Accounting reports convey both
  cash-flow news and discount-rate news}.
\newblock \bibinfo{journal}{Management Science} \bibinfo{volume}{65},
  \bibinfo{pages}{5584--5602}.
\bibitem[{Raftery et~al.(1997)Raftery, Madigan and
  Hoeting}]{HoetingMadiganRaftery_1997}
\bibinfo{author}{Raftery, A.E.}, \bibinfo{author}{Madigan, D.},
  \bibinfo{author}{Hoeting, J.A.}, \bibinfo{year}{1997}.
\newblock \bibinfo{title}{Bayesian model averaging for linear regression
  models}.
\newblock \bibinfo{journal}{Journal of the American Statistical Association}
  \bibinfo{volume}{92}, \bibinfo{pages}{179--191}.
\bibitem[{Raftery and Zheng(2003)}]{Raftery_2003}
\bibinfo{author}{Raftery, A.E.}, \bibinfo{author}{Zheng, Y.},
  \bibinfo{year}{2003}.
\newblock \bibinfo{title}{Discussion: Performance of {Bayesian} model
  averaging}.
\newblock \bibinfo{journal}{Journal of the American Statistical Association}
  \bibinfo{volume}{98}, \bibinfo{pages}{931--938}.
\bibitem[{Sandulescu(2022)}]{Sandulescu_2022}
\bibinfo{author}{Sandulescu, M.}, \bibinfo{year}{2022}.
\newblock \bibinfo{title}{How integrated are corporate bond and stock markets?}
\newblock \bibinfo{note}{Working Paper, Ross School of Business}.
\bibitem[{Schervish(1995)}]{Schervish_1995}
\bibinfo{author}{Schervish, M.J.}, \bibinfo{year}{1995}.
\newblock \bibinfo{title}{Theory of Statistics}.
\newblock Springer Series in Statistics, \bibinfo{publisher}{Springer-Verlag}.
\newblock \URLprefix \url{http://books.google.com/books?id=F9A9af4It10C}.
\bibitem[{Sharpe(1964)}]{Sharpe_1964}
\bibinfo{author}{Sharpe, W.F.}, \bibinfo{year}{1964}.
\newblock \bibinfo{title}{Capital asset prices: A theory of market equilibrium
  under conditions of risk}.
\newblock \bibinfo{journal}{The Journal of Finance} \bibinfo{volume}{19},
  \bibinfo{pages}{425--442}.
\bibitem[{Stambaugh and Yuan(2017)}]{StambaughYuan_2017}
\bibinfo{author}{Stambaugh, R.F.}, \bibinfo{author}{Yuan, Y.},
  \bibinfo{year}{2017}.
\newblock \bibinfo{title}{Mispricing factors}.
\newblock \bibinfo{journal}{The Review of Financial Studies}
  \bibinfo{volume}{30}, \bibinfo{pages}{1270--1315}.
\bibitem[{Vuolteenaho(2002)}]{Vuolteenaho_2002}
\bibinfo{author}{Vuolteenaho, T.}, \bibinfo{year}{2002}.
\newblock \bibinfo{title}{What drives firm-level stock returns?}
\newblock \bibinfo{journal}{The Journal of Finance} \bibinfo{volume}{57},
  \bibinfo{pages}{233--264}.
\bibitem[{Zviadadze(2021)}]{Zviadadze_2021}
\bibinfo{author}{Zviadadze, I.}, \bibinfo{year}{2021}.
\newblock \bibinfo{title}{Term structure of risk in expected returns}.
\newblock \bibinfo{journal}{The Review of Financial Studies}
  \bibinfo{volume}{34}, \bibinfo{pages}{6032--6086}.

\end{thebibliography}
\end{document}